\documentclass[twocolumn,trackchanges]{aastex63}
\usepackage{amssymb}
\usepackage{graphicx}
\usepackage{grffile}
\usepackage{epsfig}
\usepackage{epstopdf}
\usepackage[T1]{fontenc}

\usepackage{url}
\usepackage{newtxtext,newtxmath}
\usepackage[T1]{fontenc}

\usepackage{graphicx}	
\usepackage{amsmath}	
\usepackage{amssymb}
\usepackage{amsmath}
\usepackage{tablefootnote}
\usepackage{hyperref}
\usepackage[flushleft]{threeparttable}
\usepackage{booktabs}
\usepackage{float}

\usepackage{savesym}
\savesymbol{tablenum}
\usepackage{siunitx}
\restoresymbol{SIX}{tablenum}

\shorttitle{Direct gas metallicity and C/O abundance at $z\sim4$}
\shortauthors{A.\ Citro et al.}

\turnoffeditone

\begin{document}

\defcitealias{Jaskot_Ravindranath2016}{JR16}

\title{\large {A comprehensive metallicity analysis of J0332-3557: establishing a $z\sim4$ anchor for direct gas metallicity and C/O abundance investigations}}

\correspondingauthor{Annalisa Citro}
\email{acitro@uwm.edu}

\author[0009-0000-9676-0538]{Annalisa Citro}
\affil{Minnesota Institute for Astrophysics, School of Physics and Astronomy, University of Minnesota, 316 Church Str. SE, Minneapolis, MN 55455, USA}

\author[0000-0001-9714-2758]{Danielle A. Berg}
\affiliation{Department of Astronomy, The University of Texas at Austin, 2515 Speedway, Stop C1400, Austin, TX 78712, USA}

\author[0000-0001-9714-2758]{Dawn K. Erb}
\affiliation{The Leonard E.\ Parker Center for Gravitation, Cosmology and Astrophysics, Department of Physics,\\ University of Wisconsin-Milwaukee, 3135 N Maryland Avenue, Milwaukee, WI 53211, USA}

\author{Matthew W. Auger}
\affiliation{Institute of Astronomy, University of Cambridge, Madingley Road, Cambridge, CB3 0HA, UK}

\author[0000-0003-2344-263X]{George D. Becker}
\affiliation{Department of Physics \& Astronomy, University of California, Riverside, CA 92521, USA}

\author[0000-0003-4372-2006]{Bethan L. James}
\affiliation{AURA for ESA, Space Telescope Science Institute, 3700 San Martin Drive, Baltimore, MD 21218}

\author[0000-0003-0605-8732]{Evan D. Skillman}
\affil{Minnesota Institute for Astrophysics, School of Physics and Astronomy, University of Minnesota, 316 Church Str. SE, Minneapolis, MN 55455, USA}

\newcommand{\OIIIuv}{\ion{O}{3}] $\lambda$1666}
\newcommand{\CIII}{\ion{C}{3}] $\lambda$1906,9}
\newcommand{\OIIIopt}{[\ion{O}{3}] $\lambda$5007}
\newcommand{\OIIopt}{[\ion{O}{2}] $\lambda$3727}

\newcommand{\Oratio}{[\ion{O}{3}]/[\ion{O}{2}]}
\newcommand{\OIIIHB}{[\ion{O}{3}]/H{$\beta$}}
\newcommand{\OIIHB}{[\ion{O}{2}]/H{$\beta$}}
\newcommand{\NIIHA}{[ion{N}{2}]/H{$\alpha$}}
\newcommand{\NII}{[\ion{N}{2}]}
\newcommand{\HII}{\ion{H}{2}}
\newcommand{\SII}{[\ion{S}{2}]}

\newcommand{\wa}{$\si\angstrom$}
\newcommand{\Z}{$\rm Z_{\star}$}
\newcommand{\Zg}{$\rm Z_{gas}$}
\newcommand{\MZRg}{$\rm MZR_{g}$}
\newcommand{\ewciii}{$\rm EW_{C~III]1906,9}$}
\newcommand{\lya}{Ly$\alpha$}

\begin{abstract}

We provide one of the most comprehensive metallicity studies at $z\sim4$ by analyzing the UV/optical HST photometry, and rest-frame VLT-FORS2 ultraviolet and VLT-XSHOOTER optical spectra of J0332-3557, a gravitationally lensed galaxy magnified by a factor of $20$. With a 5$\sigma$ detection of the auroral \OIIIuv\ line, we are able to derive a direct gas metallicity estimate for our target. We find \Zg\ $\rm=12+log(O/H)=8.26\pm0.06$, which is compatible with an increasing of both the gas fraction and the outflow metal loading factor from $z\sim0$ to $z\sim4$. J0332-3557 is the most metal-rich individual galaxy {at ${z\sim4}$} for which the C/O ratio has been measured. We derive a low log(C/O) = $-1.02\pm0.2$, which suggests that J0332-3557 is in the early stages of ISM carbon enrichment driven mostly by massive stars. The low C/O abundance also indicates that J0332-3557 is characterized by a low star formation efficiency, higher yields of oxygen, and longer burst duration. We find that \ewciii\ is as low as $\sim3$ \wa. The main drivers of the low \ewciii\ are the higher gas metallicity and the low C/O abundance. J0332-3557 is characterized by one diffuse and two more compact regions $\sim$ 1 kpc in size. We find that the carbon emission mostly originates in the compact knots. {Our study on J0332-3557 serves as an anchor for studies investigating the evolution of metallicity and C/O abundance across different redshifts.}
\end{abstract}


\keywords{Galaxy evolution (594), High-redshift galaxies (734), Strong gravitational lensing (1643)}



\section{Introduction}

Galaxies at the peak of the cosmic star formation history ($z\sim1-3$; \citealp{Madau_Dickinson2014}) are very useful to infer how star formation processes occur and develop inside galaxies. However, although the development of colour selection criteria \citep{Steidel+1996} has increased the number of galaxies discovered at redshift $z \simeq 1$--10, at these high $z$ galaxies are faint ($m_{\rm R}^\ast = 24.4$ at $z = 2-3$; \citealp{Steidel+1999}; \citealp{Reddy+2008}) and the S/N needed to perform detailed spectroscopic studies is difficult to achieve. A way to overcome these difficulties is to study high redshift \textit{gravitationally lensed galaxies}, where the magnification produced by the lensing provides high S/N spectra which can be analyzed in greater detail. Since the duration and efficiency of the star formation depend on the availability of gas (regulated by outflows and inflows),  and its ability to collapse and form stars, understanding the mechanisms that regulate star formation and evolution relies on our ability to measure the physical properties of a galaxy's interstellar medium (ISM) and the stellar populations. \textit{Rest-frame UV} and \textit{optical} wavelengths are particularly useful to gain insights into the physical properties of high-redshift galaxies. 

The ultraviolet continuum and photospheric absorption features can be used to derive the ``stellar metallicity'' (\Z), i.e., the abundance of metals in the atmospheres of stars. However, determining stellar metallicities typically relies on the UV/optical stellar continuum, which has remained challenging to discern in high-redshift sources due to its faintness. Only recently, \Z\ has been measured at earlier epochs using stacks of Keck Barionic Structure Survey (KBSS) and  The MOSFIRE Deep Evolution Field (MOSDEF) survey spectra (\citealp{Steidel+2016}, \citealp{Theios+2019}, \citealp{Topping+2020a}), stacked galaxies from the VANDELS survey at $z\sim2.5-5$ (\citealp{Cullen+2019}, \citealp{Calabro+2021}) and the zCOSMOS-deep survey at $z\sim1.6 -3$ \citep{Kashino+2022}. These studies have shown that \Z\ increases by a factor $\sim4$ from $z\sim3.5$ to the present day, for a given stellar mass.

Nebular emission lines can be used to infer the ``gas-phase metallicity'' (\Zg), i.e., the oxygen abundance in the ionized gas surrounding young stars. 
The most robust way to measure \Zg\ is the \textit{direct method}, which adopts temperature dependent ratios between auroral lines (e.g., [O III] $\lambda$1666) and strong lines (e.g., [O III] $\lambda$5007). {Observing auroral lines poses a significant challenge due to their low intensity, especially when signal-to-noise ratio (S/N) is low, particularly at higher metallicities. The existing set of galaxies with detected auroral lines comprises approximately 20 galaxies within the redshift range of $1.6 < z < 3.6$ (e.g., \citealp{Yuan_Kewley2009, Christensen+2012a, Christensen+2012b, Stark+2014, Bayliss+2014, James+2014, Berg+2018, Gburek+2019, Sanders+2020a}). However, the advent of the James Webb Space Telescope (JWST) has significantly expanded this dataset, allowing for detections at redshifts up to approximately $z \sim 8$ \citep[e.g.,][]{Heintz+2023, Laseter+2023, Sanders+2023c,Sanders+2024}.}

An alternative method to derive \Zg\ relies on the use of strong optical emission lines that have been calibrated with direct metallicity measurements or photoionization models. However, since strong emission lines depend on other quantities as well as metallicity, these calibrations are usually challenging. {Literature works on the gas metallicity have shown that the gas phase-mass metallicity relation evolves in a similar way as the stellar-mass metallicity (e.g., \citealp{Shapley+2005, Erb+2006, Maiolino+2008, Mannucci+2009, Zahid+2011, Zahid+2014, Guo+2016, Sanders+2018,  Cullen+2021, Henry+2021, Sanders+2021, Topping+2021}; also see the review by \citealp{MaiolinoMannucci2019}).}

Another way to infer how early star formation develops in galaxies is by means of the relative abundances of carbon and oxygen, which are produced by stars of different masses and on different timescales. Oxygen is only synthesized by massive stars ($\rm M \gtrsim\,8\,M_{\odot}$) and released into the ISM by Core-Collapse Supernova events on short timescales, while carbon can be produced by any star with mass $\gtrsim\,1\,M_{\odot}$ and released into the ISM also during the Asymptotic Giant Branch (AGB) phase on longer timescales. Therefore, the C/O abundance value builds up as stars of different mass leave the main sequence, and informs us on the evolutionary stage of a galaxy (e.g., \citealp{Mattsson2010}, see also the review by \citealp{Romano2022}). The scatter of C/O at any given \Zg\ is also related to galaxy evolutionary properties. For examaple, \citet{Berg+2019} found that the C/O ratio depends on the star formation efficiency, the amount of oxygen released into the ISM by Supernovae Type II, and the duration of the  star formation episodes. 

The equivalent width (\ewciii) of the doublet formed by the forbidden [\ion{C}{3}] $\lambda1906$ and the semi-forbidden \ion{C}{3}] $\lambda1909$ transitions (\CIII\ hereafter) is also a proxy of the galaxy's physical and evolutionary properties. Specifically, the main dependence of \ewciii\ is on \Zg: both observations and models have shown that \ewciii\ peaks at \Zg = $\rm 12+log(O/H)\sim7.75$, and decreases both below (where the metallicity is too low) and above (where the carbon cooling is very efficient) this threshold \citep[e.g.,][]{Leitherer+2011,Rigby+2015,Nakajima+2018}.  
Besides the metallicity, \ewciii\ also increases for increasing ionization parameter, decreasing age of the current ionizing stellar population, increasing optical depth, decreasing C/O abundance, and increasing sSFR of a galaxy \citep{Rigby+2015, Jaskot_Ravindranath2016, Ravindranath+2020}. However, not all galaxies fit in this straightforward picture. For example, the nearby galaxy I Zw 18 is characterized by a low \ewciii\ although being metal poor, vigorously star forming, and characterized by a highly ionized ISM \citep{Aloisi+2001,Aloisi+2007,Rigby+2015}.

In this paper, we analyze the rest-frame UV and optical spectra of FORJ0332-3557 (J0332 hereafter), a gravitationally lensed galaxy at $z\sim3.8$, previously studied by \citet{Cabanac+2008}. Exploiting UV absorption, optical absorption, and emission features, we perform one of the most comprehensive metallicity analyses of a galaxy at these redshifts. Using UV and optical spectra, we infer the metallicity of the stellar populations and of the ISM, comparing and contrasting them with those of other galaxies at different redshifts. Thanks to a $\sim5\sigma$ detection of the $\rm [O III] \lambda1666$ auroral line, we are able to provide a direct measurement of the gas metallicity of J0332. Using the UV oxygen line and \CIII, we investigate the chemical enrichment history of J0332 by means of the C/O abundance. We also conduct a spatially resolved analysis of the chemical enrichment in J0332, analyzing spatial scales of approximately $\rm 1\,kpc$.

The paper is organized as follows: in Section \ref{sect:data} we describe the photometric (HST, Section \ref{sec:hst}) and spectroscopic (FORS2, Section \ref{sec:fors2}, and XSHOOTER, Section \ref{sec:xho}) data this work is based on; in Section \ref{sec:lens}, we describe the lens model; in Section \ref{sec:sed}, we present the results obtained from the SED fitting of HST images; 
in Section \ref{sec:abs_spectrum}, we focus on UV absorption features, describing the kinematics of the ISM and its chemical composition; in Section \ref{sect:stellar_spec} we derive the stellar metallicity of J0332; 
in Section \ref{sec:ionized_gas}, we illustrate the properties of the ionized ISM, such as the gas metallicity through the direct method (Section \ref{sec:direct});
in Section \ref{sec:gas_vs_stellar_Z}, we compare the gas and stellar metallicity of J0332; 
in Section \ref{sect:co}, we investigate the relative abundance of carbon and oxygen in J0332; 
in Section \ref{sec:ewciii}, we discuss the carbon emission in terms of its equivalent width and dependencies on physical and evolutionary properties; 
in Section \ref{sec:spatially_resolved}, we show the spatially resolved study performed on the equivalent width of the
\CIII\ 
line on spatial scales of $\sim$ 1 kpc; we outline our conclusions in Section \ref{sec:conclusions}. 

Throughout this paper, we adopt a flat Planck 18 cosmology \citep{Planck+2020}, with $\Omega_{\mbox{m}}$ = 0.31 and H$_0$ = 67.7 km s$^{-1}$ Mpc$^{-1}$. The adopted solar metallicity scale is that of \citet{Asplund+2021}, where $\rm 12 + log(O/H)_{\odot} = 8.69$ (corresponding to $Z=0.014$) for instances related to the ISM metallicity. However, when we derive the stellar metallicity of J0332, a solar metallicity $\rm 12 + $log(O/H) = 8.83 (corresponding to $Z=0.02$) is adopted, since this is the solar metallicity assumed by the Binary Population and Spectral Synthesis (BPASS) \citep{Eldridge+2016} used for the derivation. We assume (C/O)$_\odot = -0.23$.

\begin{figure}[h]
	\includegraphics[width=\columnwidth]{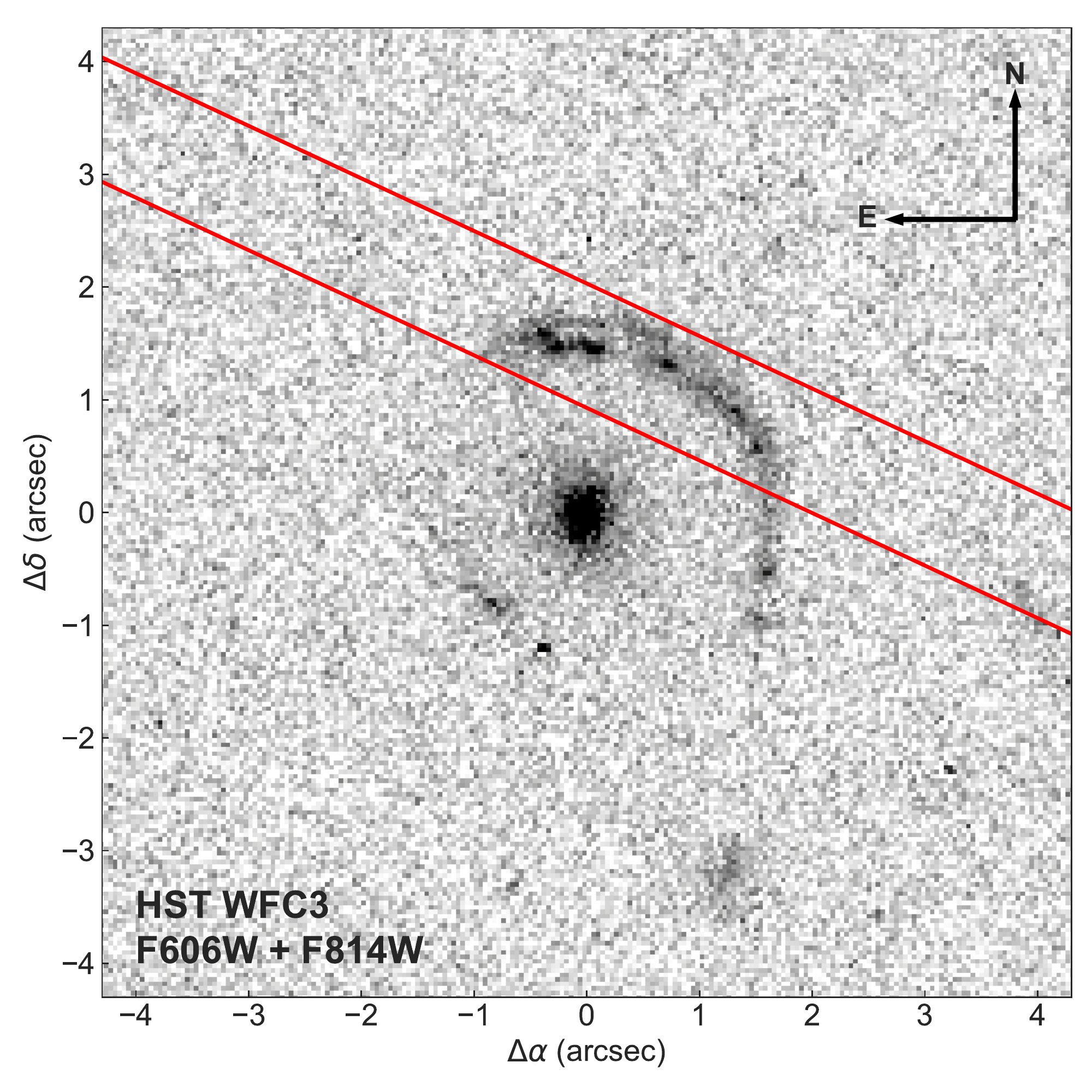}
      \caption{HST/WFC3 F606W+F814W image of J0332. The FORS2 slit position is oriented to maximize the light entering the slit from the lensed galaxy.}
   \label{fig:slit}
\end{figure}

\begin{figure*}
\includegraphics[scale=0.6]{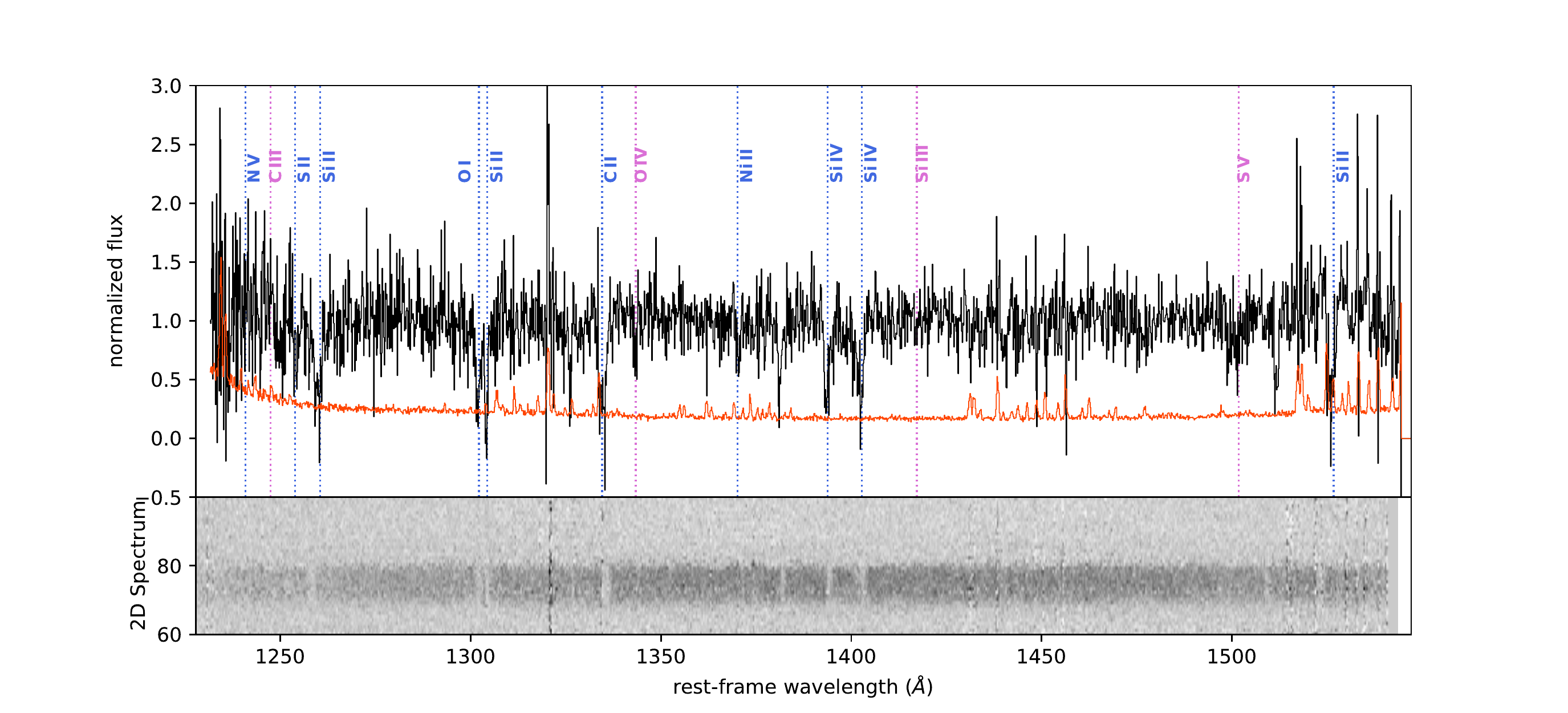}
\includegraphics[scale=0.6]{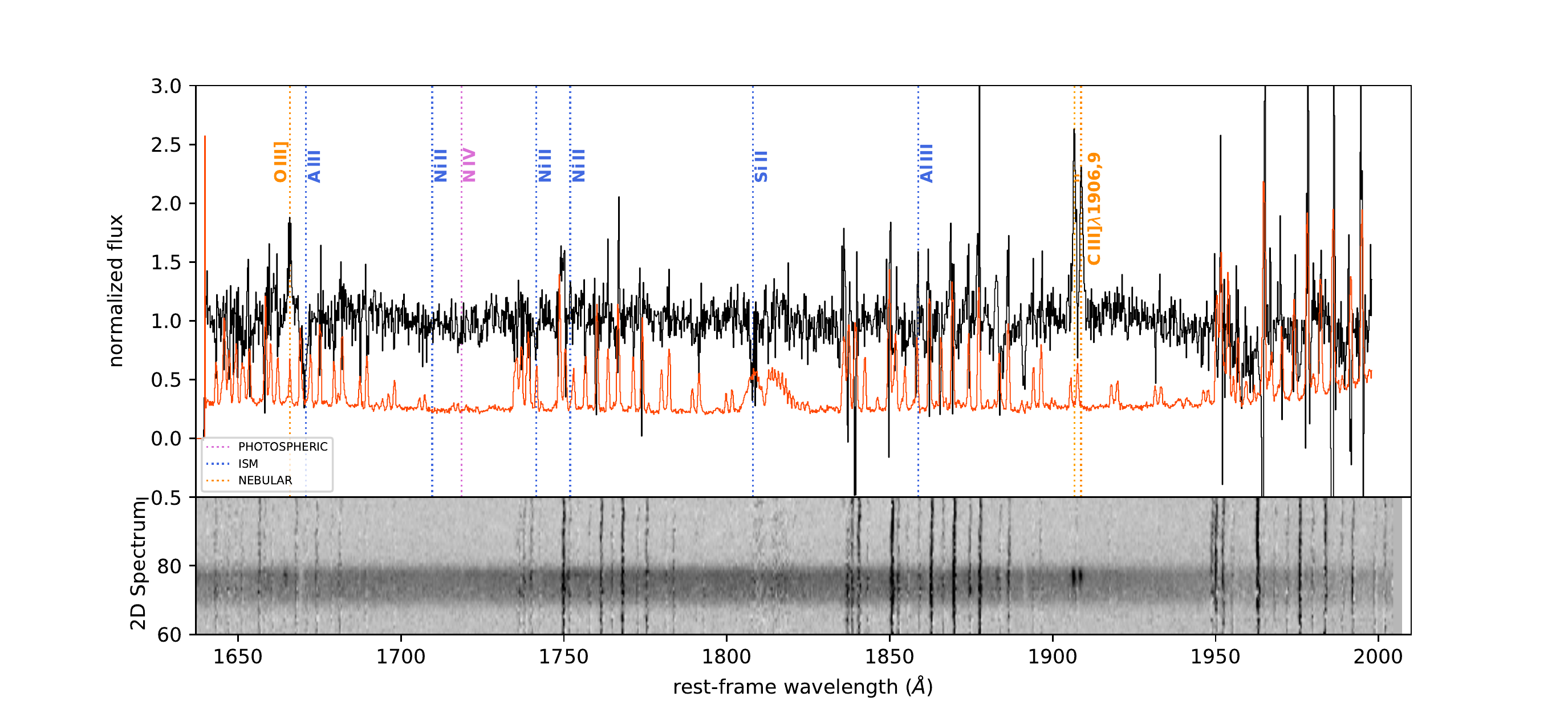}
      \caption{ UV FORS2 spectrum of J0332. The science spectrum is shown in black, the error spectrum is shown in orange. The narrower bottom panels illustrate the 2D spectrum. Nebular emission, ISM, and stellar absorption lines are marked by the vertical lines in different colors (pink: photospheric lines, blue: interstellar absorption lines, orange: nebular lines).}
   \label{fig:spec1}
\end{figure*}

\begin{figure*}
\begin{center}
\includegraphics[width=1.\textwidth]{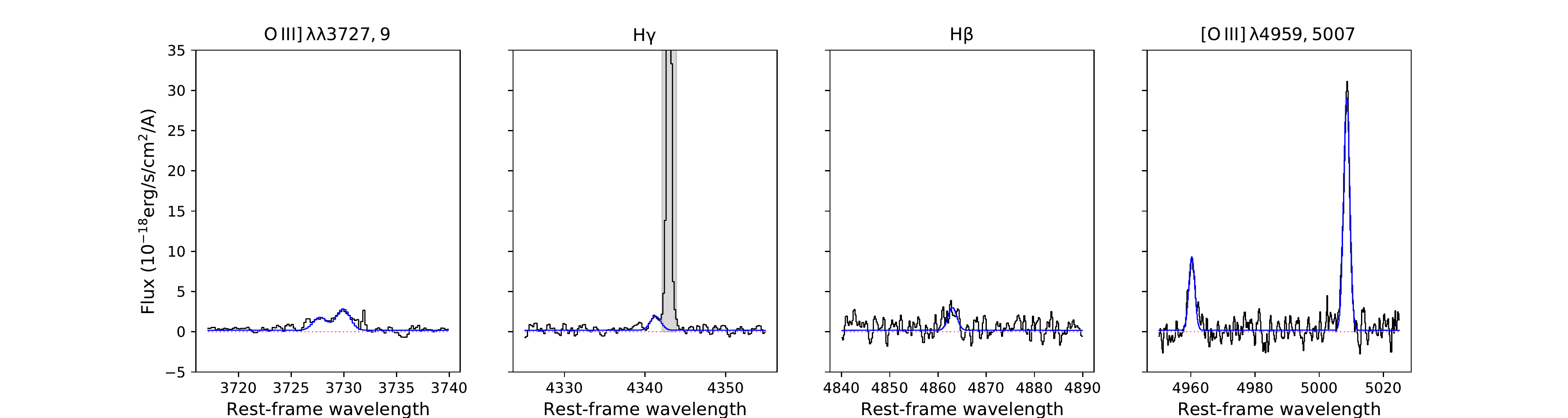}
\caption{Main optical emission lines detected in the XSHOOTER spectrum of J0332 (from left to right: \OIIopt,  $\rm H\gamma$, $\rm H\beta$ and \OIIIopt. The error spectrum is shown as the pink dotted curve. The blue curves are the fit to each line. The strong sky line close to $\rm H\gamma$ highlighted in grey has been masked out.} 
   \label{fig:xsho_spec}
   \end{center}
\end{figure*}

\section{Data}
\label{sect:data}

FORJ0332-3557 (RA=53.24198, Dec=$-$35.96458 (J2000), J0332 hereafter) is a gravitationally lensed galaxy, lensed by a foreground galaxy at  $z=0.986$. It is located in a sight line through the outskirts of the Fornax cluster, where the reddening of our Galaxy is E(B-V) = 0.02 \citep{Schlegel+1998}. Its  discovery was first reported by \citet{Cabanac+2005, Cabanac+2008}, who analyzed its VLT/FORS2 ultraviolet spectrum ($\sim 900-1700$ \wa\ in the rest-frame). Their analysis was focused on the brightest portion of the arc caught by their slit (which was oriented N-S to include both the arc and the lensing galaxy) and on the study of Ly$\alpha$ and metal absorption lines. Our study is instead based on new, higher resolution VLT/FORS2 and XSHOOTER observations, and therefore complements and extends the analysis performed by \citet{Cabanac+2008} (see Sections \ref{sec:fors2} and \ref{sec:xho}).

\subsection{HST imaging}
\label{sec:hst}
In order to improve the gravitational lensing model and study the stellar populations in the arc, we obtained WFC3 Hubble Space Telescope imaging (Program ID 14093, PI Berg). Exposure times were 1348, 1209, 1080, and 1170 s in the F606W, F814W, F125W, and F160W filters respectively. Three exposures were obtained for each filter utilizing a 3-point dither pattern to improve spatial sampling, and each set of exposures was then drizzled to a common output frame with a center defined by the centroid of the lensing galaxy. A 0\farcs04 pixel scale and Lanczos3 kernel were used for drizzling the UVIS data, and the IR data reduction employed a 0\farcs08 pixel scale and Gaussian kernel.


\subsection{Spectroscopy}\label{sec:spectra}
\subsubsection{FORS2 Spectra}\label{sec:fors2}

{Rest-frame FUV spectra were obtained using the FOcal Reducer and low 
dispersion Spectrograph 2 (FORS)} on the Very Large Telescope (VLT) of the European Southern Observatory (ESO). 
Spectra were observed for the ESO program 086.A-0035(A) on the UT dates of 2012 April 12, 
October 10, and November 18.
As shown in Figure~\ref{fig:slit}, the 1\farcs0$\times$500\arcsec\ slit was used at an angle of $65^{\circ}$ in order to encompass the most flux possible from the arc.
The 1200R and 1028z grisms were used with the GG435+81 and OG590+32 filters, respectively,
and a CCD pixel binning of $2\times2$ (0\farcs252/pixel in the spatial direction).
This combination provided an observed wavelength coverage of roughly 5770$-$7380 \AA\ and 
7830$-$9590 \AA\ at resolutions of $R=2140$ and $R=2560$, respectively.
At the redshift of J0332, this corresponds to rest-frame wavelength ranges of 
roughly $1230-1545$ \AA\ and $1640-2010$ \AA\, respectively.
The 1200R grism was used for $16\times1300$ s or 5.8 hours and  
the 1028z grism for $40\times1300$ s  or 14.4 hours, 	
summing 20.2 hours of total integration time. 
We note that the previous FORS2 spectrum was taken at a position angle of $0^{\circ}$
centered on the lens and brightest knot of the arc only \citep[see][]{Cabanac+2008},
and so was not combined with the FORS2 spectra presented here. 

\subsubsection{XSHOOTER Spectra}
\label{sec:xho}
Medium-resolution echelle spectra were obtained for J0332 using XSHOOTER on VLT
in order to obtain the rest-frame optical emission lines.
Observations were taken as part of the ESO program 094.A-0252(A) on the UT dates of 2014 
October 21 and 28. 
Each of two positions were observed for $4\times900$ s,
for a total exposure time of two hours. 
XSHOOTER\footnote{\url{https://www.eso.org/sci/facilities/paranal/instruments/xshooter/overview.html}} has three spectroscopic arms:
UVB with an observed wavelength range of roughly 3,000--5,595 \AA ,
VIS with an observed wavelength range of roughly 5,595--10,240 \AA, and
NIR with an observed wavelength range of roughly 10,240--24,800 \AA\ or 1.024--2.480 $\micron$. 
Note that the rest-frame XSHOOTER VIS wavelength coverage overlaps with the rest-frame 
FORS2 coverage, allowing us to consistently calibrate the two spectra and use emission line ratios for our analysis.
Slits of [1\farcs0, 0\farcs9, 0\farcs9] $\times$ 11\arcsec\ were used for the UVB, VIS, and NIR
arms, respectively, all positioned at $65^{\circ}$ in order to match the FORS2 spectrum.
This setup provides resolutions of $R=$~5,400, 8,900, and 5,600 for the UVB, VIS, and NIR arms, respectively.

\subsubsection{Spectral Reductions}
The initial spectral reductions for the FORS2 and XSHOOTER spectra were performed using the
\texttt{IDL} packages \texttt{FORS2\_REDUCE} and \texttt{XSHOOTER\_REDUCE}, respectively,
written by \textcolor{red}{one of the authors}.
The individual 2D frames were bias-subtracted and flat-fielded using a ``normalized" pixel flat,
with the illumination function in each order divided out. 
For the NIR arm of the XSHOOTER spectra, dark exposures are also combined and subtracted
to remove underlying structure.

Individual FORS2 exposures were combined into a single 2D frame with bad-pixel rejection. The combined frame was then analyzed using
standard \texttt{iraf} tools\footnote{\url{https://iraf-community.github.io/}}. 
Two versions of FORS2 1D spectra were extracted.
By visual inspection of the 2D spectra in the bottom panels of Figure~\ref{fig:spec1}, 
the absorption features transverse the entire continuum emission, while the emission features are more 
spatially compact.
Therefore, a wide aperture of 20 pixels, or 5\farcs04, was used to extract 99\%\ of the 
continuum light in the slit in order to optimize absorption line measurements.
Additionally, a narrow aperture of 9.6 pixels, or 2\farcs42, was used to capture 99\%\ of 
the nebular emission.
Finally, wavelength and flux calibration were performed using the known FORS2 wavelength map and 
the standard sensitivity functions provided for the 1028z and 1200R grisms.

The remaining XSHOOTER spectral reduction was performed with the aforementioned 
\texttt{IDL} package. 
Wavelength calibration was performed using the known XSHOOTER wavelength-pixel map.
The sky was then modeled using a b-spline fit and subtracted from the 2D frame in each arm.
An optimal sky subtraction was performed on each order by fitting a b-spline
to the sky counts\footnote{The \texttt{XSHOOTER\_REDUCE} package uses the optimal 
sky subtraction techniques of \citet{Kelson2003}. 
It also uses the \texttt{IDLUTILS} routines developed for SDSS spectral reduction, 
in particular the b-spline fitting routines written by Scott Burles and David Schlegel.}
Because the sky is very bright in the NIR arm, a nodded exposure was used for the primary 
sky subtraction.
Additionally, since little to no continuum was detected in the XSHOOTER-NIR observations of J0332, 
the bright [\ion{O}{2}] $\lambda\lambda3726,3729$ nebular emission doublet was used to 
determine the fiducial center of the extraction,
while the trace shape was determined from a standard star spectrum.
An extraction FWHM of 15 pixels, or 2\farcs42, was used for the VIS arm and
9 pixels, or 2\farcs23, was used for the NIR arm in order to best match the extraction 
aperture of the narrow FORS2 extraction.
Atmospheric corrections and response functions for each arm are used to simultaneously 
flux calibrate and optimally extract a single 1D spectrum by flux scaling the individual arms to match.

\subsubsection{Relative Flux Calibration}
While neither our absorption line analysis (using the normalized spectrum) or  
nebular emission line analysis (using line ratios) depends on the absolute flux calibration,
the latter does combine line fluxes from both the FORS2 and XSHOOTER spectra.
Therefore, for this work, we are primarily concerned with having a robust relative flux 
calibration between the FORS2 and XSHOOTER spectra.

For the XSHOOTER spectra, the instrumental profile of the skylines has FWHM $\sim6$ pixels,
allowing us to rebin the XSHOOTER spectra by 2 to improve the signal-to-noise.
The XSHOOTER spectra are then scaled to the narrow-aperture FORS spectrum
using the flux in the \ion{C}{3}] $\lambda\lambda1906,1909$ emission lines, 
which are common to both spectra.
The resulting set of flux-calibrated spectra, i.e., the narrow- and wide-aperture FORS spectra, and 
the XSHOOTER spectrum, have equivalent measurements of the \ion{C}{3}] $\lambda\lambda1906,1909$ flux.

The resulting FORS2 spectrum is shown in Figures \ref{fig:spec1}.
The FORS2 rest-frame UV spectrum covers a wavelength range of $\sim$1230--2000 \AA,
with a gap in the wavelength coverage from 1540--1640 \AA\ due to the use of two 
non-overlapping gratings.
Unfortunately, the FORS2 spectrum also lacks coverage of the the Ly$\alpha$ feature. 
The XSHOOTER spectrum covers the rest-frame UV and optical regimes. 
The four strongest rest-frame emission lines detected ([\ion{O}{2}] $\lambda$3727,
H$\gamma$ $\lambda4340$, H$\beta$ $\lambda4861$, and [\ion{O}{3}] $\lambda\lambda$4959,5007)
are shown in Figure~\ref{fig:xsho_spec}. 


\section{Lens model, source reconstruction and spectroscopic redshift}
\label{sec:lens}
The lens model is fitted directly to the HST imaging data. The NIR images have significantly higher signal-to-noise than the UVIS images, but they also suffer from relatively poor spatial resolution and significant confusion between the lensed Einstein ring and the lensing galaxy. We therefore use the UVIS images for the lens modelling and combine the F606W and F814W bands to improve the signal-to-noise. Our noise model is determined by adding in quadrature the empirical background (sky plus detector) noise with the Poisson noise determined from the images in conjunction with the exposure time map produced in our drizzling procedure.

We employ a singular isothermal ellipsoid plus external shear model for the lensing mass distribution. The source surface brightness distribution is described as the amplitudes of pixels on an irregular grid that adapts to the local lensing magnification \citet{Vegetti+2009}. A field star from the combined F606W+F814W image is used for a PSF model, and we use the lensing galaxy position as a starting point for our modelling but otherwise ignore the light from the foreground galaxy as it has essentially no impact on the resulting lens or source models, which yield an inferred magnification of  $\mu = 20\pm2$.

{We measure the spectroscopic redshift of J0332 from the nebular emission lines \OIIIuv,  \CIII, and \OIIIopt, which are captured by the FORS2/XSHOOTER spectra.} The redshift values obtained from each individual emission line are reported in Table \ref{tab:redshift}. 

\begin{table}[h]
\centering
\caption{Emission lines used for the determination of the spectroscopic redshift of J0332.}
\begin{tabular}{cccc}
\hline
Ion     & $\lambda_{\rm lab}$ & redshift & error \\
\hline
O\,{\sc iii}] & 1666.15 & 3.77195 & 0.00024\\\
[C\,{\sc iii}]  & 1906.68 &   3.77302 & 0.00006\\
C\,{\sc iii}] & 1908.73 &  3.77358 & 0.00012\\\
[O\,{\sc iii}] & 5008.24 &   3.77327 & 0.00003\\
\hline
\label{tab:redshift}
\end{tabular}
\end{table}

{The derived average redshift, weighted by the significance of each emission line detection, is $z=3.7732 \pm 0.0003$.} This estimate is, within the uncertainties, in agreement with the redshift reported by \citet{Cabanac+2008}. 

\section{SED fitting}
\label{sec:sed}
We model the broadband photometry of J0332 with the SED-fitting code \texttt{prospector} \citep{Johnson+2021}. In addition to the HST measurements in the F606W, F814W, F125W, and F160W filters, which span the wavelength range $\sim1200$--3200 \AA\ in the rest frame of the arc, we use ground-based $K_s$ imaging \citep{Cabanac+2008} in order to extend the photometric coverage beyond the Balmer break. The observed AB magnitudes in the five modeled bands are given in Table \ref{tab:photometry}.

{We use the Binary Population and Spectral Synthesis (BPASS, \citealt{Eldridge+2017}) v2.2 stellar population models, and assume a \citet{Chabrier2003} intial mass function and a SMC extinction law \citep{Gordon+2003}.} In order to minimize the number of free parameters, we assume a constant star formation history (SFH) and fix the metallicty of the models to 0.05 $Z_{\odot}$, the best-fit metallicity from fitting the stellar spectrum (see Section \ref{sect:stellar_metallicity} below).

The resulting best-fit stellar population properties are $\log(M_{\star}/M_{\odot}) = 9.32^{+0.33}_{-0.32}$, age $93^{+238}_{-63}$ Myr, and $E(B-V)=0.15^{+0.02}_{-0.03}$. A constant SFH then implies a star formation rate of 21.6 M$_{\odot}$ yr$^{-1}$. {We also explore the possibility of a bursty SFH. To do so, we use \texttt{DYNESTY} \citep{Speagle2020}, a dynamic nested sampling method within \texttt{prospector}. Our settings are similar to those adopted above, with the exception of the SFH. We now adopt a non-parametric model consisting of eight independent temporal bins\footnote{The number of bins was chosen following \citet{Leja+2019}.}: the most recent bin spans 0-10 Myr in lookback time, and the remaining four bins are evenly spaced in log(lookbacktime) up to $z=20$.  We derive a galaxy's total formed mass is $\log(M_{\star}/M_{\odot}) = 9.44 \pm 0.17$, with $E(B-V)=0.15 \pm 0.01$ (in agreement with the E(B-V) found assuming a continuous SFH). The SFR decreases from 10.8 to 1.5 $\rm M_{\odot}\,yr^{-1}$ from the most recent to the oldest age bin. We find that the sSFR is $\rm \lesssim 9.8\times10^{-9}\,yr^{-1}$, but increases up $\rm \sim 1\times10^{-7}\,yr^{-1}$ in the last 100 Myrs, compatible with a bursty SFH. It is worth noting that the results from both fits are largely consistent within the uncertainties, underscoring the challenge in precisely constraining the SFH of this particular object.}


\begin{table}
\centering
\caption{Observed Photometry. Magnitudes are given in AB system.}
\begin{tabular}{lc}
\hline
Band & $\rm mag_{AB} \pm \delta mag_{AB}$\\
\hline
F606W & $22.74\pm0.10$\\
F814W & $22.09\pm0.10$ \\
F125W & $21.83\pm0.10$ \\
F160W & $21.66\pm0.10$ \\
$K_s$ & $21.25\pm0.30$ \\
\hline
\label{tab:photometry}
\end{tabular}
\end{table}

\section{The UV interstellar absorption spectrum}
\label{sec:abs_spectrum}

The velocity profiles of interstellar absorption features encode information about the kinematics of the gas where the lines are produced, while their strength is related to the abundance of the elements they are produced from (to the extent that they are not saturated). In the following Sections, we focus on the UV absorption features in the FORS2 spectrum of J0332 and derive the kinematics and chemical composition of its ISM. 

\subsection{Gas kinematics}
\label{sec:kinematics}
We identify several absorption lines in the FORS2 spectrum of J0332 (see Figure \ref{fig:spec1}). These lines are produced by a variety of ions in different ionization states, from $\rm O\,I\,\lambda1302$  to $\rm Si\,IV\,\lambda\lambda1393,1402$.

\begin{figure}[h]
	\includegraphics[width=1.1\columnwidth]{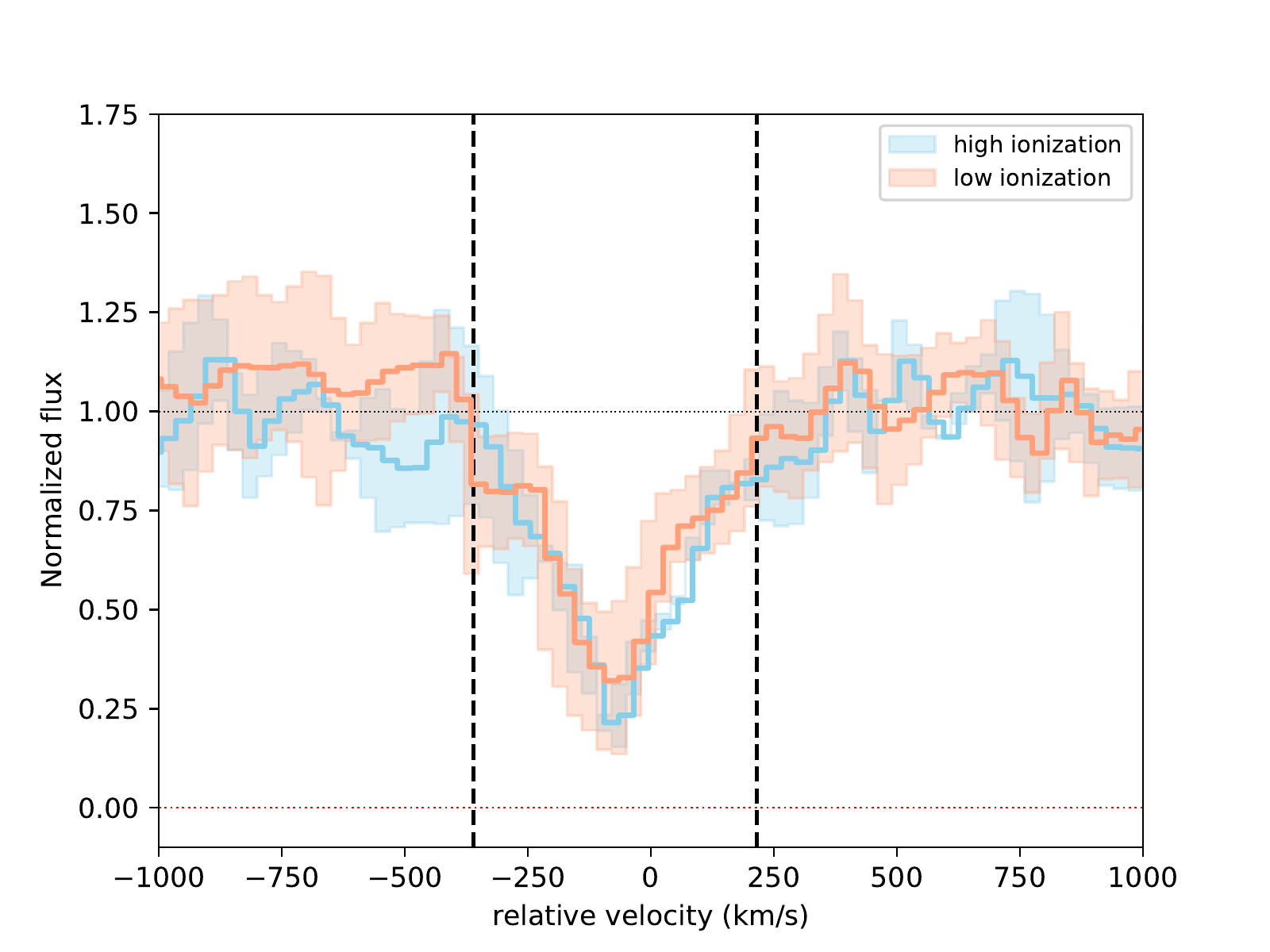}
      \caption{Average velocity profiles for high and low ionization lines (from FORS2) with the respective uncertainties. The dashed vertical lines mark the range $\rm -360$ to $\rm +215\,km\,s^{-1}$.}
   \label{fig:avg_prof}
\end{figure}

Figure \ref{fig:avg_prof} shows the average velocity profiles for the high and the low ionization lines. 
The velocity profiles are obtained by separately averaging the individual velocity profiles of the low ionization lines ($\rm O\,I\,\lambda1302$, $\rm Si\,II\,\lambda1304$, $\rm Al\,III\,\lambda1670$,  $\rm Si\,II\,\lambda1526$, $\rm Si\,II\,\lambda1808$) and the high ionization lines ($\rm Si\,IV\,\lambda1393$ and $\rm Si\,IV\,\lambda1402$)\footnote{The average high ionization profile has larger uncertainties since only 2 lines are used to obtain it.}. Moreover, we masked the portion of the lines which are contaminated by stellar features (where resolved). When spectral features are close to each other, we alternately include one of them in the average while masking the other.
 From the average profiles, we note the presence of blue-shifted absorption extending to velocities up to $v_{\rm outflow}\sim -360\rm ~km~s^{-1}$. 
 
\begin{figure}
	\includegraphics[width=1.1\columnwidth]{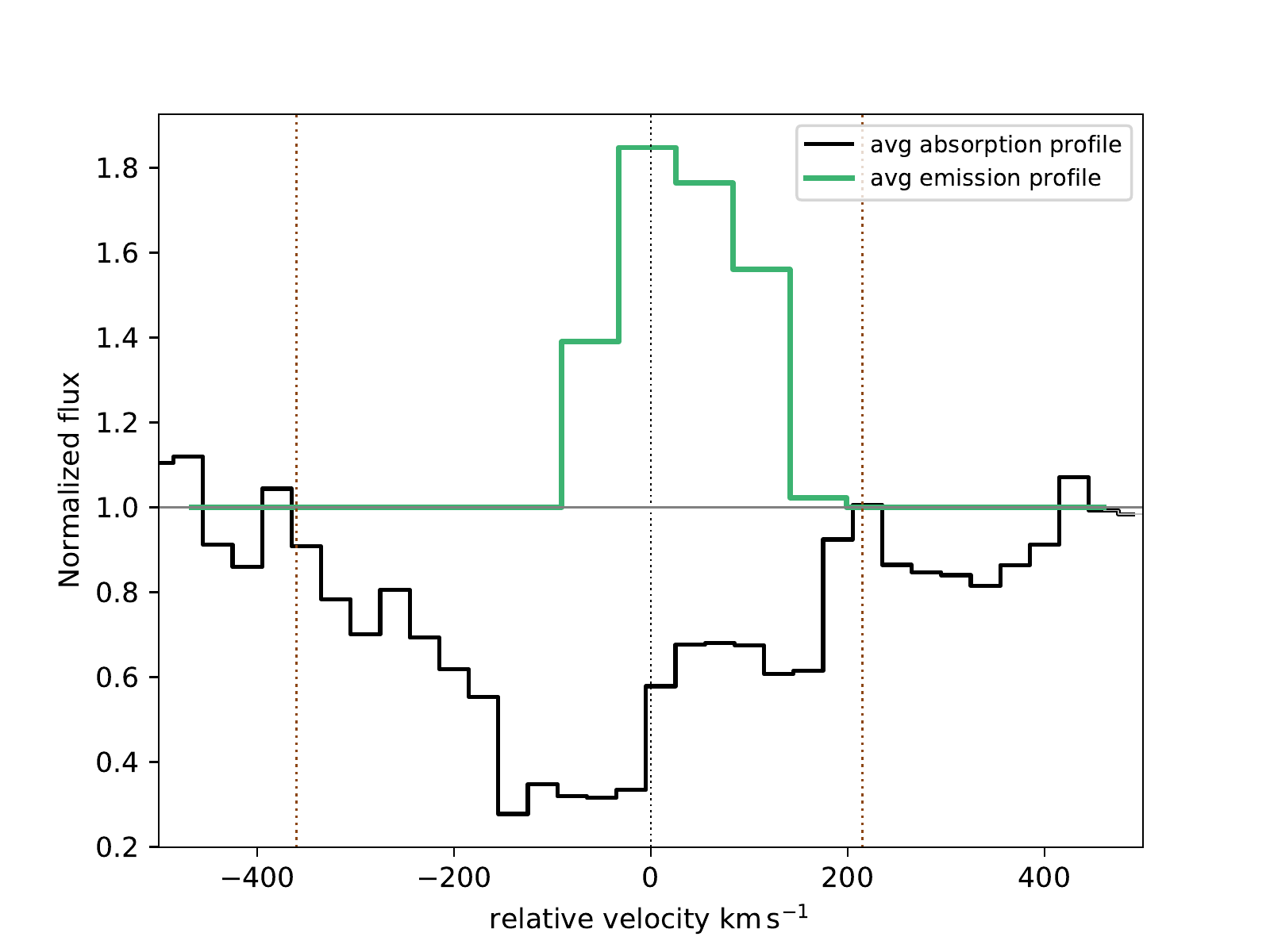}
      \caption{Comparison between the average velocity absorption profile (obtained from both low and high ionization lines) and average emission profile (obtained from the \CIII\ doublet) in J0332.}
   \label{fig:abs_em_comp}
\end{figure}

Figure \ref{fig:abs_em_comp} compares the average absorption profile obtained by averaging all the UV absorption lines (both high and low ionization) to the average emission profile obtained by averaging together the individual lines of the \CIII\ doublet. Also in this case, we alternately mask the doublet line not being used in the average. We observe that the absorption profile extends to higher negative velocities than the emission profile (which has a velocity dispersion $\sigma \sim 30\,\rm km\,s^{-1}$).  In contrast, the emission and absorption profiles share the same velocity range at positive velocities (they both extend to $\rm +215\,km\,s^{-1}$), suggesting the absence of inflows.

\subsection{The mechanism of outflow production}

Outflows are a common feature in star-forming galaxies at high as well as low redshifts \citep{Pettini+2001, Shapley+2003,Steidel+2010,MarquesChaves+2020}. The relationship between outflow and star formation properties has been investigated by several theoretical and observational studies \citep[e.g.,][]{Ferrara_Ricotti+2006,Steidel+2010, Murray+2011, Sharma_Nath2012}. These studies have suggested the presence of two main mechanisms of outflow production: (1) mechanical energy injected into the ISM by Supernovae or (2) momentum injected through radiation pressure from massive stars acting on dust grains or (3) a combination of the two (e.g., \citealp{Xu+2022}).

At low redshift ($z\lesssim0.3$), there is general agreement that the $v_{\rm outflow}$ - SFR relationship is a weak power law, with $v_{\rm outflow}\propto \rm SFR^{0.15-0.35}$ \citep{Martin+2005, Weiner+2009, Chisholm+2016,Trainor+2015, Sugahara+2017}. At higher redshift ($z\gtrsim2$), there is instead uncertainty about the significance of this relation. 

Recently, \citet{Weldon+2022} have investigated the relation between $v_{\rm outflow}$ and the SFR of a sample of 155 typical star-forming galaxies at $z\sim2$ drawn from the MOSFIRE Deep Evolution Field (MOSDEF - \citealp{Shapley+2015}) survey. They found that $\rm log(|v_{\rm outflow}|) = 2.51 + 0.24\times log(SFR)$. This weak dependence (characterized by a large scatter) is compatible with the mechanical energy injection scenario.
Given the H$\rm \beta$ - derived SFR for J0332, $\rm SFR =4.55\pm0.46\,M_{\odot}\,yr^{-1}$ (see Section \ref{sec:Te}), and the outflow velocity $v_{\rm outflow}\rm \sim-360\,km\,s^{-1}$, we find that J0332 actually lies within the scatter of the \citet{Weldon+2022} relation, favoring the mechanical energy injection scenario.

The fact that we do not detect any inflows in J0332 is a common finding in high redshift galaxies. To our knowledge, only two gravitationally lensed galaxies show signatures of inflows in their spectra: the Cosmic Eye (\citealp{Quider+2010}), and J1059+4251 \citep{Citro+2021}. The reason why inflows of accreting cold gas at high redshift are very elusive and difficult to observe is that they are often obscured by outflows or by absorption from the galaxy’s ISM \citep{Steidel+2010}. Nevertheless, alternative factors could play a significant role. For instance, a reduced covering fraction in comparison to the outflows, or lower metallicities. 

\subsection{Chemical composition}
\label{sect:chemical_composition}
\begin{figure*}[h]
	\includegraphics[width=\textwidth]{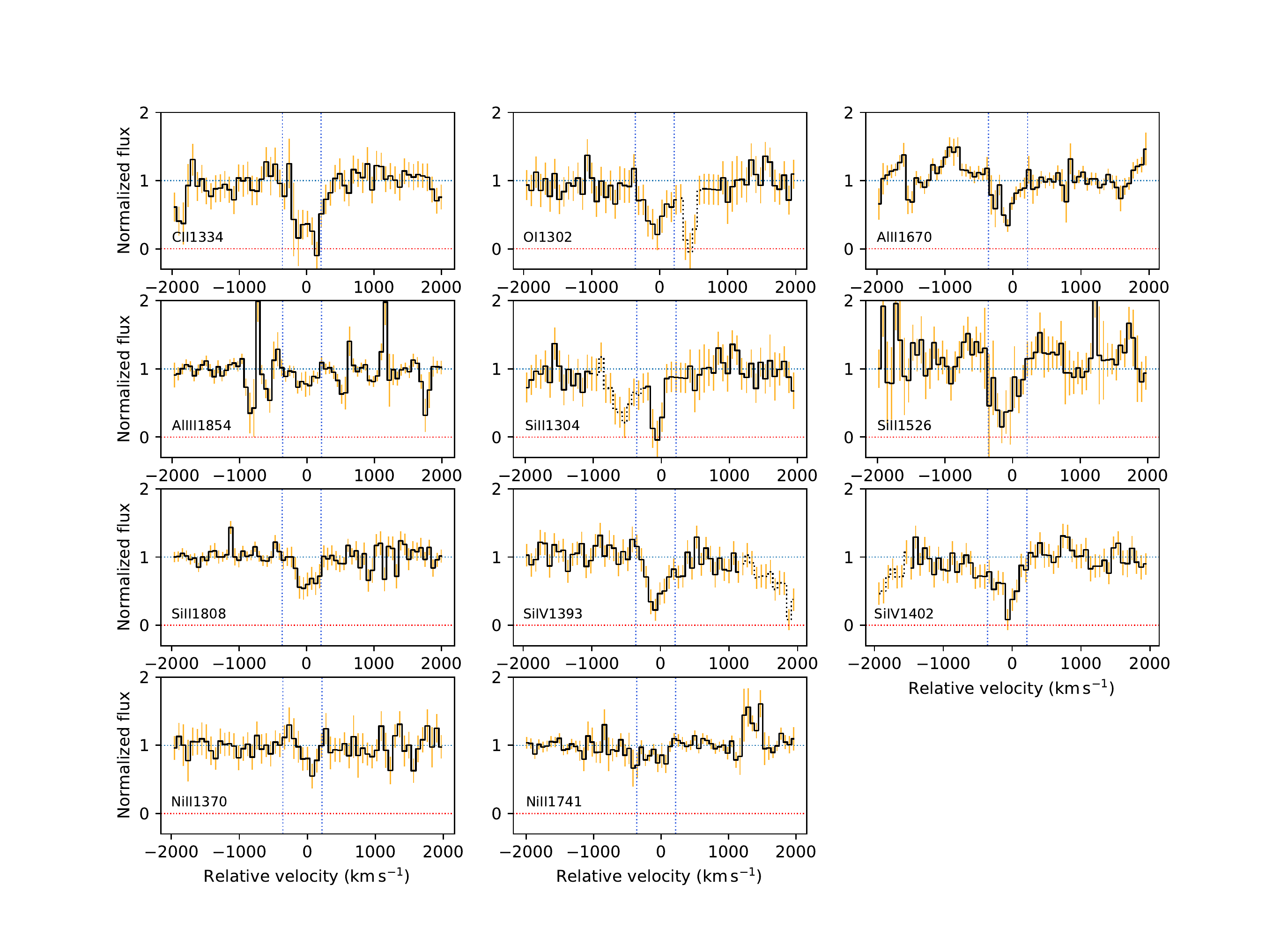}
      \caption{Velocity profiles of the UV ISM absorption lines in J0332 (black) and corresponding uncertainties (yellow). In each panel, strong spectral features which are close to the one of interest are marked as  dotted portions in the spectrum. The vertical lines indicate the average velocity extent of the absorption lines.}
   \label{fig:ism_abs}
\end{figure*}

In this Section, we use the UV absorption lines in the FORS2 UV rest-frame spectrum of J0332 to derive the chemical composition of its neutral ISM. Figure \ref{fig:ism_abs} shows the velocity profiles of the identified ISM absorption lines. The first step consists in  deriving the column densities of the absorption lines. In order to do so, we apply the apparent optical depth (AOD) method proposed by \cite{SavageSembach1991}. This approach allows us to recognize cases where narrow saturated components are hidden by overlapping broader ones; this circumstance is very likely to happen in J0332, given the shape of the velocity profiles and the fact that the observed absorption lines are a combination of many unresolved sight-lines.

In the AOD method, the column density of an ion per velocity bin, $N_{\rm a}(v)$ $\rm [cm^{-2} (km\,s^{-1})^{-1}]$, is related to the apparent optical depth in that bin, $\tau_{\rm a}(v)$, by the expression:
\begin{equation}
    N_{\rm a}(v)=3.768\,\times\,10^{14}\frac{\tau_{\rm a}(v)}{\lambda\,\textit{f}},
\label{eq:AOD}
\end{equation}
where $\lambda$ and \textit{f} are, respectively, the wavelength (in \AA) and oscillator strength of the atomic transition. When lines with differing $f$-values absorbing from the same ground state of an ion produce discordant values of $N_{\rm a}(v)$, then hidden saturation is present (since, by definition, there is only one value of the column density for that ground state). 

Partial, as opposed to complete, coverage of the stars by the absorbing gas would produce a similar effect. However, in our case we do not expect this to be a significant complication, since our strongest lines reach the zero flux level (i.e.,  $\rm Si\,II\,\lambda1304$, $\rm C\,II\,\lambda1334$ and $\rm Si\,II\,\lambda1526$). We apply the AOD method to the three silicon lines available, which are characterized by very different $f$-values. We find that $\rm Si\,II\,\lambda1526$ (as expected) and $\rm Si\,II\,\lambda1304$ are saturated. $\rm Si\,II\,\lambda1808$, the weakest of the three, is instead not saturated and therefore the only one able to provide a reliable estimate of the Si column density. 

We expect $\rm O\,I\,\lambda1302$ to be saturated, since O I has a ionization level which is very similar to that of Si II.  Moreover, Figure \ref{fig:ism_abs} shows that $\rm Al\,II\,\lambda1670$ has the same depth as $\rm Si\,II\,\lambda1808$, despite its much higher $f$-value (see Table \ref{tab:table_na}). This suggests that $\rm Al\,II\,\lambda1670$ is also saturated. $\rm Ni\,II\,\lambda1370$ and $\rm Ni\,II\,\lambda1741$ appear not saturated. Instead, $\rm Si\,IV\,\lambda\lambda1393,1402$ result to be both saturated. 
Lastly, we are not able to verify the $\rm Al\,III\,\lambda1854$ saturation level, as the $\rm Al\,III\,\lambda1854$ transition is the only one we detect. Summarizing, we rely on the $\rm Si\,II\,\lambda1808$, $\rm Ni\,II\,\lambda1370$ and $\rm Ni\,II\,\lambda1741$ transitions as representative of the abundances of $\alpha$-capture and iron peak elements. \textcolor{black}{We confirm that our measured EWs and $N_{\rm a}$ values align with those obtained from the J0332 spectrum analyzed by \citet{Cabanac+2008} (obtained through private communication) when assuming the same velocity range for integration. However, it is worth noting that the EWs and column densities reported in their paper exceed ours, likely due to the adoption of a larger velocity range for integration in our analysis (see Section \ref{sect:data}).}

\begin{figure}
	\includegraphics[width=1.1\columnwidth]{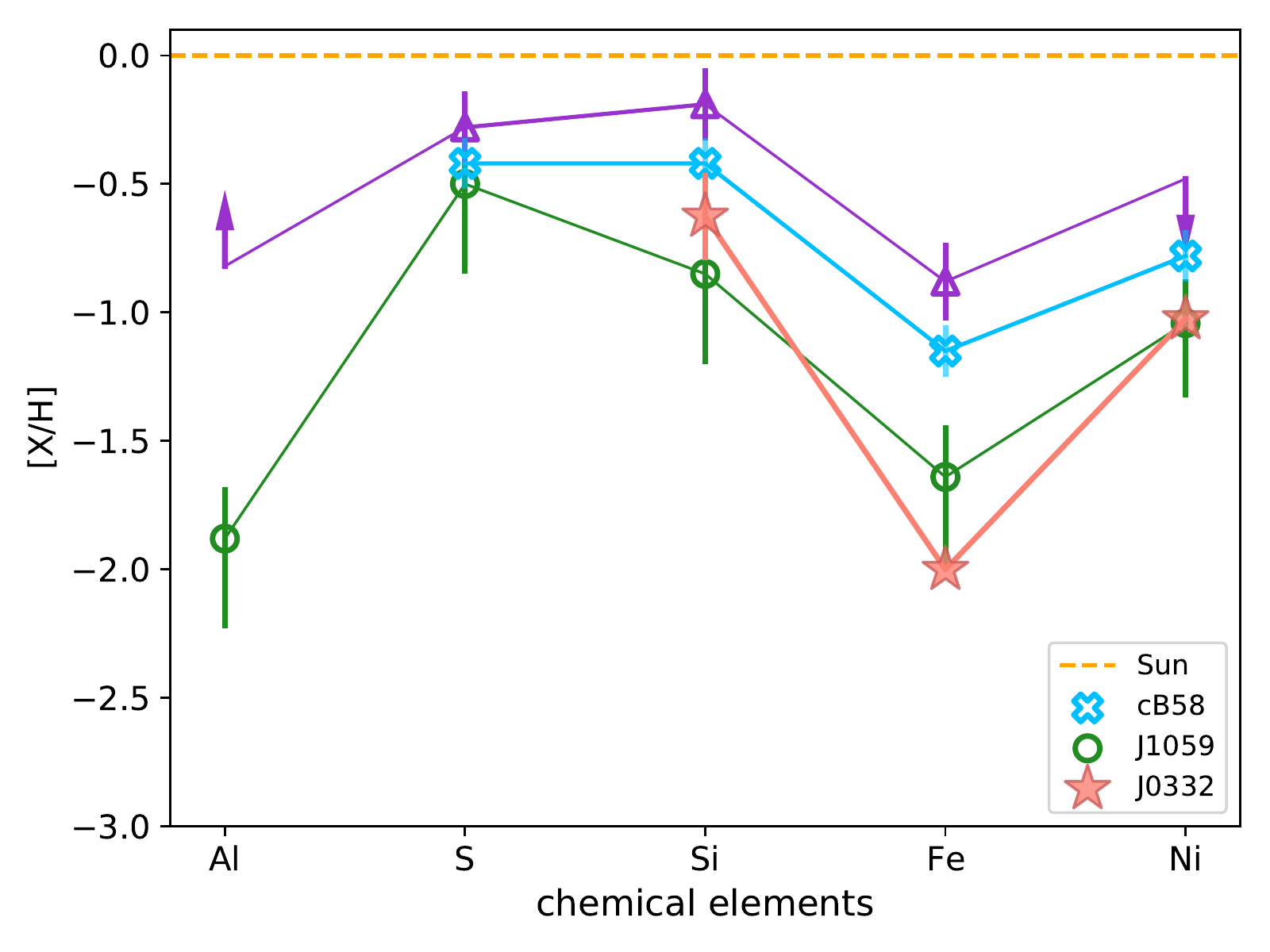}
      \caption{Element abundances in the interstellar gas of J0332 (pink) compared with those of three other well-studied lensed galaxies: MS15-cB58 (cyan; \citealp{Pettini+2002}), the 8 O’Clock arc (magenta; \citealp{Dessauges-Zavadsky+2010}) and J1059 (green; \citealp{Citro+2021}). Note that the iron abundance for J0332 has been taken from \citet{Cabanac+2008}, who measured it from the Fe II $\lambda$1608 line.}
   \label{fig:abd}
\end{figure}

Figure \ref{fig:abd} shows the chemical abundances derived from the AOD method compared with those obtained for other lensed galaxies at high redshift. For these calculations, we assume $Z_{\odot}=0.014$ \citep{Asplund+2021}, corresponding to 12+log(O/H) = 8.69. Moreover, we assume a neutral hydrogen column density $\rm N(HI)=10^{21.4}\,cm^{-2}$, which is the value derived by \citet{Cabanac+2008} through fitting the damped Ly$\alpha$ profile in J0332. Together with our measurements for Si and Ni, we also show the Fe abundance that \citet{Cabanac+2008} derived from the Fe II $\lambda$1608 line. We note that Si is more abundant than Fe by 1.3 dex. Instead, given the large uncertainty on the Ni abundance, we cannot draw definitive conclusions about the Si/Ni relative abundance.
Our findings suggest that the ISM in J0332 is $\rm \alpha$-enhanced. This is indicative of a rapid star formation, where iron peak elements such as Fe and Ni, which are produced on longer timescales ($\sim$ 1 Gyr) by Supernova Type Ia events, have not had time to dilute the $\alpha$-elements (such as Si) produced on shorter timescales ($\sim30$ Myr) by Supernova type II explosions. 

However, we remind that Fe is strongly affected by dust depletion, which can be another reason for it to be weak. In addition, our findings could potentially be influenced by the omission of ionization corrections and dust depletions, parameters for which we lack adequate quantitative data in this study. In this regard, as suggested by \citet{Hernandez+2020}, the ionization correction for Si II tends to be positive (ranging between 0.009 and 0.052) for $\rm log(N_{HI})= 21.4 \pm0.2$. This indicates that the discrepancy observed in Si compared to Ni might be attributed to the ionization correction factor.

\begin{table*}
	\centering
	\caption{Rest frame equivalent widths ($\rm EW$) and column densities $\rm N_a$ [$\rm cm^{-2}$] of the absorption lines described in the text.}
	\label{tab:table_abs_lines}
	\begin{threeparttable}
	\begin{tabular}{lllccccc} 
		\hline
		Ion & ~~~~~$\lambda^{\tnote{a}}$ & ~~~${f}^{\tnote{a}}$ & $\rm EW$ & $\rm \delta\,EW$ & $\log\,(N/\rm cm^{-2})$ & $\log (\delta N/\rm cm^{-2})$\\
		
			 & ~~~~\,$(\si{\angstrom})$ & & $~(\si{\angstrom})$& $~(\si{\angstrom})$ &  &  \\
		\hline

C\,{\sc ii}& 1334.53 $^{\tnote{b}}$   & 0.129 & 1.4 & 0.3 & 15.04 & 0.03\\
O\,{\sc i} & 1302.17  $^{\tnote{b}}$ & 0.048 & 1.1 & 0.2 & 15.395 & 0.002\\ 
Al\,{\sc ii} & 1670.78 $^{\tnote{b}}$ & 1.74 & 0.8 & 0.2 & 13.43 & 0.02\\
Al\,{\sc iii}& 1854.72 $^{\tnote{c}}$& 0.561 &  0.4& 0.1 & 13.50&  0.02\\
Si\,{\sc ii} &1260.42 $^{\tnote{b}}$& 1.20 & 1.3& 0.2 &  14.10 & 0.02\\
Si\,{\sc ii} &1304.37 $^{\tnote{b}}$ & 0.091 & 0.3 & 0.1 & 14.73 & 0.09\\
Si\,{\sc ii}& 1526.72 $^{\tnote{b}}$~~ & 0.144~~ & 1.2& 0.4 & 14.81 & 0.05\\ 
Si\,{\sc ii}& 1808.00 & 0.00245  &  0.8 & 0.1 &  16.27 & $0.02$\\
Si\,{\sc iv} & 1393.76 $^{\tnote{b}}$ & 0.513  & 0.99 & $0.07$ &  14.3 & 0.04\\
Si\,{\sc iv}& 1402.77 $^{\tnote{b}}$ & 0.254  & 1.19& $0.08$ & 14.6 &  0.04\\
Ni\,{\sc ii} &  1370.13 & 0.0811 & 0.3& $0.1$ & 14.58 & 0.02\\
Ni\,{\sc ii} & 1741.55 & 0.0488  & 0.4 &  $0.1$ & 14.58 & $0.02$\\

		\hline
	\end{tabular}
	\begin{tablenotes}
  \item[a] Rest wavelengths and $f$-values from \cite{Cashman+2017}.
  \item[b] Saturated line.
  \item[c] We cannot tell its level of saturation because it is the only transition of Al$^{2+}$ that we detect.
\end{tablenotes}
   \end{threeparttable}
\end{table*}

\begin{table*}
	\centering
	\caption{Observed and dust corrected fluxes of nebular lines and absolute uncertainties. Fluxes and uncertainties are in units of $\rm 10^{-18}~erg~s^{-1}~cm^{-2}$.}
	\label{tab:table_na}
	\begin{threeparttable}
	\begin{tabular}{lccccc} 
		\hline
		Ion & Rest $\lambda$ (\si\angstrom) & Flux (observed)$^{\tnote{a}}$ & $\delta$F (observed) &  Flux (dust corrected) &  $\delta$F (dust corrected) \\
		
		\hline

He\,{\sc ii}$^{\tnote{b}}$& 1640.42 & 5.9 & 0.3 & 15.3&  0.9 \\\
O\,{\sc iii}]& 1660.81 & 4.5 &  1.0 & 11.8 & 2.3 \\\
O\,{\sc iii}]& 1666.15&  8.5 & 1.5 & 21.7  & 4.0 \\\
N\,{\sc iii}& 1751.91 & 6.3 & 1.3 & 16.1 & 3.3 \\\
Si\,{\sc ii}& 1883.00 & 3.6 & 1.0 & 9.5 & 2.3 \\\
[C\,{\sc iii}] &1906.68& 15.2 & 1.4 & 40.5 & 3.8\\\
C\,{\sc iii}] & 1908.73 & 11.9 & 1.2 & 31.7 & 3.1 \\\
[O\,{\sc ii}] & 3726.03 & 29.2 & 3.2 & 51.8 & 5.7 \\\
[O\,{\sc ii}]& 3728.81& 48.0 & 4.1 & 85.1 & 7.3 \\\
H$\gamma$& 4340.47 & 31.1 & 3.8 & 51.3 & 6.3 \\\
H$\beta$ &4861.33 & 69.6 & 7.0 & 107.4 & 11.0\\\
[O\,{\sc iii}]& 4958.91 & 225.5 & 18.1 & 343.9 & 27.7 \\\
[O\,{\sc iii}]& 5006.84 & 749.7 & 54.4 & 1137.3 & 82.6\\

		\hline
	\end{tabular}
	\begin{tablenotes}
 \item[a] Flux corrected for Galactic extinction.
  \item[b] Obtained by multiplying the He II  flux and error from \citet{Cabanac+2008} by 1.30, which is the conversion factor between our \OIIIuv\ and Cabanac's \OIIIuv\ lines (see Sect. \ref{subsec:dust}).
 
\end{tablenotes}
   \end{threeparttable}
\end{table*}

\section{Properties of the stellar populations}
\label{sect:stellar_spec}

{After analyzing the ISM properties of J0332, we now shift our focus to its UV continuum to determine the metallicity of its stellar populations.}

\subsection{Stellar metallicity}
\label{sect:stellar_metallicity}

The stellar metallicity of a galaxy ($\rm Z_{\star}$) serves as an indicator of the abundance of metals, primarily iron, within the photospheres of its stars. Typically derived from the stellar continuum and absorption lines, $\rm Z_{\star}$ represents an average value across all stellar populations contributing to the integrated light of the galaxy within the considered spectral range. Therefore, stellar metallicities derived from ultraviolet (UV) data reflect the iron abundance of young, massive O- and B-type stars, whereas stellar metallicities derived from optical/infrared (IR) data are indicative of the metal content of older stars.

We fit the UV spectrum of J0332 using the Binary Population and Spectral Synthesis (BPASS) models v2.0 \citep{Eldridge+2016}, which include massive binary stars. This older version of the BPASS models was adopted since it includes synthetic spectra with a continuous SFH, which is a reasonable representation of the SFH of high redshift star forming galaxies, and also provides a good fit to the observed photometry (see Section \ref{sec:sed}). We adopt a 100 Myr old continuous SFH and stellar metallicities in the range $Z=0.001-0.04$. We assume a Salpeter IMF \citep{Salpeter+1955} with an upper mass cut-off of 100 $M_{\odot}$. 
The 100 Myr old SFH is chosen to ensure that the synthetic UV spectra are stable against the fast evolution of very massive stars, which occurs on timescales shorter than $\sim$ 30 Myr. 

We compare the synthetic spectra and the data through a $\chi^{2}$ minimization. Before this step, we match the velocity dispersion, $\rm {\sigma}$, of the models with that of the stars in J0332. Specifically, we measure $\rm {\sigma}$ by fitting the \OIIIopt\, line in the available XSHOOTER spectrum with a Gaussian profile and obtain a value $\rm \sigma = 30 \pm 3 \,km\,s^{-1}$ (corrected for instrumental resolution). We attribute this small value of $\sigma$ to the fact that the nebular emission might be dominated only by one clump of star formation (as we will see for the \CIII\ emission in Sect. \ref{sec:ewciii}), while the stellar mass measurements use all the light. Since the BPASS models are designed to reproduce only the stellar components of a galaxy spectrum, we mask out the interstellar and nebular features from the data. We define the $\chi^{2}$ as:
\begin{equation}
   \rm \chi^2=\sum_{\lambda}{(O_{\lambda}-M_{\lambda})^2/e_{\lambda}^2}\, ,
    \label{eq:eq_chi2}
\end{equation}
where $\rm O_{\lambda}$ is the observed spectrum, $\rm M_{\lambda}$ is the model considered for the fit and $\rm e_{\lambda}$ is the error on the observed spectrum. {From 500 re-simulations of the J0332 spectrum, we find that only the two lowest metallicity models ($Z=0.001$ and $Z=0.002$) are  chosen as best fit models, therefore we conclude that the stellar metallicity of J0332 is in the range $Z=0.001-0.002$, which corresponds to  \Z $= 5-10\,\%\,Z_{\odot}$ (assuming the BPASS solar metallicity $Z_{\odot}=0.02$). In the following, we assume the value $Z_\star =0.0015$.}

\subsection{The stellar mass-stellar metallicity relation}

In the local universe, a tight relation between stellar mass and \Z\ has been established, mostly exploiting the vast spectral database provided by the Sloan Digital Sky Survey (SDSS; \citealp{York+2000}). According to this relation, less massive galaxies are characterized by lower stellar metallicities than more massive ones. The existence of this relation reflects the complex interplay between inflows, outflows and enrichment rate (e.g.,  \citealp{Cullen+2019}). 

While in the local Universe ground-based facilities provide high quality spectra from which the stellar metallicities can be easily derived, performing this type of study at high redshift is hampered by the low spectral S/N of distant sources. For this reason, the majority of high redshift stellar metallicity measurements have so far been inconclusive. Recently, \citet{Steidel+2016}, \citet{Cullen+2019}, \citet{Theios+2019}, \citet{Topping+2020a}, \citet{Calabro+2020}, \citet{Cullen+2021} and \citet{Kashino+2022} performed high-$z$ stellar metallicity analyses on \textit{composite} spectra. \citet{Cullen+2021} found that the metallicities increase from $\rm Z/Z_{\odot} < 0.09$ to $\rm Z/Z_{\odot} = 0.27$ across the stellar mass range $\rm 8.5 < log(M/M_\odot) < 10.2$ in the redshift range $z\sim3.5-4$ . Moreover, they observed a decrease of \Z\ by $\sim0.6$ dex at this redshift relative to local galaxies of similar mass.
Our result is one of the few which is obtained from an \textit{individual} spectrum at $z\sim4$.

\begin{figure}[h]
	\includegraphics[width=1.1\columnwidth]{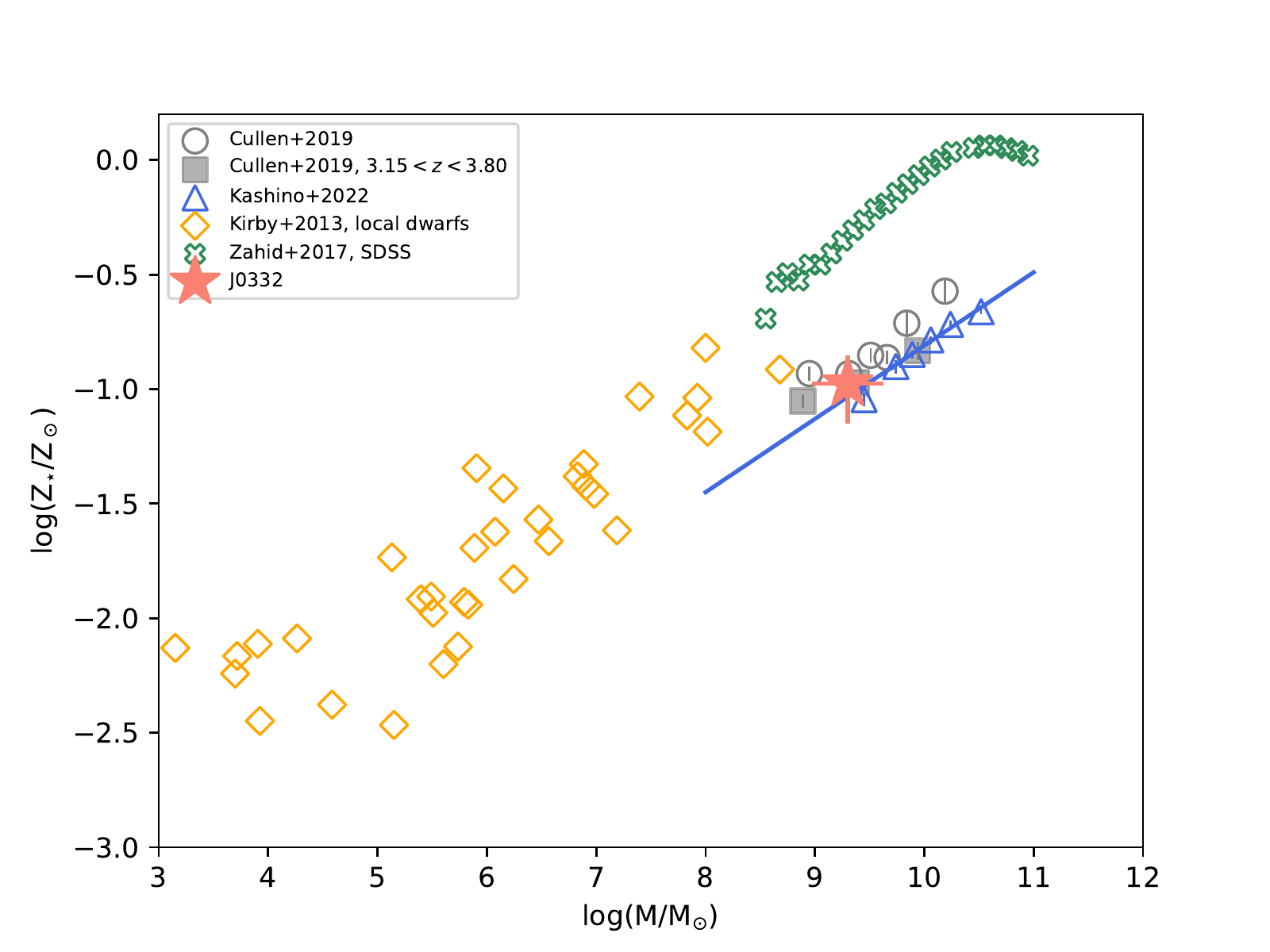}
      \caption{Stellar mass-\Z\ relation. The green crosses and yellow diamonds are local \Z\ values from \citet{Kirby+2013} and \citet{Zahid+2017}. Grey circles and blue triangles are the \citet{Cullen+2019} mass and mass+redshift bins, respectively (see text for further details). J0332 is marked as a pink star.}
   \label{fig:Zmass}
\end{figure}

Figure \ref{fig:Zmass} shows the stellar mass-\Z\ plane for J0332 and literature galaxies at different redshifts. Specifically, we compare J0332 to the local samples analyzed by \citet{Kirby+2013} and \citet{Zahid+2017} and to the the mass stacks of VANDELS galaxies at $2.5<z< 5$ by \citet{Cullen+2019}. The stellar metallicity of J0332 is compatible, within the uncertainties, with the results by \citet{Cullen+2019}. These authors also created stacks in bin of mass \textit{and} redshift. Our metallicity estimate is in agreement with their stacks in the range $3.15<z<3.80$ and $\rm 8.3 \lesssim log(M/M_{\odot}) \lesssim 10.8$. This result supports the hypothesis that the stellar mass-stellar metallicity relation does not strongly evolve over the redshift range $z\sim2.5-5$. 

Compared to the nearby samples by \citet{Kirby+2013} and \citet{Zahid+2017}, we find that the stellar metallicity of J0332 is $\sim0.65$ dex lower than that of local galaxies of similar mass and the difference is compatible with that found by \citet{Cullen+2019}. As mentioned before, UV-based and optical-based stellar metallicities do not necessarily trace the same stellar populations. Therefore, following \citet{Cullen+2019}, we compare J0332 to the HST Faint Object Spectrograph (FOS) and the Goddard High Resolution Spectrograph (GHRS) samples \citep{Leitherer+2011}. In particular, we consider the three galaxies He2-10a, He2-10b, and NGC4670. These galaxies have stellar masses (as measured by \citealp{Cullen+2019} from their $\rm M_{K}$) similar to J0332 and a high S/N per resolution element of 14, 6, and 12, respectively. Fitting the UV spectra of He2-10a, He2-10b, and NGC4670 using the same procedure described in Section \ref{sect:stellar_metallicity}, we derive that their metallicity is $\rm log(Z/Z_{\odot})=0.3$. We observe that J0332 lies 1.4 dex below the local metallicity values.

The evolution of the stellar mass-stellar metallicity relation with redshift can be explained in terms of the gas consumption timescale (which describes how efficiently gas is transformed into stars) and the mass loading factor (which describes the efficiency of the outflows). In particular, \citet{Cullen+2019} found that a high loading factor, more than a long depletion timescale, can explain the low stellar metallicities of VANDELS galaxies at $2.5<z<5$. The stellar metallicity of J0332 is compatible with this scenario.

\section{Properties of the H II regions}
\label{sec:ionized_gas}

We now move our analysis to the properties of the ionized gas in J0332 and the scaling relations characterizing it.

\subsection{Dust extinction correction}
\label{subsec:dust}
The findings presented in the subsequent Sections are based on the analysis of optical and UV emission lines. Given that UV lines are especially influenced by dust extinction, it is crucial to correct them for dust extinction. \textcolor{black}{First, we correct our emission lines for Galactic extinction. We assume a \citet{Cardelli+1989} extinction curve and use the E(B-V) produced by the Bayestar 3D dust map \citep{Green+2018} which, at the coordinates of J0332, produces a E(B-V) = 0.024.}
After applying this correction, we correct the emission line fluxes for the dust within J0332. Since the $\rm H\alpha$ line is not captured by the XSHOOTER spectrum, we cannot use the Balmer decrement $\rm H\alpha/H\beta$ to estimate the dust extinction. We instead use the ratio $\rm H\gamma/H\beta$. Adopting the PyNeb v1.1.15 \citep{Luridiana+2015} \textit{get\_emissivity} task and the values of electron temperature in the low ionization zone $\rm T_{e, low}$ and $\rm n_{e, low}$ (see Section \ref{sec:Te}), we  derive an intrinsic $\rm H\gamma/H\beta=0.47\pm0.0004$ (our observed value is $\rm H\gamma/H\beta=0.44\pm0.07$). Adopting the PyNeb \textit{RedCorr} task and assuming a \citet{Cardelli+1989} extinction curve, we derive a color excess $E(B-V)=0.13\pm0.30$\footnote{Note that the large uncertainty is probably due to the relatively small wavelength range between $\rm H\gamma$ and $\rm H\beta$}. This value is in agreement with that obtained from the SED fitting (see Section \ref{sec:sed}). Note that, based on the large equivalent widths of the emission lines, we assume no underlying absorption in H$\beta$ and H$\gamma$. 

\subsection{Electron temperature, density and star formation rate}
\label{sec:Te}
We adopt the 1.1.15 version of PyNeb and the \textit{get\_temden} task to derive the electron temperature of the high ionization region from the \OIIIuv\ and \OIIIopt\ lines. We obtain $T_{e,{\rm high}} = 13,307 \pm635\,K$. As a second step, we assume the $T_e-T_e$ relationships by \citet{Stasinska1982} to derive the electron temperature in the low ionization region, finding a value $T_{e,{\rm low}} = 12,315 \pm444\,K$. 
We derive the electron density of the high ionization region using the  $\rm C III] \lambda 1909 / C III] \lambda 1906$ ratio, obtaining $\rm log(n_{e,{\rm high, C III]}})= 3.8 \pm 0.6$.

The SFR is derived from the dust and magnification-corrected $\rm H\beta$ line using the \citet{Theios+2019} SFR - L(H$\alpha$) relation (their Eq. 6), adopting L(H$\alpha$) = 2.8~L(H$\beta$). This relation assumes t = $10^8 yr$, $Z_{\star} = 0.002\,Z_{\odot}$ and an IMF with a high mass cutoff $\rm M= 100\,M_{\odot}$. We find that J0332 has a $\rm SFR(\rm H\beta)=4.55\pm 0.46\,M_{\odot}\,yr^{-1}$. This corresponds to a sSFR of $\rm 2.1 \pm 0.1\,Gyr^{-1}$ (log(sSFR) = $\rm -8.66\pm0.32\,yr^{-1}$). This value is compatible with the expected sSFR at $z\sim4$ \citep{Khusanova+2020, Lehnert+2015}. 

\subsection{Ionization parameter}
Optical and UV lines can be used to infer the ionization state of a galaxy, which informs us about the excitation level of the H II regions. The ionization state is usually parameterized by means of the ionization parameter ($ U \propto q/R$), where $q$ is the emission of ionizing photons and $R$ is the size of the H II region. In the optical regime, a variety of diagnostics have been defined to derive $U$. For J0332, we derive log(U) by adopting the \Oratio\ ratio and the photoionization models fit coefficients derived by \citet{Berg+2019}. We find a high value of $\rm log$(U) $=-2.0\pm0.08$. This result suggests the presence of a hard ionizing spectrum in J0332, which aligns with its young age and low stellar metallicity (see Sections \ref{sec:sed} and \ref{sect:stellar_metallicity}).

\subsection{Direct-metallicity estimate}
\label{sec:direct}
The metallicity of the ionized gas in a galaxy is usually parameterized through the abundance of oxygen, which is the third most abundant element after hydrogen and helium. The most robust method to derive the gas-phase metallicity is through the ``direct method'', which adopts the flux ratio of auroral to strong lines as a proxy of the gas electron temperature. In turn, since metals are the primary coolants in \HII\ regions, $\rm T_{e}$ is strictly linked to the gas metallicity, with hotter electron temperatures corresponding to lower metallicities.

In J0332, we detect the auroral $\rm OIII]~\lambda1666$ line with $\rm S/N\sim 5$, and we are therefore able to apply the direct method. Specifically, we use the ratio between \OIIIuv\ and \OIIIopt\ to determine the electron temperature (see Section \ref{sec:Te}), and then the strengths of \OIIopt\ and \OIIIopt\ to obtain the O$^+$/H$^+$ and  O$^{2+}$/H$^+$ ionic abundances, respectively. We adopt the ionization corrections derived by \citet{Berg+2019} to account for contributions of ions in different ionization states, in particular O$^{3+}$. This contribution is usually considered to be negligible since O$^{3+}$ has a high ionization energy of 54.9 eV. However, the presence of the high ionization He II line in the spectrum of J0332 (as seen by \citealp{Cabanac+2008}) suggests that this correction must be taken into account. We use the atomic data by \citet{Aggarwal+1999} to define the radiative and collisional transition probabilities. From the direct method, we derive a gas-phase metallicity $\rm 12+log(O/H)=8.26\pm0.06$. {This value corresponds to 0.37 $Z_{\odot}$ (if the \citealp{Asplund+2021} solar abundance $Z_\odot = 12 + $log(O/H) = 8.69 is assumed).}

\subsection{Strong line estimate}

\textcolor{black}{An alternative way to derive \Zg\ consists in using calibrated strong optical line ratios such as the $\rm [O III] \lambda 5007/[O II] \lambda\lambda3726,9$ (O32) and the $\rm R23=[O III] \lambda5007 + [O III] \lambda 4959/H\beta$ ratio. The calibrations can be empirical (i.e., based on direct metallicity measurements - e.g.,  \citealp{Pettini_Pagel2004}; \citealp{Marino+2013}; \citealp{Pilyugin_Grebel2016}), theoretical (i.e., based on photoionization models - e.g.,  \citealp{Kobulnicky_Kewley2004}; \citealp{Tremonti+2004}; \citealp{Dopita+2013}, \citeyear{Dopita+2016}), or a combination of the two.} Unfortunately, the calibrations are usually not perfect, and the metallicities estimated through different calibrations present large discrepancies, even for the same sample of objects, with variations up to $\sim$ 0.6 dex (\citealp{kewley_ellison2008}; \citealp{Moustakas+2010}). This is, in part, due to the fact that strong emission lines are hardly pure metallicity indicators, being often also probes of the ionization parameter. 
At high redshift, one significant source of uncertainty in calibrations arises from the varying physical conditions that generate strong emission lines in galaxies across cosmic time. 
As a result, the applicability of calibrations derived in the local universe to high-redshift regimes cannot be assured. 
{Nevertheless, before the advent of the James Webb Space Telescope (JWST), using local analogs of high-redshift galaxies, presumed to share similar physical characteristics with their distant counterparts, was the primary approach \citep{Bian+2018, Jiang+2019}. Fortunately, the launch of JWST has considerably augmented the dataset of high-redshift galaxies available for calibration purposes \citep[e.g.,][]{Heintz+2023, Laseter+2023, Sanders+2023c, Sanders+2024}, bringing us closer to the direct calibration of strong methods in high-redshift galaxies.}

Here we derive the gas metallicity of J0332 using the O32 ratio and the empirical calibrations by \citet{Maiolino+2008}, \citet{Curti+2017}, and \citet{Bian+2018}. We do not adopt the R23 index since our value of metallicity is around the transition region between the low and high metallicity branches of the R23 calibration (e.g.,  \citealp{Nagao+2006}), making it not very sensitive to metallicity in this range. Moreover, our wavelength range does not include either the H$\alpha$ or the [N II] $\lambda$ 6584 lines. {By comparing the strong line and direct gas metallicities, we derive $\rm \Delta[logO/H_{O32}- log(OH)_{T_e}]_{Bian}=-0.33$ dex, $\rm \Delta[logO/H_{O32}- log(OH)_{T_e}]_{Curti}=0.50$ dex and $\rm \Delta[logO/H_{O32}- log(OH)_{T_e}]_{Maiolino}=0.76$ dex (see Table \ref{tab:metallicities}). These findings suggest that various calibrations can yield significantly divergent values for gas metallicity. Furthermore, they support the scenario that the physical conditions of galaxies at earlier epochs could be different, implying that calibrations employed at lower redshifts may not be applicable at higher redshifts.}

\begin{table*}[]
\caption{{Oxygen abundances 12+log(O/H) inferred for J0332 throughout this paper.}}
\label{tab:metallicities}
\begin{threeparttable}
\begin{tabular}{@{}ccccc@{}}
\toprule
stellar &
  gas phase  (direct) &
  \begin{tabular}[c]{@{}c@{}}gas phase (O32)\tnote{b}\\ \citet{Bian+2018}\end{tabular} &
  \begin{tabular}[c]{@{}c@{}}gas phase (O32)\tnote{c}\\ \citet{Curti+2017}\end{tabular} &
  \begin{tabular}[c]{@{}c@{}}gas phase (O32)\tnote{c}\\ \citet{Maiolino+2008}\end{tabular} \\ \midrule
7.52-7.83$^{\tnote{a}}$ &
  $8.26 \pm0.06$ &
  $7.93 \pm0.06$ &
  $7.76 \pm 0.07$ &
  $7.5 \pm 0.07$ \\ \bottomrule
 
\end{tabular}
 \begin{tablenotes}
 \item[a] These two values correspond to 5 and 10 $\%$ of the solar metallicity assumed by the BPASS models $Z=0.02$ (which corresponds to 12+log(O/H)=8.83).
 \item[b] {O32 is defined as [\ion{O}{3}] $\lambda$4959,5007/[\ion{O}{2}] $\lambda$3727,9.}
 \item[c] {O32 is defined as [\ion{O}{3}] $\lambda$5007/[\ion{O}{2}] $\lambda$3727,9.}
\end{tablenotes}
\end{threeparttable}

\end{table*}

\subsection{Ionizing source}
{Over the past few decades, there has been a growing interest in UV diagnostics due to their ability to provide insights into the ionization source of  galaxies at high redshift, where standard optical diagnostic diagrams struggle to differentiate between stellar and AGN activity at higher redshifts \citep[e.g.,][]{Groves+2006, Coil+2015, Feltre+2016, Hirschmann+2017}.}

\begin{figure}[h]
	\includegraphics[width=1.\columnwidth]{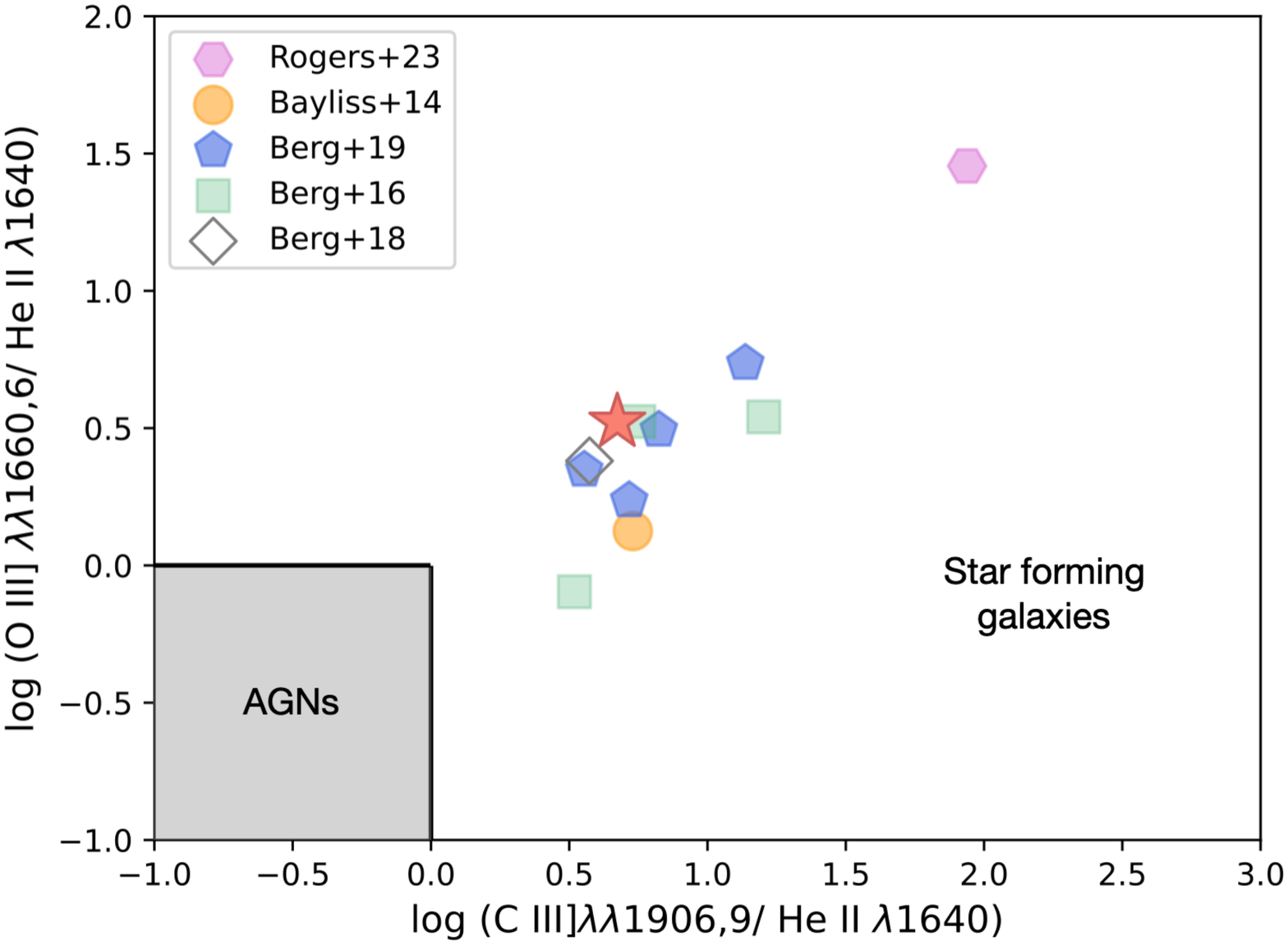}
      \caption{\ion{O}{3}]/\ion{He}{2} vs. \ion{C}{3}]/\ion{He}{2} plane. Besides J0332 (pink star), we show results for galaxies at different redshifts studied by \citet{Bayliss+2014,Berg+2016,Berg+2018, Berg+2019, Rogers+2023}. The grey shaded area marks the region where AGNs lie according to \citet{Feltre+2016}.}
   \label{fig:UV}
\end{figure}

{Specifically, the  UV diagram \ion{O}{3}] $\lambda\lambda1660,6$/\ion{He}{2} $\lambda1640$ vs. \ion{C}{3}] $\lambda\lambda$1906,9/\ion{He}{2} $\lambda1640$ proves particularly valuable in distinguishing between star formation and AGN-powered sources. This is attributed to He II having a higher ionization potential than \ion{O}{3}] and \ion{C}{3}], and therefore requiring the more powerful ionization background found in active galaxies to be produced.}
{In this diagram, AGNs tend to exhibit \ion{C}{3}]/\ion{He}{2} $<$ 0 and \ion{O}{3}]/\ion{He}{2} $<$ 0, whereas star-forming galaxies display higher values for these ratios ($>$ 0, \citealp{Feltre+2016}). Figure \ref{fig:UV} illustrates a compilation of data for galaxies at various redshifts.} {Due to the gap between the two gratings (see Sect. \ref{sec:fors2}) that precludes the detection of \ion{He}{2} emission, we estimate the ratio \ion{C}{3}]/\ion{He}{2} in J0332 using the \ion{He}{2} flux determined by \citet{Cabanac+2008} and rescaling it on the \ion{O}{3}] emission that the two spectra have in common.
We observe that, based on this diagnostic, J0332 appears to be powered by star formation (see Figure \ref{fig:UV}).}

{Benefitting from the broad wavelength range covered in our observations, we can compare the results derived from UV diagnostics with those obtained through optical diagnostics. Specifically, we measure the  O32 ([\ion{O}{3}] $\lambda$4959 + [\ion{O}{3}] $\lambda$5007/ \ion{O}{2} $\lambda\lambda$3726,9) and the R23 ([\ion{O}{3}] $\lambda$4959 + [\ion{O}{3}] $\lambda$5007 + \ion{O}{2} $\lambda\lambda$3726,9/H$\beta$) ratios for J0332 (though we caution about the lower sensitivity of R23 to metallicity in the considered range). We find that J0332 has O32 $\sim10.8\pm0.11$ and R23 $\sim15\pm0.12$. These elevated values support the findings from the UV, suggesting that J0332 is characterized by a hard ionizing continuum. The presence of a hard ionizing continuum in this galaxy may be attributed to the presence of Wolf-Rayet stars, which are capable of elevating O32 beyond 1 \citep{Barrow+2020}. This interpretation gains further support from the detection of the high ionization line \ion{He}{2}. It is important to highlight, as we conclude this Section, that these findings are influenced by the challenge faced by many photoionization models, which are often unable to accurately replicate intense nebular \ion{He}{2} emission \citep{Berg+2018, Kehrig+2018, Nanayakkara+2019, Saxena+2020}. However, it is worth mentioning that newer models may present advancements in this regard \citep[e.g.,][]{Lecroq+2024}.}

\subsection{Stellar mass-gas metallicity relation}
\label{sect:mzrg}

 The stellar mass–\Zg\ relation (\MZRg) is one of the most important scaling relations observed in the local and high-redshift Universe, and shows the existence of a trend between galaxy stellar mass and gas-phase metallicity, with \Zg\ increasing for increasing stellar mass. In the local Universe, the correlation between stellar mass and oxygen abundance has been defined with a scatter of only 0.1 dex in log(O/H) (e.g.,  \citealp{Tremonti+2004}) and extends for over five orders of magnitude in stellar mass, from $M_{\star} = 10^6-10^{11} M_{\odot}$ \citep{Lee+2006, Berg+2012}.
The stellar mass-\Zg\ relation has been observed out to $z\sim10$, and evolves such that O/H decreases with increasing redshift at fixed stellar mass \citep{Mannucci+2010, Steidel+2014, Troncoso+2014, Sanders+2015, Sanders+2020a, Hunt+2016, Onodera+2016, Suzuki+2017, Langeroodi+2022, Nakajima+2023}.  Up to $z\sim2.5$, literature studies agree in finding a slow evolution of the \MZRg, with O/H $\sim0.3$ dex lower than at $z\sim0$ at a fixed stellar mass (e.g., \citealp{Erb+2006a}; \citealp{Steidel+2014}; \citealp{Sanders+2015}). However, a general consensus on the evolution rate of the \MZRg\ at $z>2.5$ is yet to be reached. Some authors find a rapid decrease of the gas metallicity above these redshifts, with a drop $\rm \Delta_{12+log(O/H)}\sim-0.7$ dex between $z\sim0$ and $z\sim3$, and a drop of $0.3-0.4$ dex between $z\sim2.5$ and $z\sim3.5$ (\citealp{Maiolino+2008}; \citealp{Mannucci+2010}; \citealp{Troncoso+2014}; \citealp{Onodera+2016}). On the contrary, other studies show very little evolution between $z\sim2$ and $z\sim3.2$ and a general shallower decrease of log(O/H) over the whole redshift range $z\sim 0-3.5$ (e.g.,  \citealp{Suzuki+2017}, \citealp{Sanders+2021}). {Thanks to JWST, the redshift evolution of $\rm MZR_{g}$ has been delineated up to $z\sim4-10$ \citep[e.g.,][]{Langeroodi+2022, Curti+2023b, Nakajima+2023}. In comparison to local $\rm MZR_{g}$, galaxies at these higher redshifts exhibit noticeably lower metallicity levels for a given stellar mass. The reduction is typically around $\sim$ 0.5 dex for stellar masses $\sim10^{9}~M_{\odot}$, but it diminishes at the lower-mass range to $\sim$ 0.3 dex. In addition, \citet{Langeroodi+2023c} find a decrease of the gas metallicty of $\sim0.9$ dex at $z\sim8$ compared to the local one.}

\begin{figure}
\centering
\includegraphics[width=1.1\columnwidth]{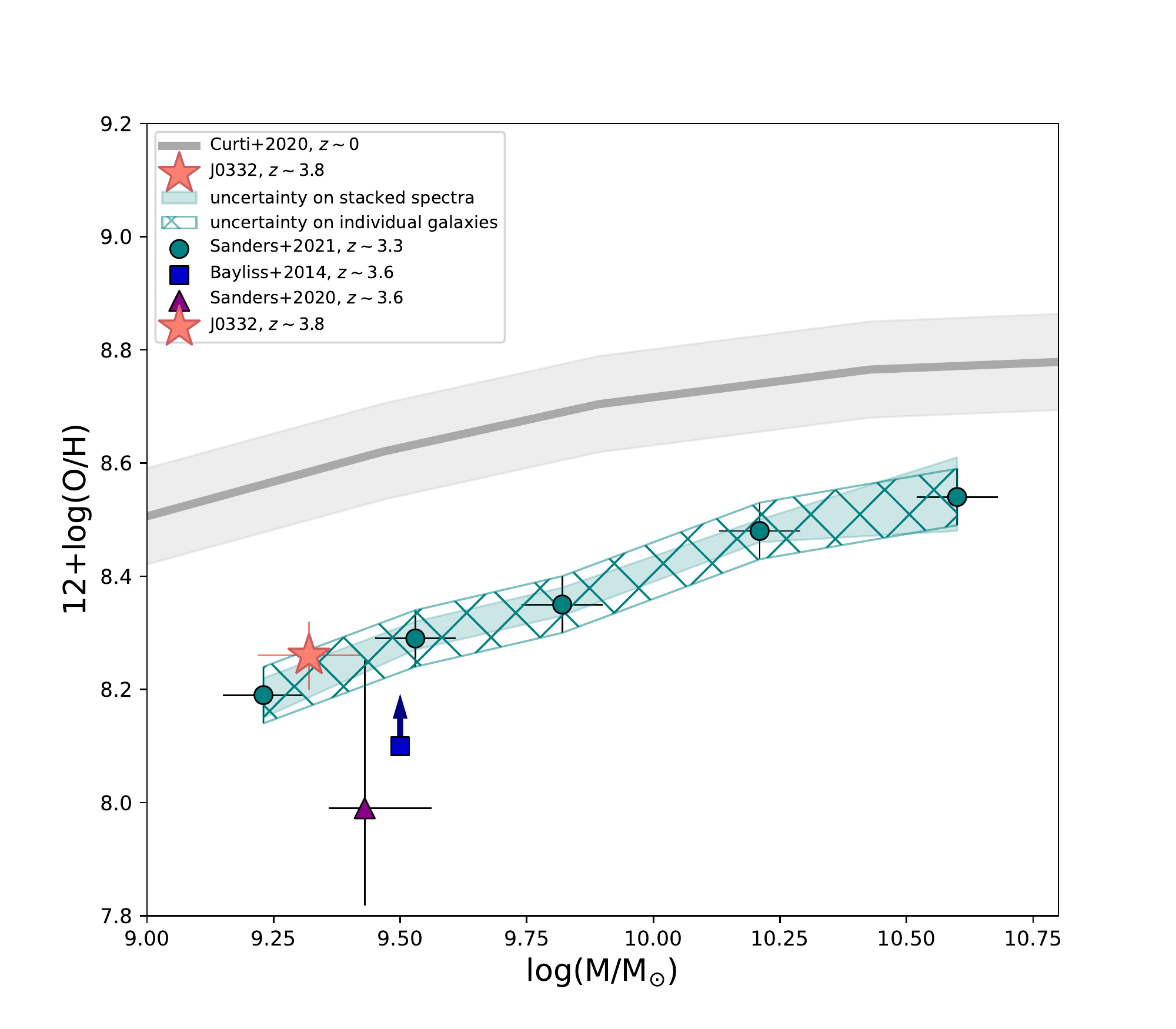}
      \caption{Stellar mass-\Zg\ relation at different redshifts. The grey curve and shaded area is the direct-method calibrated $z\sim0$ \MZRg\ by \citet{Curti+2020}. The pink star is J0332; the green circles are the stacked spectra measurements by \citet{Sanders+2021}. The shaded area shows the uncertainty on the stacked spectra, while the hatched area shows the results on the individual 150 galaxies of their sample. The square and triangle are the literature results by \citet{Bayliss+2014} and \citet{Sanders+2020a}, respectively. Note that the value by \citet{Bayliss+2014} is a lower limit.}
   \label{fig:mzr}
\end{figure}

Figure \ref{fig:mzr} shows J0332 within the \MZRg\ plane. We compare J0332 with the local relation by \citet{Curti+2020}, which is based on strong-line diagnostics calibrated on $\rm T_{e}$-based measurements. The plot also illustrates the results from \citet{Sanders+2021}. These authors stacked a subsample of $\sim$ 150 galaxies at $z\sim3.3$ from the MOSDEF survey in mass bins. They calibrated their strong-line metallicity measurements with direct ones employing the $z\sim2$ local analogue-based calibrations by \citet{Bian+2018}. We also show the direct metallicity estimate obtained for SGAS J105039.6+001730, a gravitationally lensed galaxy at $z=3.6252$ \citep{Bayliss+2014}, and that for COSMOS-23895, a gravitationally lensed galaxy at $z\sim3.3$ \citep{Sanders+2020a}. 

We find that J0332 lies $\sim$0.3 dex below the \citet{Curti+2020} local relation. This result is compatible, within the uncertainties, with what found by \citet{Bayliss+2014},  \citet{Sanders+2020a} and \citet{Sanders+2021}.  
The decrease of \Zg\ with redshift can be traced back to the interplay between three main factors: (i) the fraction of gas $\rm \mu_{gas}$ {(defined as the ration between the gas and stellar mass $M_{\rm gas}/M^{\star}$)}, which is indicative of how diluted the metals are, (ii) the star formation efficiency $\rm \epsilon=SFR/M_{gas}$ - i.e.,  how much of the gas (and metals) is turned into stars and consequently returned to the ISM, and (iii) the outflow metal loading factor - i.e.,  the amount of metals ejected in outflow events.

{The gas fraction $\mu_{\rm gas}$  has been recently observed to strongly evolve with redshift as $\mu_{\rm gas}\propto (1+z)^{2.5}$ \citep{Tacconi+2018}.
The star formation efficiency $\epsilon$ has been observed to increase with redshift as well, but the rate of its evolution is less clear. For example, \citet{Gribel+2017} find that $\epsilon$ is almost constant within the redshift range $\sim 3.5-20$, while it decreases rapidly for $z<3.5$. Conversely, \citet{Genzel+2015} find $\epsilon \propto (1+z)^{0.34}$, while \citet{Onodera+2016} find no evolution of $\epsilon$.
Cosmological models tend to disfavor a weakly evolving $\rm\epsilon$, since such an evolution seems to underpredict the observed metallicities at $z\sim1.5-2.5$ (\citealp{Mannucci+2010}, \citealp{Sanders+2018}, \citealp{Curti+2020}). From the direct method, we infer a \Zg\ for J0332 which is only $\sim$ 0.3 dex lower than that of local galaxies at similar masses. This shallow \Zg\ vs. redshift evolution is consistent with a strong evolution of $\rm \epsilon$, which compensates for the higher $\rm \mu_{gas}$ at high redshifts.}

The location of J0332 within the stellar mass-gas metallicity plane is also compatible with the scenario recently presented by \citet{Sanders+2021},  where the redshift evolution of \Zg\ is attributed to an increase of \textit{both} the  gas fraction \textit{and} the outflow efficiency. It is important to point out that the models from \citet{Sanders+2021} assume that gas inflows are negligible. If inflows are included, the redshift evolution of the mass-metallicity relation can be explained without increasing the outflow loading factor (the inflow scenario was first suggested by \citealp{Dave+2011}). However, it is worth noting that observations can only be reproduced by an increase of $\rm Z_{inflow}/Z_{ISM}$\footnote{$\rm Z_{inflow}$ is the metallicity of the inflowing gas and $\rm Z_{ISM}$ is the metallicity of the surrounding ISM.} from $\sim0$ to $\sim0.5$ from $z\sim3$ to $z\sim0$ (assuming gas is $\sim$ pristine at $z\sim3$). The $z\sim0$ is higher than  observed in HI clouds around the Milky Way ($\rm Z_{inflow}/Z_{ISM}\sim0.1$ - \citealp{Sancisi+2008}).\\

\subsection{Fundamental metallicity relation}
\label{sect:fmr}

{In addition to the scaling relation between stellar mass and gas metallicity, a secondary dependence of the \MZRg\ on star-formation rate (SFR) has been observed at $z\sim0$. The first evidence for an anti-correlation between O/H and (s)SFR at fixed stellar mass was reported by \citet{Ellison+2008}. Later, the full concept of a ``Fundamental Metallicity Relation'' was introduced by \citet{LaraLopez+2010} and \citet{Mannucci+2010}}. This secondary dependence suggests that, at a fixed stellar mass, galaxies with higher SFRs are characterized by lower O/H. The most common interpretation for the existence of the FMR is the inflow of pristine gas from the intergalactic medium, which increases the SFR while diluting the metallicity of the ISM. Even though these observations are supported by both semi-analytic models and numerical simulations (\citealp{Yates+2012}; \citealp{Torrey+2018}; \citealp{DeLucia+2020}), some literature studies argue that the local \MZRg\ has a stronger secondary dependence on the gas content rather than the SFR \citep{Bothwell+2013}, and that the SFR-defined FMR is a projection of this more fundamental relation (e.g., \citealp{Brown+2018}). The evolution rate of the FMR with redshift has not been clearly defined yet. \citet{Mannucci+2010} found that galaxies up to $z\sim2.5$ lie on the same FMR as local galaxies, and their result is supported by recent work based on larger samples and more uniform analyses of metallicity up to $z\sim2.5$ (\citealp{Sanders+2018}; \citealp{Cresci+2019}; \citealp{Curti+2020}) and $z\sim3.3$ \citep{Sanders+2021}.

However, other authors find a significant evolution of the FMR over the redshift range $z\sim0 - z\sim3$: for example, \citet{Troncoso+2014} analyzed a sample of 40 star-forming galaxies at $z\sim3.4$ from the AMAZE and LSD ESO programs and found that a significant fraction of these galaxies are located up to a factor of ten below the local FMR. \citet{Onodera+2016} found that $3<z<3.7$ galaxies are a factor 5 more metal poor than local galaxies of similar masses.
The strong redshift evolution of the FMR could be a physical phenomenon. However, another possible explanation could be the different calibrations used to derive \Zg\ at different redshifts (see \citealp{Sanders+2021}). In fact, theoretical calibrations based on photoionization models tend to yield metallicities that are $\sim0.25$ dex higher than direct-method calibrations (e.g., \citealp{kewley_ellison2008}).

\begin{figure}

	\includegraphics[width=1.1\columnwidth]{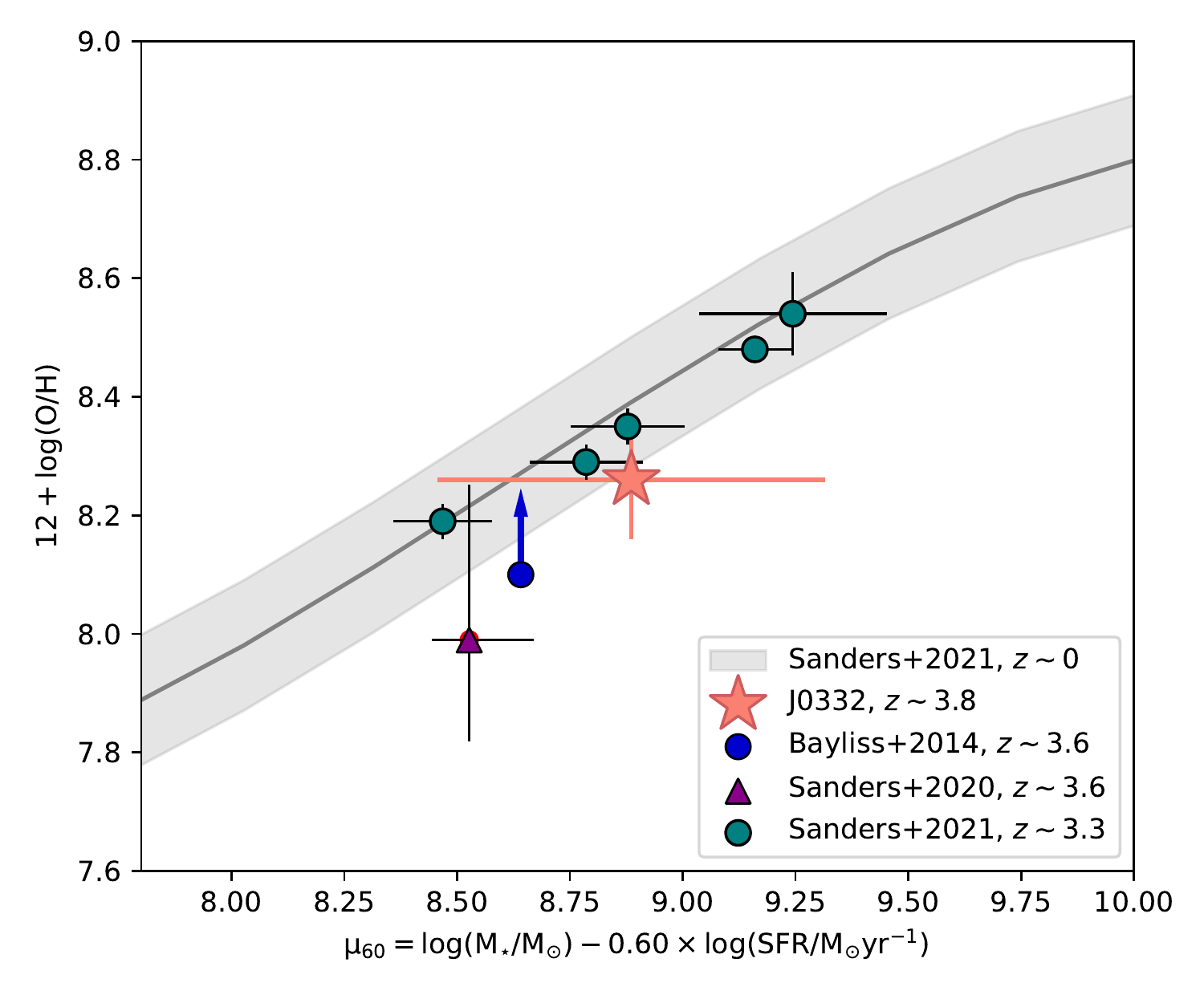}
      \caption{Projection of the fundamental metallicity relation at $z\sim0$ (from \citealp{Sanders+2021}) compared with data at higher redshift. Green circles are the stacked spectra by \citet{Sanders+2021} at $z\sim3.3$ and $2,3$, respectively. The pink star is our direct metallicity estimate for J0332. The blue and violet circles are the results obtained for SGAS J105039.6+001730 and COSMOS-23895, respectively. Note that the value from \citet{Bayliss+2014} is a lower limit.}
   \label{fig:fmr}
\end{figure}

Figure \ref{fig:fmr} shows the projection of the local FMR relation as O/H vs. $\rm \mu_{0.60} = log(M/M_{\odot}) - 0.60\times log(SFR/M_{\odot}\,yr^{-1})$ taken from \citet{Sanders+2021}. 
These authors derived the SFR from Balmer lines ($\rm H\alpha$ or $\rm H\beta$ at higher redshift). They parametrize the $z\sim0$ FMR using the method described in \citet{Mannucci+2010}, i.e.,  defining:
\begin{equation}
   \rm \mu_{\alpha}=(M/M_{\star}) - \alpha\times log(\frac{SFR}{M_{\odot}yr^{-1}}) \, ,
   \end{equation}
where $\alpha=0.60$ is the value that minimizes the scatter in $\rm O/H$ at fixed $\mu_{\alpha}$.

Adopting the SFR derived from $\rm H\beta$ (SFR = 4.55 $\pm$ 0.46 $\rm M_{\odot}\,yr^{-1}$), we find that the location of J0332 is compatible, within the uncertainties, with the local FMR and also with SGAS J105039.6+001730 \citep{Bayliss+2014} and COSMOS-23895 \citep{Sanders+2020a}. This result suggests a redshift invariant FMR up to $z\sim4$.
According to the standard intepretation of the FMR relation, the invariance implies that galaxies with same stellar mass and SFR have similar fractions of pristine gas. According to \citet{Sanders+2021}, the invariance of the FMR indicates instead that galaxies with same mass and SFR have \textit{both} similar gas fractions \textit{and} similar outflow metal loading factors.

\section{Gas-phase vs. stellar metallicity}
\label{sec:gas_vs_stellar_Z}

In the previous Sections, we have derived the stellar and the gas-phase metallicity of J0332 using the UV full-spectrum fitting and the direct method, respectively. While the stellar metallicity is a measurement of the iron abundance in the photospheres of young, massive, O- and B-type stars in the galaxy, the direct \Zg\ traces the abundance of oxygen surrounding the same young stellar populations. We find that the two metallicities differ by a factor 3-4, with the gas metallicity being higher than the stellar one\footnote{We remind that the stellar metallicity is derived using BPASS models, which assume a solar metallicity $Z_{\odot}=0.02$ (i.e.,  $\rm 12+log(O/H)=8.83.$)}. Discrepancies between ISM and stellar metallicities have been observed both in the local \citep{GonzalezDelgad0+2014} and high redshift universe \citep{Lian+2017}. However, it is worth reminding that some authors do not find such discrepancies (e.g.,\ \citealp{Kudritzki+2014}, \citeyear{Kudritzki+2015}).

One potential explanation for the observed discrepancy might be an overabundance of oxygen, leading to a higher gas-phase metallicity. Such an overabundance can be traced through oxygen or other $\alpha$-elements (such as carbon, neon, magnesium, silicon, and sulfur) produced via the same $\alpha$-capture process as oxygen. \citet{Steidel+2016} discovered evidence of oxygen enrichment, observing $\rm (O/Fe) \sim 4-5 \times (O/Fe){\odot}$ in galaxies at $z \sim 2.4$. Similarly, \citet{Cullen+2019} identified enhanced O/Fe ratios around $\rm (O/Fe) \sim 1.8 \times (O/Fe){\odot}$ in star-forming galaxies at $z \sim 2.5$. Evidence of $\alpha$ enhancement at $z \sim 2.3$ has also been reported by \citet{Topping+2020a} and \citet{Topping+2020b}. Additionally, \citet{Becker+2012} identified enhanced O/Fe ratios in Damped Ly$\alpha$ systems at $z \gtrsim 6$. Recently, \citet{Cullen+2021} analyzed the combined FUV + optical spectra of a sample of 33 star-forming galaxies from the NIRVANDELS survey (VANDELS+MOSFIRE) within the redshift range $2.95 < z < 3.80$ and found evidence for $\rm (O/Fe)=2.54 \pm 0.38 \times (O/Fe)_{\odot}$, with no apparent dependence on the stellar mass. Our findings, described in Section \ref{sect:chemical_composition} and illustrated in Figure \ref{fig:abd}, suggest that J0332 is $\alpha-$enhanced, with Si being more abundant than Fe and Ni. Therefore, we can conclude that the discrepancy between stellar and gas metallicity is due to J0332 being oxygen-enriched.

\begin{table}[]

\centering
\caption{Summary of the J0332 properties derived in this work.}
 \begin{threeparttable}
\begin{tabular}{ll}

\hline
\multicolumn{1}{c}{Property} & \multicolumn{1}{l}{value} \\ \hline
\hline
redshift                       &        $3.7732\pm0.0002$                    \\
\vspace{1mm}
age [Myr]\tnote{a}                           &         $93^{+238}_{-63}$                   \\
\vspace{1mm}
$\rm log(M/M_{\odot})$\tnote{a}                 &          $9.32^{+0.33}_{-0.32}$                  \\
\vspace{1mm}
$\rm O^{+}/H^+$ ($10^5$)           &           $2.11\pm0.43$             \\
\vspace{1mm}
$\rm O^{++}/H^+$  ($10^5$)          &     $17.20\pm3.77$                   \\
\vspace{1mm}
$\rm C^{++}/O^++$           &            $0.09\pm0.07$            \\
\vspace{1mm}
C ICF          &               $0.97 \pm 0.20$         \\

\vspace{1mm}
12+log(O/H) (direct)               &          $8.26 \pm 0.06$                  \\
\vspace{1mm}
log(C/O)                      &        $-1.02\pm0.2$                   \\
\vspace{1mm}
SFR($\rm H\beta$) [$\rm M_{\odot}\,yr^{-1}$]                           &             $4.55\pm0.46$   \\
\hline
\end{tabular}
\begin{tablenotes}
\item[a] Calculated assuming a constant SFH.
\end{tablenotes}
 \end{threeparttable}
\end{table}

\section{The carbon-to-oxygen ratio}
\label{sect:co}
As explained in Section \ref{sec:gas_vs_stellar_Z}, a useful way to study the chemical enrichment and SFH of a galaxy is by means of the relative abundances of elements produced by stars of different mass, since they trace different timescales for star formation. Among these elements, carbon and oxygen are particularly relevant to trace the \textit{early} SFH. In fact, oxygen is almost entirely produced by massive stars ($\rm M>\,8\,M_{\odot}$) and ejected into the interstellar medium via Core-Collapse Supernovae explosions; carbon can instead be released into the ISM by both massive stars ($\rm M>\,8\,M_{\odot}$) through Type II Supernovae explosions and by low-intermediate-mass stars ($\rm 1< M < \,8\,M_{\odot}$) through the convective dredge-up of freshly-synthesized carbon during the AGB phase (see review by \citealp{Nomoto+2013}). The relative C/O abundance has been investigated for many years as a function of the gas metallicity both for stars in the Milky Way (e.g.,\ \citealp{Mattsson2010}) and galaxies \citep[e.g.,][]{Berg+2016, Berg+2019}; however, the relative contribution of the two carbon production/release channels (massive vs. low/intermediate mass stars) has not been clearly defined yet \citep[e.g.,][]{Chiappini+2003, Mattsson2010}. In fact, at any given metallicity, the observed C/O abundance is a snapshot of the carbon and oxygen produced and released into the galaxy's ISM up to that point. The difficulty in distinguishing between the two paths derives from the fact that at low redshift galaxies are old enough to have produced carbon through both channels. 

An increasing trend of C/O with 12+log(O/H) has been observed in the past decades \citep[e.g.,][]{Chiappini+2003, Mattsson2010, Berg+2019}. Both production channels can explain this behaviour: if carbon is mostly produced by intermediate mass stars and released into the ISM on longer timescales than oxygen, then the C/O abundance  might build up as O/H does. However, if carbon is mostly supplied by massive stars through metallicity-dependent stellar winds \citep[e.g.,][]{Garnett+1999, Henry+2000a, Chiappini+2003}, than the C/O abundance increases with O/H as well. Since the launch of James Webb Space Telescope (JWST), the opportunity to derive the individual contributions from massive and low-intermediate mass stars to the carbon production has become more concrete, thanks to the possibility of observing rest-UV spectra of galaxies at extremely high redshifts. At such early epochs, galaxies are too young to be carbon-enriched by low-intermediate mass stars, and the massive star contribution to carbon can be safely isolated. 

\begin{figure*}
    \centering

 \includegraphics[width=1\linewidth, trim = {2em 0em 0em 0em}]{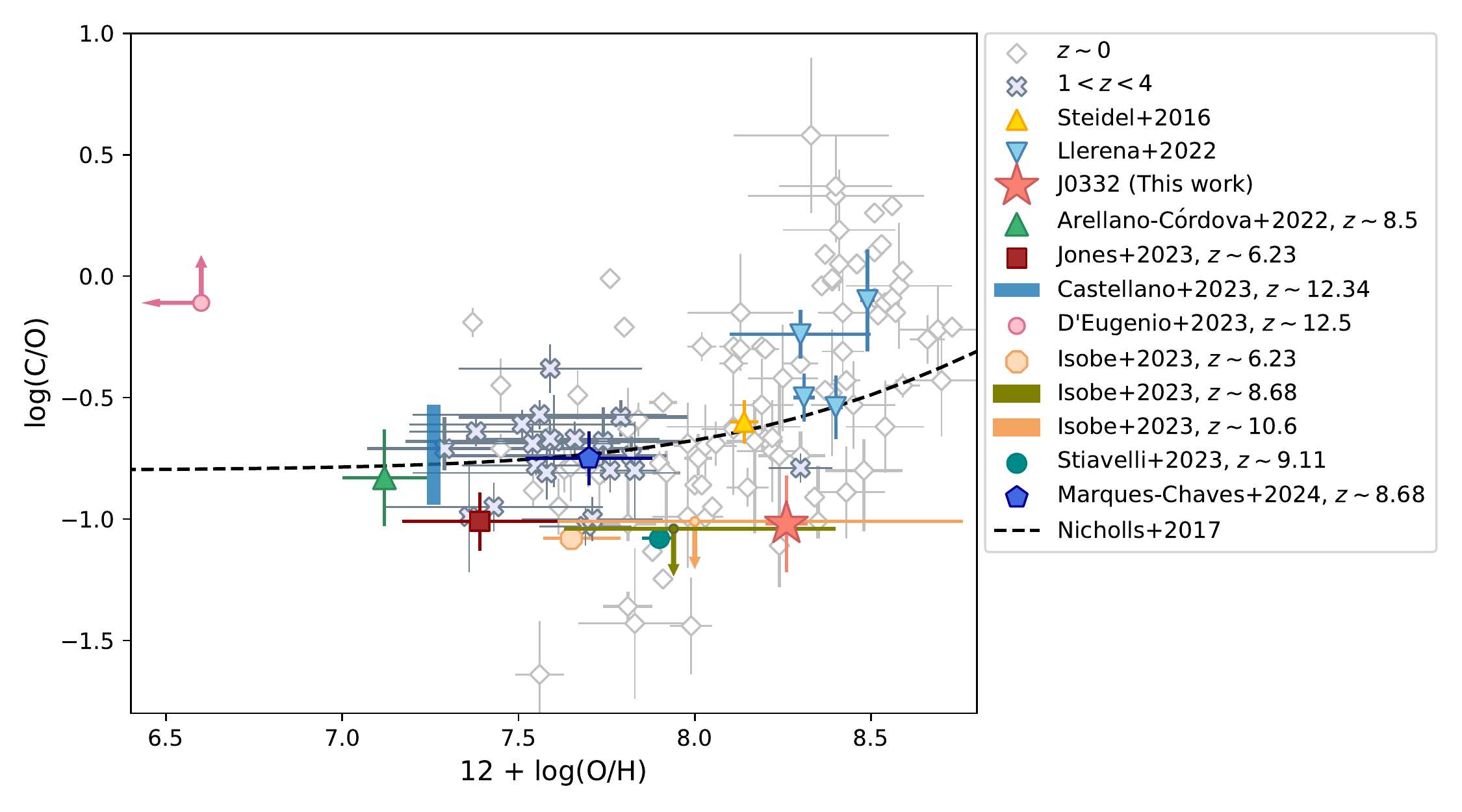}

      \caption{log(CO) vs. 12+log(O/H) diagram. J0332 is marked as a pink star. Results at $z\sim0$ are shown as white diamonds (\citealp{Tsamis+2003}, \citealp{Esteban+2004}, \citeyear{Esteban+2009}, \citeyear{Esteban+2014},  \citealp{GarciaRojas+2004}, \citeyear{GarciaRojas+2005}, \citeyear{GarciaRojas+2007}, \citealp{Peimbert+2005}, \citealp{LopezSanchez+2007},  \citealp{Berg+2016}, \citeyear{Berg+2019},  \citealp{ToribioSanCipriano+2016},  \citeyear{ToribioSanCipriano+2017}, \citealp{PenaGuerrero+2017}, \citealp{Senchyna+2017}, \citealp{Ravindranath+2020}, \citealp{Senchyna+2021}). Results at $z\sim1.5-4$ are shown as light blue  crosses (\citealp{Erb+2010}, \citealp{Christensen+2012a}, \citealp{Bayliss+2014}, \citealp{James+2014}, \citealp{Stark+2014}, \citealp{Steidel+2016}, \citealp{Amorin+2017}, \citealp{Mainali+2020}, \citealp{Matthee+2021}, \citealp{Iani+2023}). {Note that the results obtained from stacked spectra are those from \citet{Steidel+2016} (yellow upward triangle) and \citet{Llerena+2022} (light blue downward triangles). Colored symbols represent JWST galaxies with C/O measurements (note that individual galaxies may have multiple measurements from different authors). JWST galaxies are shown as a red square (\citealp{Jones+2023} - GLASS150008, $z\sim6.23$), a green triangle (\citealp{ArellanoCordova+2022} -   s04590, $z\sim8.5$), a blue pentagon (\citealp{MarquesChaves2024} - CEERS1019, $z\sim8.68$), a teal circle (\citealp{Stiavelli+2023} - MACS1149-JD1, $z\sim9.11$), a beige octagon, a beige downward arrow, and an olive green downward arrow (\citealp{Isobe+2023} - GLASS150008 at $z\sim6.23$, GNz-11 at $z\sim10.6$, and CEERS1019 at $z\sim8.68$, respectively), pink arrows (\citealp{deugenio+2023} - GS-z12 at $z\sim12.5$) and a light blue vertical bar (\citealp{Castellano+2024} - GHZ2 at $z\sim12.34$). The black curve is the fit to data by \citet{Nicholls+2017}. The C/O derived for J0332 is the first individual one at $z\sim4$ and at high metallicity.}}
   \label{fig:logCO_logOH}
\end{figure*}

So far, studies of the C/O abundance across different redshifts have been mostly focused on gas metallicities $\rm 12+log(O/H)< 7.5$ \citep[e.g.,][]{Berg+2019, ArellanoCordova+2022,Jones+2023}. {Deriving the C/O abundance in a galaxy such as J0332, represents an anchor $z\sim4$ for redshift evolution studies of C/O. Moreover, it provides the first individual C/O abundance measurement  \textit{individual} measurement obtained at $\rm {12+log(O/H)> 8}$ and ${z\sim4}$.} In order to obtain the C/O abundance in J0332, we first derive the ionic abundances $\rm O^+/H^+ = 2.11\pm0.43 \times 10^{-5}$ from the [\ion{O}{2}] $\lambda$3727 line and $\rm O^{++}/H^+ = 17.2\pm 3.77 \times 10^{-5} $ from the [\ion{O}{3}] $\lambda$5007 line. 
From the [\ion{C}{3}] $\lambda$1909 and [\ion{O}{3}] $\lambda1666$, we derive $\rm C^{++}/O^{++}= 0.09 \pm 0.07$. Applying the ICF correction fraction from \citet{Berg+2019}, we derive a low log(C/O)= $-1.02\pm0.2$.  Figure \ref{fig:logCO_logOH} illustrates where J0332 lies within the log(C/O) vs. 12+log(O/H) diagram with respect to galaxies at different redshifts. We believe that the low C/O ratio of J0332 might arises from the fact that J0332 is a young system (its age is $\rm 93^{+238}_{-63}$ Myrs - see Section \ref{sect:co}) which has not yet developed a significant population of low-intermediate mass stars experiencing the AGB phase. This conclusion is supported by the chemical evolution models by \citet{Mattsson2010}, which predict a contribution to the carbon abundance from low-intermediate mass stars $<5\%$ at ages $<$ 500 Myr. Even though studies at higher redshifts are needed to safely isolate massive star carbon production at $\rm 12+log(O/H)>8$, J0332 suggests that the trend of the C/O abundance with O/H might flatten out over the whole metallicity range $\rm 7 < 12+log(O/H)<9$ as we go towards higher and higher redshifts, where carbon is released by massive stars only.

Another interesting aspect of the log(C/O) vs. 12+log(O/H) diagram is its scatter at any given metallicity. A detailed study of the scatter has been performed by \citet{Berg+2019}. In particular, they modelled the C/O abundance using the OPENDISK chemical evolution code \citep{Henry+2000b}, which assumes a galaxy to be a single, well-mixed zone (see \citealp{Tinsley1980}). The free parameters of their models are: the number of star formation episodes characterizing the galaxy SFH, the duration of the bursts, and the amount of oxygen which is re-injected into the ISM through Supernovae Type II outflows. They find that galaxies with low C/O abundance might be characterized by longer burst duration (i.e.,  a larger yield of oxygen from Supernova Type II), a smaller amount of oxygen expelled through outflows, and a smaller number of SF episodes (i.e.,  lower star formation efficiency). Given its low C/O, J0332 might be characterized by similar properties. {However, an independent and more detailed study of the SFH of J0332 (derived from its rest-frame optical continuum) would be needed in order to confirm this scenario.}

\section{Equivalent width of the [C III] $\lambda$1906,9 emission line}
\label{sec:ewciii}

\begin{figure*}
  \centering
	\includegraphics[]{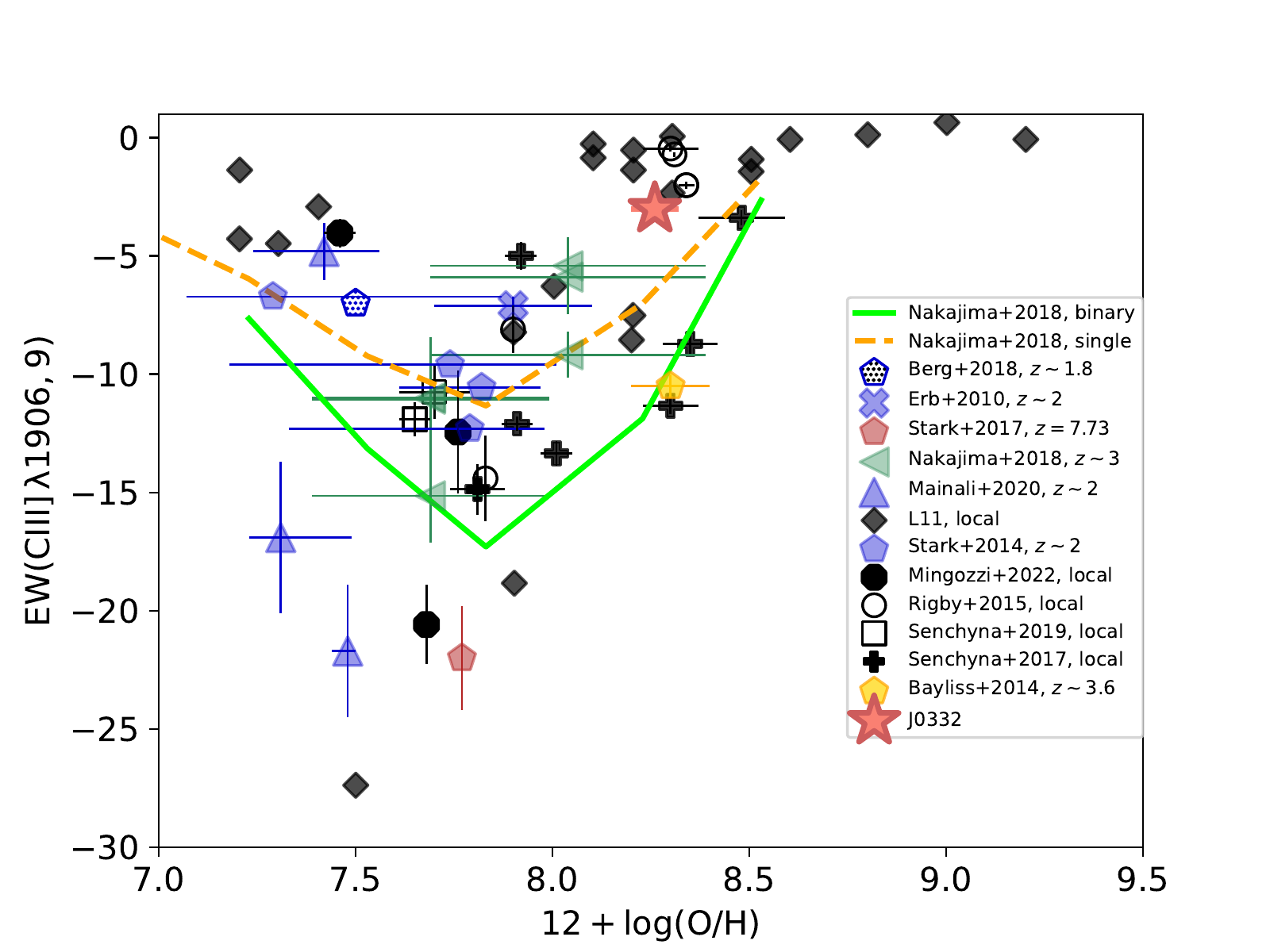}
   
   \caption{\ewciii\ as a function of the gas metallicity. Black diamond symbols are local galaxies from \citet{Leitherer+2011} (L11), \citet{Senchyna+2017}, \citet{Senchyna+2019} and \citet{Mingozzi+2022}, while colored markers indicates galaxies at redshift $1.8<z<2$ (\citealp{Berg+2018}, \citealp{Erb+2010}, \citealp{Stark+2014}, \citealp{Rigby+2015}, \citealp{Mainali+2020}), $3<z<3.6$ (\citealp{Bayliss+2014}, \citealp{Nakajima+2018}), $z\sim7.73$ \citep{Stark+2017}. The curves are photionization models computed by \citet{Nakajima+2018} using BPASS models with (green solid) and without (orange dashed) binary stars.}
   \label{fig:ciii_em}
\end{figure*}

J0332 shows an \ewciii\ $= -3.0\pm0.2$ \wa, which is weaker compared to that of other galaxies of similar redshift and gas metallicity (see Figure \ref{fig:ciii_em}). The \ewciii\ is related to many physical properties, such as ionization parameter, gas metallicity, C/O abundance ratio, gas optical depth, dust extinction, age, and sSFR. In the following, we explore in more detail the effect of these factors on \ewciii, and explain which are the main drivers of the low \ewciii in J0332.

The first factor affecting \ewciii\ is the age of the stellar populations. \citet{Jaskot_Ravindranath2016} found that \ewciii\ peaks at very early ages (< 10 Myr) and decreases afterwards, stabilizing at $\sim$ 10 \wa\ around 20 Myr (assuming a continuous star-formation history). This happens because, by this time, an equilibrium is reached between the birth and the death of the most massive stars and the increase of the 1909 \wa\ continuum flux from the growing stellar populations is the only factor lowering the \ewciii. From the SED fitting, we derive that the age of J0332 is $\rm 93^{+238}_{-63}$ Myr. We exclude that age is the main driver of the low \ewciii\ in J0332. In fact, according to the \citet{Jaskot_Ravindranath2016} models, we would expect its \ewciii\ to be higher than what we observe (i.e.,  $\sim10$ \wa).

Another contributing factor influencing \ewciii\ is the ionization parameter, which dictates the population of excited levels within an atom or ion, thereby impacting the production of specific emission lines. However, we discount the ionization parameter as the main driver of the low \ewciii\ observed in J0332. In fact, for stellar metallicities and ionization parameter similar to those of J0332, the \citet{Jaskot_Ravindranath2016} models predict a higher \ewciii\ values around $\sim7-10$ \wa.

According to \citet{Jaskot_Ravindranath2016} and \citet{Ravindranath+2020}, density-bounded, optically thin systems tend to exhibit weaker \ewciii\ for a given ionization parameter. This is attributed to the lower absorption of C$^{+}$ ionizing radiation, resulting in a decrease in the amount of C$^{2+}$. However, in Section \ref{sect:chemical_composition}, we noted that many of the transitions identified in the ISM of J0332 are saturated, suggesting that, at least some, of its ISM is optically thick. Therefore, the observed low \ewciii\ in J0332 does not appear to be due to the fact that J0332 is density-bounded.

A higher amount of dust might preferentially attenuate the ionizing continuum over the stellar continuum, reducing the emission line fluxes \citep{CharlotFall2000, Shapley+2003}. Assuming that the line emission and the continuum are produced within the same location, J0332 is characterized by a low dust extinction (E(B-V)$\sim$0.13 - see Section \ref{subsec:dust}). Therefore, we exclude that dust extinction can be the reason behind the low \ewciii.

A high sSFR is indicative of a phase of rapid stellar mass growth over the last 100 Myr (i.e.,  the time-scale probed by the UV continuum luminosity) of a galaxy's life-time. This rapid mass growth can provide a stronger and harder ionizing continuum, favoring the production of UV emission lines and increasing \ewciii. However, not all the galaxies with a high sSFR show a high \ewciii. For example, as pointed out in \citet{Rigby+2015}, the lensed galaxy RCS0327 at $z\sim1.7$ is characterized by a high sSFR $\rm\sim 5\,Gyr^{-1}$ but a low \ewciii. \citet{Stark+2014} and \citet{Rigby+2015} suggested that a possible explanation for these puzzling observations is that the sSFR must be coupled with \textit{both} a higher ionization parameter \textit{and} a lower metallicity in order to boost the \CIII\ emission. We find that J0332 is characterized by a sSFR of $\rm 2.1 \pm 0.1\,Gyr^-1$, which is  common at $z\sim4$. Therefore, we do not expect the sSFR to be the main driver of the low \ewciii.

Figure \ref{fig:ciii_em} illustrates the dependence of \ewciii\ on the gas metallicity. Notably, \ewciii\ increases, peaks around $\rm12+log(O/H)\sim8$ and then decreases at larger metallicities (note that negative values are indicative of emission). This trend is explained by the fact that, as the gas metallicity increases, the amount of carbon in the ISM increases and the \CIII\ emission increases as a consequence. However, at even higher metallicities, which act as coolants, the electron temperature decreases, and so does the probability of ionization of C$^{+}$ ions. Therefore, the observed low \ewciii\ in J0332 can be attributed to its higher gas metallicity. Additionally, our estimate of \ewciii\ aligns with predictions from the \citet{Jaskot_Ravindranath2016} photoionization models, particularly for their lowest assumed C/O abundance ratio of 0.04 (it is noteworthy that these models assume single bursts rather than a continuous star formation history). Considering all the galaxy's physical properties that impact \ewciii, we conclude that the primary factors driving the low \ewciii\ in J0332 are its elevated gas-phase metallicity and the reduced carbon abundance relative to oxygen. 

\section{Physical properties along the arc}
\label{sec:spatially_resolved}
\begin{figure*}
\label{fig:trend}
	\includegraphics[width=\textwidth]{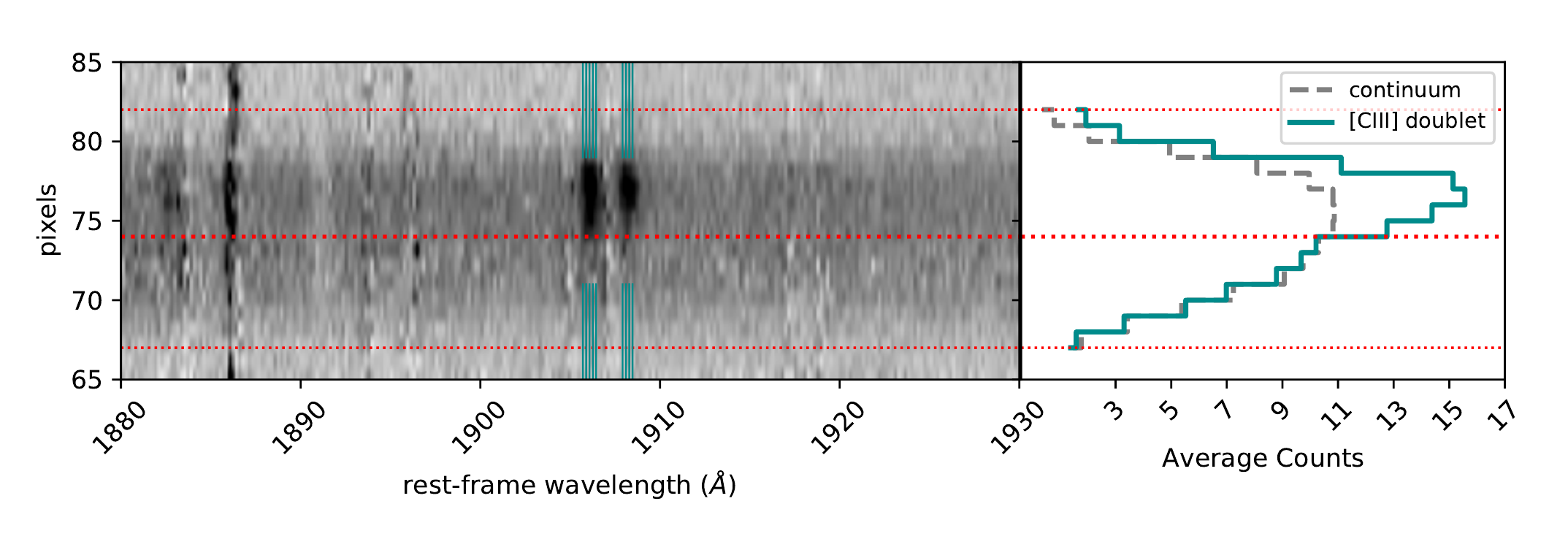}
      \caption{Left panel: J0332 2D spectrum around the \CIII\ doublet. Right panel: average continuum (grey) and \CIII\ counts (green) along the spatial direction of the 2D spectrum. The green vertical lines in the left panel show the slices used to define the average \CIII\ counts. The continuum counts were defined along the whole spectral direction avoiding regions with strong emission/absorption lines. The red horizontal lines on the left panel (extending to the right panel) show the extent of the 2D trace and the cut used to define the two regions. This cut corresponds to the y-pixel where the \CIII\ counts separate from the continuum counts. 
      }
\end{figure*}

In the previous Sections, we have inferred the physical properties of J0332 using its integrated spectra, and we have provided information about its average properties. In the following, we perform a UV spatially resolved analysis of J0332 by extracting the FORS2 spectrum within two different apertures. This analysis is motivated by the fact that the \CIII\ lines do not cover the entire spatially resolved continuum (see Figure \ref{fig:spec1}, bottom panel).
We first define several slices along the slit spatial direction, and we measure the average counts in the continuum as a function of the position along the slit, together with the average spatial \CIII\  emission counts (defining slices corresponding to the carbon emission - see Figure \ref{fig:trend}). We then move from the bottom to the top of the trace (this direction corresponds to the SW to the NE direction in Figure \ref{fig:spec1}) and define two regions along the trace. We divide the two regions at the pixel where the carbon emission counts start separating from the spatial distribution of the continuum counts (Figure \ref{fig:trend}). In the following, we refer to the upper portion of the spectral trace as the \textit{bright region}, where the carbon emission is strong, and the lower portion as the \textit{faint region}.

\begin{figure}
	\includegraphics[width=\columnwidth]{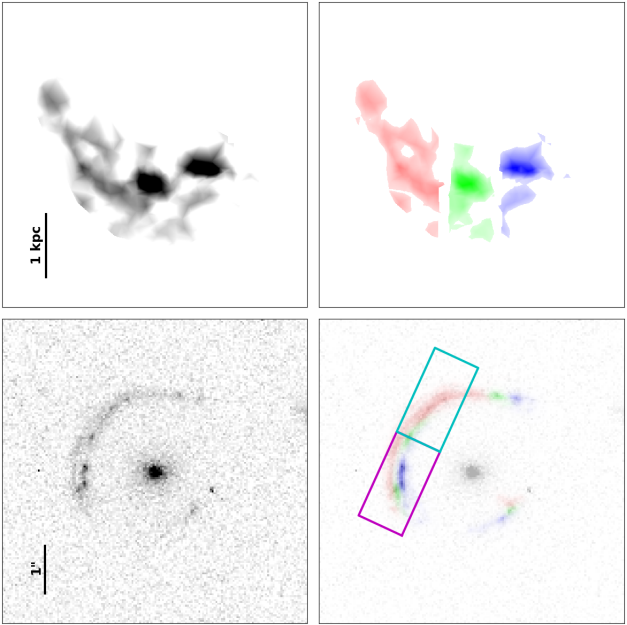}
      \caption{{Upper panels: Co-added F606W and F814W images showing the image plane (left) and source plane (right) surface brightness distributions of J0332. Lower panels:  three distinct regions in the source plane (right) have been colored red, green, and blue to show where these regions lie in the image plane (left). The left panel also shows the two extraction apertures for our FORS2 spectra in magenta and cyan. The magenta aperture is dominated by the bottom part of the source (colored blue, also in the source plane in the bottom right panel) and the cyan aperture is a blend of the high-surface-brightness emission in the center of the source as well as the lower-surface-brightness, more diffuse emission from the upper region (colored red, also in the source plane in the bottom right panel).}}
   \label{fig:regions_arc}
\end{figure}

\begin{figure*}
	\includegraphics[width=\textwidth]{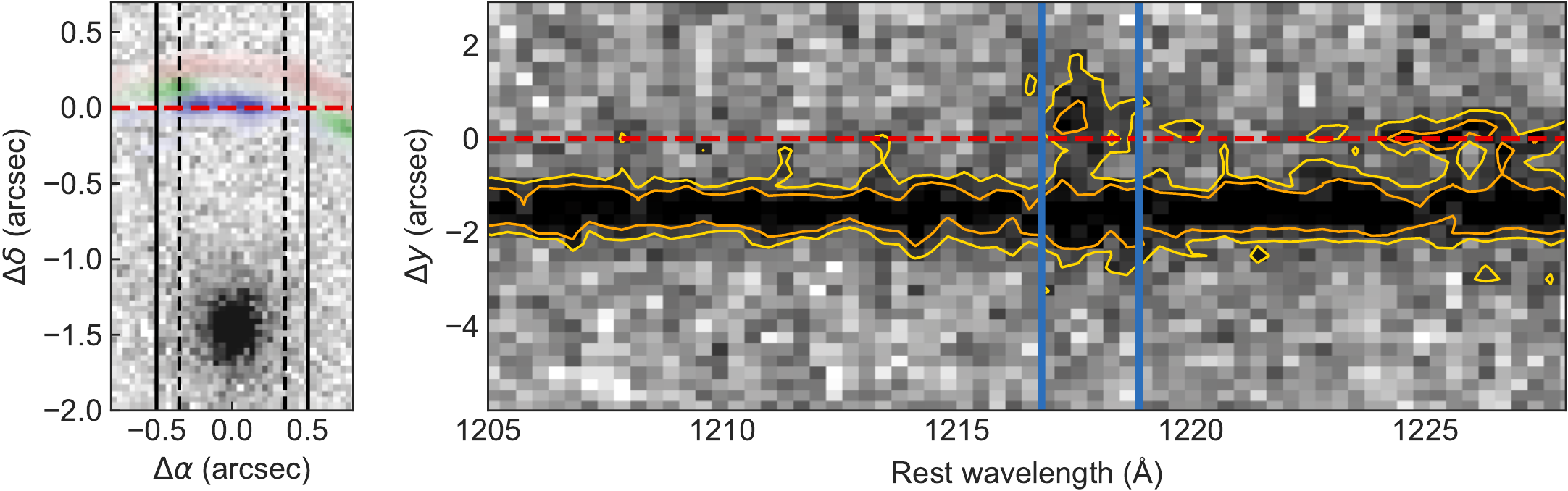}
      \caption{Spatial offset of the \lya\ emission. The left panel shows a portion of the HST image of J0332, with the foreground lensing galaxy at the bottom and the regions of the arc color-coded as in Figure \ref{fig:regions_arc}. The vertical black lines mark the two slits used by \citet{Cabanac+2008}, and the horizontal red dashed line marks the highest surface brightness emission from the blue clump. The right panel shows the 2D spectrum in the \lya\ region. The bright continuum source at about $-1.5$\arcsec\ in the spatial direction is the lensing galaxy, and the dashed red line again marks the highest surface brightness continuum emission from the arc, visible on the right of the figure at wavelengths $\gtrsim1225$ \AA. The \lya\ emission, marked by the vertical blue lines, is superimposed on a damped absorption profile (the damped profile is visible in the J0332 UV spectrum shown in \citet{Cabanac+2008}) and is spatially offset to the north of the continuum marked by the red line. The offset of the peak \lya\ emission is comparable to the offset of the diffuse region marked in red in the left panel, and it appears that the \lya\ emission arises from this region.}
   \label{fig:lya_regions}
\end{figure*}

Figure \ref{fig:regions_arc} shows the portions of the arc captured by these two extractions and the corresponding regions in the source plane. We find that the aperture centered on the brighter half of the trace (magenta box) is dominated by the emission from the brighter and more compact regions in the source (blue and green knot in lower right panel of Figure \ref{fig:regions_arc}), while the aperture centered on the lower and fainter half of the trace (cyan box) consists mostly of light from a more diffuse region (red area in the lower left panel of Figure \ref{fig:regions_arc}). However, we note that some level of blending is present between the two regions.
\citet{Cabanac+2008} found that J0332 is characterized by \lya\ emission superimposed to the damped Ly$\alpha$ profile in absorption. This emission is shifted both spatially (0.5 arcsec with respect to the UV continuum) and in velocity space (by $\rm \sim +\,830\,km\,s^{-1}$ with respect to the centroid of the low ionization absorption lines). \citet{Cabanac+2008} interpreted this feature as probably due to an expanding outflow.  
With our study, we are able to reconstruct in more detail where the Ly$\alpha$ emission described in \citet{Cabanac+2008} originates (as illustrated in Figure \ref{fig:lya_regions}). Specifically, we observe that the spatial offset of the Ly$\alpha$ emission in the \citet{Cabanac+2008} 2D spectrum is roughly equivalent to the distance between the blue color-coded knot and the diffuse (red) region. Although we cannot exclude that a contribution from the bright knot to the Ly$\alpha$ emission is present, it is reasonable to assume that the Ly$\alpha$ emission is mostly coming from the diffuse region. This is compatible with many literature studies which find extended Ly$\alpha$ emission in high redshift galaxies \citep[e.g.,][]{Steidel+2011, Erb+2018, Leclercq+2017, Leclercq+2020}.


\subsection{Spatially resolved [C III] $\lambda$1906,9 equivalent width}

\begin{figure}
	\includegraphics[width=\columnwidth]{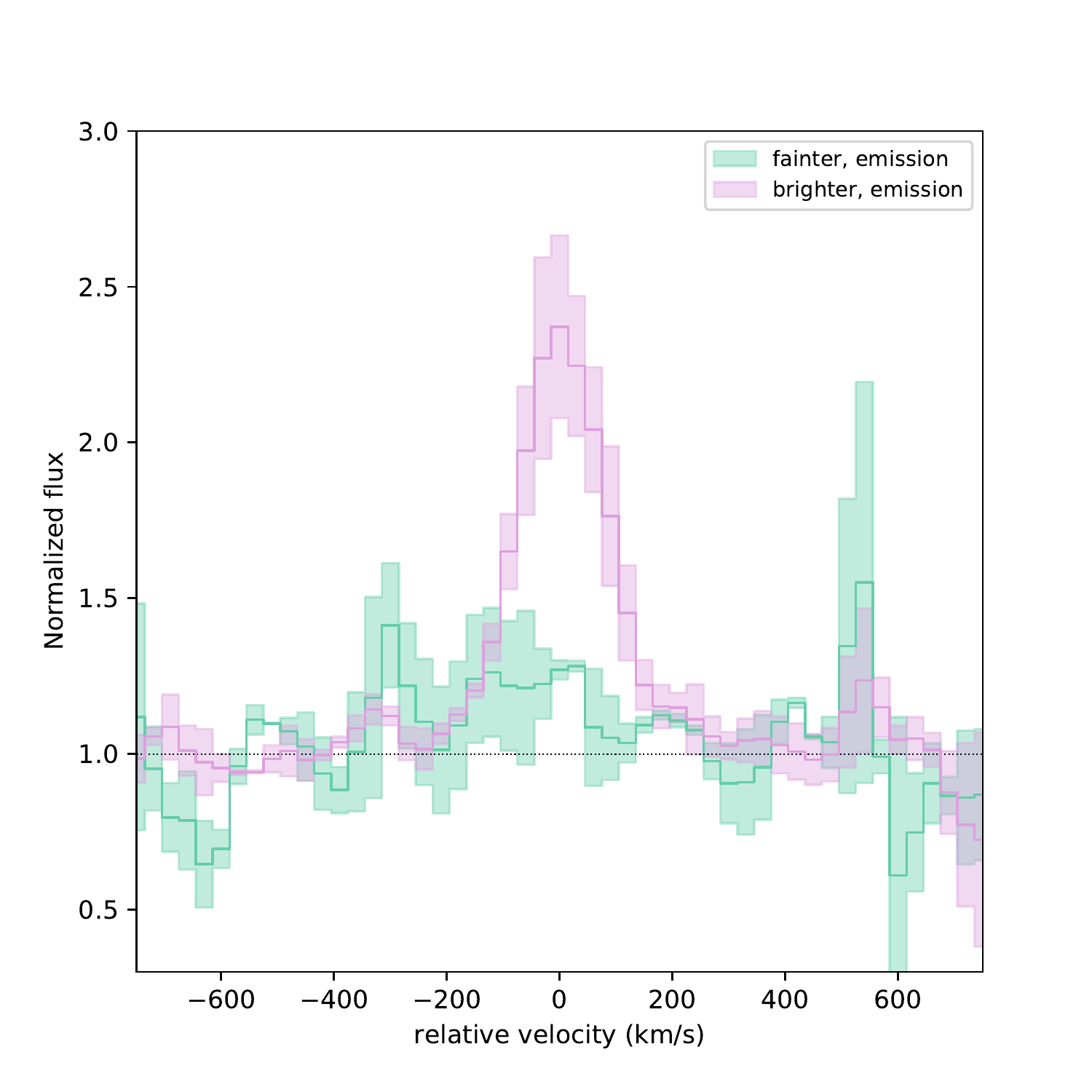}
      \caption{Average emission profiles for the faint and bright region. Note that we alternately masked out the \CIII\ doublet line not included in the average.}
   \label{fig:regions_velprof}
\end{figure}

{We quantify the strength of the UV emission lines from the two extractions by measuring their equivalent widths. We find that $\rm EW_{[CIII], bright} = -3.67\pm0.18$ \wa, while $\rm EW_{[CIII], faint} = -0.44 \pm0.19$ \wa\ (see Figure \ref{fig:regions_velprof} and Table \ref{table:ewtworegions}).}
We note that the value obtained from the integrated spectrum (see Section \ref{sec:ewciii}) is slightly lower than $\rm EW_{[CIII], bright}$, probably due to the fact that more continuum emission is included in the integrated spectrum. Our finding that the \CIII\ emission mostly comes from a bright knot inside J0332 is consistent with what found by \citet{Micheva+2020} at a much lower redshift in the galaxy Haro~11\footnote{In their analysis, \citet{Micheva+2020} do not exclude that the \CIII\ emission in the more diffuse region is not detected only because of the low S/N of their data}. Similar gradients in \CIII\ emission have been found in other high redshift galaxies, such as the Cosmic Horseshoe \citep{James+2018}, and indicate that the physical conditions in the two analyzed regions of J0332 might be different.

We utilize the measured \ewciii\ in the two apertures alongside the EW vs. 12 +log(O/H) relations defined by \citet{Mingozzi+2022} to estimate the gas metallicity of the two regions. We determine $\rm 12+log(O/H)_{bright} = 8.17\pm0.18$\footnote{{The uncertainty of 0.18 dex accounts for the intrinsic scatter of the relation noted in \citet{Mingozzi+2022}.}} and $\rm 12+log(O/H)_{faint} = 8.61\pm0.18$. Comparatively, the metallicity predicted from the integrated \ewciii\ is $\rm 12+log(O/H)_{bright} = 8.18\pm0.18$. This result aligns, within the uncertainties, with the gas metallicity determined through the direct method, likely dominated by the bright region.

\section{Summary and conclusions}
\label{sec:conclusions}

In this paper, we analyze the UV and optical spectrum of FORJ0332-3557 (J0332), a gravitationally lensed galaxy at redshift $z\sim3.8$. Exploiting the wide wavelength range offered by the rest-frame UV FORS2 and the rest-frame optical XSHOOTER spectra, we are able to provide one of the most comprehensive metallicity analyses at such high redshift. {J0332 stands out as the sole iindividual galaxy at $z\sim4$ with both a C/O abundance measurement and a direct metallicity assessment. These characteristics place it as a crucial anchor in investigations concerning the evolution of chemical abundances and scaling relations, such as the mass-metallicity relation, throughout cosmic epochs.}

We focus on the stellar metallicity (derived from the UV stellar continuum), the ISM abundances (from the UV absorption lines), the gas-phase metallicity (derived through the direct method), and the relative carbon/oxygen abundance (derived from nebular emission lines). Our main results can be summarized as follow:

\vspace{2mm}
\noindent
$\bullet$ From the SED fitting of the HST F606W, F814W, F125W, and F160W filters + ground based $K_{s}$ imaging, assuming a constant SFH, we derive a stellar mass $\log(M/M_{\odot})=9.32^{+0.33}_{-0.32}$, an age of $93^{+238}_{-63}$ Myr, and $E(B-V)=0.15^{+0.02}_{-0.03}$. We derive a star formation rate of 21.6 M$_{\odot}$ yr$^{-1}$.

\vspace{2mm}
\noindent
$\bullet$  From the spectral fitting to the FORS2 UV spectrum, we find that the stellar populations in J0332 are metal poor, with \Z\ $\rm \sim5-10\%\,Z_{\odot}$.

\vspace{2mm}
\noindent
$\bullet$  The kinematic analysis of the ISM absorption lines shows that J0332 is characterized by an outflow, with speeds up to $\rm \sim -360\,km\,s^{-1}$. Our results are compatible with the scenario where the outflows are produced by the mechanical energy released by Type II Supernovae explosions. We do not find evidences of inflows.

\vspace{2mm}
\noindent
$\bullet$  Thanks to the detection of the auroral O III] $\rm\lambda1666$ line, we are able to infer the metallicity of the ionized gas in J0332 with the direct method, obtaining $\rm 12+log(O/H)=8.26\pm0.06$ (0.37 $Z_{\odot}$).

\vspace{2mm}
\noindent
$\bullet$  Comparing the gas to the stellar metallicity, we find the first to be higher than the second by a factor $\sim$ 3-4. This discrepancy can be explained by J0332 having an enhanced O/Fe ratio. 

\vspace{2mm}
\noindent
$\bullet$  The direct gas metallicity of J0332 is compatible with that of other galaxies at similar redshifts and with similar masses, and it is $\rm\sim0.3$ dex lower than that of local galaxies at the same mass. This decrease in metallicity as a function of redshift is consistent with an increase of both the gas fraction and the outflow efficiency. The direct metallicity inferred for J0332 also supports the scenario of a redshift-invariant FMR on the redshift range $0<z<4$.

\vspace{2mm}
\noindent
$\bullet$  From the UV carbon and oxygen lines, we derive a carbon/oxygen abundance log(C/O) $=-1.02\pm0.2$. This value places J0332 on the lower envelope of the log(C/O) vs. 12+log(O/H) distribution at $\rm 12+log(O/H)>8$. The low C/O abundance can be explained by the fact that J0332 is young enough to have mostly only massive stars contributing to the carbon abundance, alongside an oxygen-enriched ISM.

\vspace{2mm}
\noindent
$\bullet$  We perform a spatially resolved study of J0332. The spatial scales of this analysis are $\rm\sim1\,kpc$ (based on the size of the compact knots in Figure \ref{fig:regions_arc}). We observe \ewciii\ to be $-3.67\pm0.18$ \wa\ in two compact knots within the source plane, contrasting with \ewciii\ of $-0.44\pm0.19$ in a more diffuse region. This suggests that the carbon emission predominantly originates from the compact knots.

\begin{table}

\caption{Rest-frame equivalent widths of the emission lines in the faint and the bright region identified along the arc. }
\begin{center}
\begin{tabular}{lccccc}
\hline
Ion &  $\lambda$ & $\rm EW_{faint}$ & $\rm \delta EW_{faint}$ & $\rm EW_{bright}$ & $\rm \delta EW_{bright}$\\
\hline

O\,{\sc iii}]& 1666 & -0.48 & 0.13 &  -1.11 & 0.08\\\
[C\,{\sc iii}]& 1906 & -0.16 & 0.14 & -2.05 & 0.15\\\
C\,{\sc iii}]& 1909 & -0.28 & 0.14 & -1.62 & 0.11\\
\hline
\end{tabular}
\end{center}
\label{table:ewtworegions}
\end{table}

As a concluding note, we reiterate that J0332 is one of the highest redshift galaxies studied at $\rm 12+log(O/H)>8.2$ and allows us to investigate the chemical enrichment history at these metallicities as early as $\sim1.5$ Gyr after the Big bang. Another important consideration is that thanks to the James Webb Telescope (JWST), the UV spectral features  used to perform our analysis on J0332 at $ z\sim 4$ are now accessible for galaxies at $z>7$. Therefore, studies such as the one we performed on J0332 represent a bridge towards the exploration of the chemical composition of galaxies back to when the universe was only a few hundred million years old.

\acknowledgements
 The authors wish to thank the referee for constructive comments that improved the paper substantially. The authors thank Prof. R. Cabanac for providing the J0332 data from the 2008 work. A.C thanks H. Williams for providing insights on implementing non-parametric SFHs in \texttt{prospector}. AC also thanks Dr. F. Cullen for the star formation rate results from the NIRVANDELS data, Prof. H. Nakajima for providing the photoionization models describing the evolution of the \ewciii\ as a function of metallicity, and Prof.\ Max Pettini for helpful conversation and discussion. AC and DKE were supported by the US National Science Foundation (NSF) through Astronomy $\&$ Astrophysics grant AST-1909198. GB was supported by the NSF through grants AST-1615814 and AST-1751404. This work is based on observations collected at the European Organisation for Astronomical Research in the Southern Hemisphere under ESO programmes 086.A-0035(A)
and 094.A-0252(A). This research is also based in part on observations made with the NASA/ESA Hubble Space Telescope obtained from the Space Telescope Science Institute, which is operated by the Association of Universities for Research in Astronomy, Inc., under NASA contract NAS 5-26555. These observations are associated with program 14093.

\bibliographystyle{aasjournal}
\bibliography{bibliography.bib}

\begin{thebibliography}{}
\expandafter\ifx\csname natexlab\endcsname\relax\def\natexlab#1{#1}\fi
\providecommand{\url}[1]{\href{#1}{#1}}
\providecommand{\dodoi}[1]{doi:~\href{http://doi.org/#1}{\nolinkurl{#1}}}
\providecommand{\doeprint}[1]{\href{http://ascl.net/#1}{\nolinkurl{http://ascl.net/#1}}}
\providecommand{\doarXiv}[1]{\href{https://arxiv.org/abs/#1}{\nolinkurl{https://arxiv.org/abs/#1}}}

\bibitem[{Aggarwal \& Keenan(1999)}]{Aggarwal+1999}
Aggarwal, K.~M., \& Keenan, F.~P. 1999, The Astrophysical Journal Supplement
  Series, 123, 311, \dodoi{10.1086/313232}

\bibitem[{{Aloisi} {et~al.}(2001){Aloisi}, {Tosi}, \& {Greggio}}]{Aloisi+2001}
{Aloisi}, A., {Tosi}, M., \& {Greggio}, L. 2001, \apss, 276, 421,
  \dodoi{10.1023/A:1017595115910}

\bibitem[{{Aloisi} {et~al.}(2007){Aloisi}, {Clementini}, {Tosi}, {Annibali},
  {Contreras}, {Fiorentino}, {Mack}, {Marconi}, {Musella}, {Saha}, {Sirianni},
  \& {van der Marel}}]{Aloisi+2007}
{Aloisi}, A., {Clementini}, G., {Tosi}, M., {et~al.} 2007, \apjl, 667, L151,
  \dodoi{10.1086/522368}

\bibitem[{{Amor{\'\i}n} {et~al.}(2017){Amor{\'\i}n}, {Fontana},
  {P{\'e}rez-Montero}, {Castellano}, {Guaita}, {Grazian}, {Le F{\`e}vre},
  {Ribeiro}, {Schaerer}, {Tasca}, {Thomas}, {Bardelli}, {Cassar{\`a}},
  {Cassata}, {Cimatti}, {Contini}, {de Barros}, {Garilli}, {Giavalisco},
  {Hathi}, {Koekemoer}, {Le Brun}, {Lemaux}, {Maccagni}, {Pentericci}, {Pforr},
  {Talia}, {Tresse}, {Vanzella}, {Vergani}, {Zamorani}, {Zucca}, \&
  {Merlin}}]{Amorin+2017}
{Amor{\'\i}n}, R., {Fontana}, A., {P{\'e}rez-Montero}, E., {et~al.} 2017,
  Nature Astronomy, 1, 0052, \dodoi{10.1038/s41550-017-0052}

\bibitem[{{Arellano-C{\'o}rdova} {et~al.}(2022){Arellano-C{\'o}rdova}, {Berg},
  {Chisholm}, {Haro}, {Dickinson}, {Finkelstein}, {Leclercq}, {Rogers},
  {Simons}, {Skillman}, {Trump}, \& {Kartaltepe}}]{ArellanoCordova+2022}
{Arellano-C{\'o}rdova}, K.~Z., {Berg}, D.~A., {Chisholm}, J., {et~al.} 2022,
  \apjl, 940, L23, \dodoi{10.3847/2041-8213/ac9ab2}

\bibitem[{{Asplund} {et~al.}(2021){Asplund}, {Amarsi}, \&
  {Grevesse}}]{Asplund+2021}
{Asplund}, M., {Amarsi}, A.~M., \& {Grevesse}, N. 2021, \aap, 653, A141,
  \dodoi{10.1051/0004-6361/202140445}

\bibitem[{{Barrow} {et~al.}(2020){Barrow}, {Robertson}, {Ellis}, {Nakajima},
  {Saxena}, {Stark}, \& {Tang}}]{Barrow+2020}
{Barrow}, K. S.~S., {Robertson}, B.~E., {Ellis}, R.~S., {et~al.} 2020, \apjl,
  902, L39, \dodoi{10.3847/2041-8213/abbd8e}

\bibitem[{{Bayliss} {et~al.}(2014){Bayliss}, {Rigby}, {Sharon}, {Wuyts},
  {Florian}, {Gladders}, {Johnson}, \& {Oguri}}]{Bayliss+2014}
{Bayliss}, M.~B., {Rigby}, J.~R., {Sharon}, K., {et~al.} 2014, \apj, 790, 144,
  \dodoi{10.1088/0004-637X/790/2/144}

\bibitem[{{Becker} {et~al.}(2012){Becker}, {Sargent}, {Rauch}, \&
  {Carswell}}]{Becker+2012}
{Becker}, G.~D., {Sargent}, W. L.~W., {Rauch}, M., \& {Carswell}, R.~F. 2012,
  \apj, 744, 91, \dodoi{10.1088/0004-637X/744/2/91}

\bibitem[{{Berg} {et~al.}(2018){Berg}, {Erb}, {Auger}, {Pettini}, \&
  {Brammer}}]{Berg+2018}
{Berg}, D.~A., {Erb}, D.~K., {Auger}, M.~W., {Pettini}, M., \& {Brammer}, G.~B.
  2018, \apj, 859, 164, \dodoi{10.3847/1538-4357/aab7fa}

\bibitem[{{Berg} {et~al.}(2019){Berg}, {Erb}, {Henry}, {Skillman}, \&
  {McQuinn}}]{Berg+2019}
{Berg}, D.~A., {Erb}, D.~K., {Henry}, R. B.~C., {Skillman}, E.~D., \&
  {McQuinn}, K. B.~W. 2019, \apj, 874, 93, \dodoi{10.3847/1538-4357/ab020a}

\bibitem[{{Berg} {et~al.}(2016){Berg}, {Skillman}, {Henry}, {Erb}, \&
  {Carigi}}]{Berg+2016}
{Berg}, D.~A., {Skillman}, E.~D., {Henry}, R. B.~C., {Erb}, D.~K., \& {Carigi},
  L. 2016, \apj, 827, 126, \dodoi{10.3847/0004-637X/827/2/126}

\bibitem[{{Berg} {et~al.}(2012){Berg}, {Skillman}, {Marble}, {van Zee},
  {Engelbracht}, {Lee}, {Kennicutt}, {Calzetti}, {Dale}, \&
  {Johnson}}]{Berg+2012}
{Berg}, D.~A., {Skillman}, E.~D., {Marble}, A.~R., {et~al.} 2012, \apj, 754,
  98, \dodoi{10.1088/0004-637X/754/2/98}

\bibitem[{{Bian} {et~al.}(2018){Bian}, {Kewley}, \& {Dopita}}]{Bian+2018}
{Bian}, F., {Kewley}, L.~J., \& {Dopita}, M.~A. 2018, \apj, 859, 175,
  \dodoi{10.3847/1538-4357/aabd74}

\bibitem[{{Bothwell} {et~al.}(2013){Bothwell}, {Maiolino}, {Kennicutt},
  {Cresci}, {Mannucci}, {Marconi}, \& {Cicone}}]{Bothwell+2013}
{Bothwell}, M.~S., {Maiolino}, R., {Kennicutt}, R., {et~al.} 2013, \mnras, 433,
  1425, \dodoi{10.1093/mnras/stt817}

\bibitem[{{Brown} {et~al.}(2018){Brown}, {Cortese}, {Catinella}, \&
  {Kilborn}}]{Brown+2018}
{Brown}, T., {Cortese}, L., {Catinella}, B., \& {Kilborn}, V. 2018, \mnras,
  473, 1868, \dodoi{10.1093/mnras/stx2452}

\bibitem[{{Cabanac} {et~al.}(2005){Cabanac}, {Valls-Gabaud}, {Jaunsen},
  {Lidman}, \& {Jerjen}}]{Cabanac+2005}
{Cabanac}, R.~A., {Valls-Gabaud}, D., {Jaunsen}, A.~O., {Lidman}, C., \&
  {Jerjen}, H. 2005, \aap, 436, L21, \dodoi{10.1051/0004-6361:200500115}

\bibitem[{{Cabanac} {et~al.}(2008){Cabanac}, {Valls-Gabaud}, \&
  {Lidman}}]{Cabanac+2008}
{Cabanac}, R.~A., {Valls-Gabaud}, D., \& {Lidman}, C. 2008, \mnras, 386, 2065,
  \dodoi{10.1111/j.1365-2966.2008.13157.x}

\bibitem[{{Calabr{\`o}} {et~al.}(2020){Calabr{\`o}}, {Castellano},
  {Pentericci}, {Fontanot}, {Menci}, {Cullen}, {McLure}, {Bolzonella},
  {Cimatti}, {Marchi}, {Talia}, {Amor{\'\i}n}, {Cresci}, {De Lucia}, {Fynbo},
  {Fontana}, {Franco}, {Hathi}, {Hibon}, {Hirschmann}, {Mannucci}, {Santini},
  {Saxena}, {Schaerer}, {Xie}, \& {Zamorani}}]{Calabro+2020}
{Calabr{\`o}}, A., {Castellano}, M., {Pentericci}, L., {et~al.} 2020, arXiv
  e-prints, arXiv:2011.06615.
\newblock \doarXiv{2011.06615}

\bibitem[{{Calabr{\`o}} {et~al.}(2021){Calabr{\`o}}, {Castellano},
  {Pentericci}, {Fontanot}, {Menci}, {Cullen}, {McLure}, {Bolzonella},
  {Cimatti}, {Marchi}, {Talia}, {Amor{\'\i}n}, {Cresci}, {De Lucia}, {Fynbo},
  {Fontana}, {Franco}, {Hathi}, {Hibon}, {Hirschmann}, {Mannucci}, {Santini},
  {Saxena}, {Schaerer}, {Xie}, \& {Zamorani}}]{Calabro+2021}
---. 2021, \aap, 646, A39, \dodoi{10.1051/0004-6361/202039244}

\bibitem[{{Cardelli} {et~al.}(1989){Cardelli}, {Clayton}, \&
  {Mathis}}]{Cardelli+1989}
{Cardelli}, J.~A., {Clayton}, G.~C., \& {Mathis}, J.~S. 1989, \apj, 345, 245,
  \dodoi{10.1086/167900}

\bibitem[{{Cashman} {et~al.}(2017){Cashman}, {Kulkarni}, {Kisielius},
  {Ferland}, \& {Bogdanovich}}]{Cashman+2017}
{Cashman}, F.~H., {Kulkarni}, V.~P., {Kisielius}, R., {Ferland}, G.~J., \&
  {Bogdanovich}, P. 2017, \apjs, 230, 8, \dodoi{10.3847/1538-4365/aa6d84}

\bibitem[{{Castellano} {et~al.}(2024){Castellano}, {Napolitano}, {Fontana},
  {Roberts-Borsani}, {Treu}, {Vanzella}, {Zavala}, {Arrabal Haro},
  {Calabr{\`o}}, {Llerena}, {Mascia}, {Merlin}, {Paris}, {Pentericci},
  {Santini}, {Bakx}, {Bergamini}, {Cupani}, {Dickinson}, {Filippenko},
  {Glazebrook}, {Grillo}, {Kelly}, {Malkan}, {Mason}, {Morishita},
  {Nanayakkara}, {Rosati}, {Sani}, {Wang}, \& {Yoon}}]{Castellano+2024}
{Castellano}, M., {Napolitano}, L., {Fontana}, A., {et~al.} 2024, arXiv
  e-prints, arXiv:2403.10238, \dodoi{10.48550/arXiv.2403.10238}

\bibitem[{{Chabrier}(2003)}]{Chabrier2003}
{Chabrier}, G. 2003, \pasp, 115, 763, \dodoi{10.1086/376392}

\bibitem[{{Charlot} \& {Fall}(2000)}]{CharlotFall2000}
{Charlot}, S., \& {Fall}, S.~M. 2000, \apj, 539, 718, \dodoi{10.1086/309250}

\bibitem[{{Chiappini} {et~al.}(2003){Chiappini}, {Romano}, \&
  {Matteucci}}]{Chiappini+2003}
{Chiappini}, C., {Romano}, D., \& {Matteucci}, F. 2003, \mnras, 339, 63,
  \dodoi{10.1046/j.1365-8711.2003.06154.x}

\bibitem[{{Chisholm} {et~al.}(2016){Chisholm}, {Tremonti}, {Leitherer}, {Chen},
  \& {Wofford}}]{Chisholm+2016}
{Chisholm}, J., {Tremonti}, C.~A., {Leitherer}, C., {Chen}, Y., \& {Wofford},
  A. 2016, \mnras, 457, 3133, \dodoi{10.1093/mnras/stw178}

\bibitem[{{Christensen} {et~al.}(2012{\natexlab{a}}){Christensen}, {Laursen},
  {Richard}, {Hjorth}, {Milvang-Jensen}, {Dessauges-Zavadsky}, {Limousin},
  {Grillo}, \& {Ebeling}}]{Christensen+2012a}
{Christensen}, L., {Laursen}, P., {Richard}, J., {et~al.} 2012{\natexlab{a}},
  \mnras, 427, 1973, \dodoi{10.1111/j.1365-2966.2012.22007.x}

\bibitem[{{Christensen} {et~al.}(2012{\natexlab{b}}){Christensen}, {Richard},
  {Hjorth}, {Milvang-Jensen}, {Laursen}, {Limousin}, {Dessauges-Zavadsky},
  {Grillo}, \& {Ebeling}}]{Christensen+2012b}
{Christensen}, L., {Richard}, J., {Hjorth}, J., {et~al.} 2012{\natexlab{b}},
  \mnras, 427, 1953, \dodoi{10.1111/j.1365-2966.2012.22006.x}

\bibitem[{{Citro} {et~al.}(2021){Citro}, {Erb}, {Pettini}, {Auger}, {Becker},
  \& {James}}]{Citro+2021}
{Citro}, A., {Erb}, D.~K., {Pettini}, M., {et~al.} 2021, \apj, 922, 187,
  \dodoi{10.3847/1538-4357/ac24a2}

\bibitem[{{Coil} {et~al.}(2015){Coil}, {Aird}, {Reddy}, {Shapley}, {Kriek},
  {Siana}, {Mobasher}, {Freeman}, {Price}, \& {Shivaei}}]{Coil+2015}
{Coil}, A.~L., {Aird}, J., {Reddy}, N., {et~al.} 2015, \apj, 801, 35,
  \dodoi{10.1088/0004-637X/801/1/35}

\bibitem[{{Cresci} {et~al.}(2019){Cresci}, {Mannucci}, \&
  {Curti}}]{Cresci+2019}
{Cresci}, G., {Mannucci}, F., \& {Curti}, M. 2019, \aap, 627, A42,
  \dodoi{10.1051/0004-6361/201834637}

\bibitem[{{Cullen} {et~al.}(2019){Cullen}, {McLure}, {Dunlop}, {Khochfar},
  {Dav{\'e}}, {Amor{\'\i}n}, {Bolzonella}, {Carnall}, {Castellano}, {Cimatti},
  {Cirasuolo}, {Cresci}, {Fynbo}, {Fontanot}, {Gargiulo}, {Garilli}, {Guaita},
  {Hathi}, {Hibon}, {Mannucci}, {Marchi}, {McLeod}, {Pentericci}, {Pozzetti},
  {Shapley}, {Talia}, \& {Zamorani}}]{Cullen+2019}
{Cullen}, F., {McLure}, R.~J., {Dunlop}, J.~S., {et~al.} 2019, \mnras, 487,
  2038, \dodoi{10.1093/mnras/stz1402}

\bibitem[{{Cullen} {et~al.}(2021){Cullen}, {Shapley}, {McLure}, {Dunlop},
  {Sanders}, {Topping}, {Reddy}, {Amor{\'\i}n}, {Begley}, {Bolzonella},
  {Calabr{\`o}}, {Carnall}, {Castellano}, {Cimatti}, {Cirasuolo}, {Cresci},
  {Fontana}, {Fontanot}, {Garilli}, {Guaita}, {Hamadouche}, {Hathi},
  {Mannucci}, {McLeod}, {Pentericci}, {Saxena}, {Talia}, \&
  {Zamorani}}]{Cullen+2021}
{Cullen}, F., {Shapley}, A.~E., {McLure}, R.~J., {et~al.} 2021, \mnras, 505,
  903, \dodoi{10.1093/mnras/stab1340}

\bibitem[{{Curti} {et~al.}(2017){Curti}, {Cresci}, {Mannucci}, {Marconi},
  {Maiolino}, \& {Esposito}}]{Curti+2017}
{Curti}, M., {Cresci}, G., {Mannucci}, F., {et~al.} 2017, \mnras, 465, 1384,
  \dodoi{10.1093/mnras/stw2766}

\bibitem[{{Curti} {et~al.}(2020){Curti}, {Mannucci}, {Cresci}, \&
  {Maiolino}}]{Curti+2020}
{Curti}, M., {Mannucci}, F., {Cresci}, G., \& {Maiolino}, R. 2020, \mnras, 491,
  944, \dodoi{10.1093/mnras/stz2910}

\bibitem[{{Curti} {et~al.}(2023){Curti}, {Maiolino}, {Curtis-Lake},
  {Chevallard}, {Carniani}, {D'Eugenio}, {Looser}, {Scholtz}, {Charlot},
  {Cameron}, {{\"U}bler}, {Witstok}, {Boyett}, {Laseter}, {Sandles}, {Arribas},
  {Bunker}, {Giardino}, {Maseda}, {Rawle}, {Rodr{\'\i}guez Del Pino}, {Smit},
  {Willott}, {Eisenstein}, {Hausen}, {Johnson}, {Rieke}, {Robertson},
  {Tacchella}, {Williams}, {Willmer}, {Baker}, {Bhatawdekar}, {Egami},
  {Helton}, {Ji}, {Kumari}, {Perna}, {Shivaei}, \& {Sun}}]{Curti+2023b}
{Curti}, M., {Maiolino}, R., {Curtis-Lake}, E., {et~al.} 2023, arXiv e-prints,
  arXiv:2304.08516, \dodoi{10.48550/arXiv.2304.08516}

\bibitem[{{Dav{\'e}} {et~al.}(2011){Dav{\'e}}, {Oppenheimer}, \&
  {Finlator}}]{Dave+2011}
{Dav{\'e}}, R., {Oppenheimer}, B.~D., \& {Finlator}, K. 2011, \mnras, 415, 11,
  \dodoi{10.1111/j.1365-2966.2011.18680.x}

\bibitem[{{De Lucia} {et~al.}(2020){De Lucia}, {Xie}, {Fontanot}, \&
  {Hirschmann}}]{DeLucia+2020}
{De Lucia}, G., {Xie}, L., {Fontanot}, F., \& {Hirschmann}, M. 2020, \mnras,
  498, 3215, \dodoi{10.1093/mnras/staa2556}

\bibitem[{{Dessauges-Zavadsky} {et~al.}(2010){Dessauges-Zavadsky}, {D'Odorico},
  {Schaerer}, {Modigliani}, {Tapken}, \& {Vernet}}]{Dessauges-Zavadsky+2010}
{Dessauges-Zavadsky}, M., {D'Odorico}, S., {Schaerer}, D., {et~al.} 2010, \aap,
  510, A26, \dodoi{10.1051/0004-6361/200913337}

\bibitem[{{D'Eugenio} {et~al.}(2023){D'Eugenio}, {Maiolino}, {Carniani},
  {Curtis-Lake}, {Witstok}, {Chevallard}, {Charlot}, {Baker}, {Arribas},
  {Boyett}, {Bunker}, {Curti}, {Eisenstein}, {Hainline}, {Ji}, {Johnson},
  {Looser}, {Nakajima}, {Nelson}, {Rieke}, {Robertson}, {Scholtz}, {Smit},
  {Venturi}, {Tacchella}, {Uebler}, {Willmer}, \& {Willott}}]{deugenio+2023}
{D'Eugenio}, F., {Maiolino}, R., {Carniani}, S., {et~al.} 2023, arXiv e-prints,
  arXiv:2311.09908, \dodoi{10.48550/arXiv.2311.09908}

\bibitem[{{Dopita} {et~al.}(2016){Dopita}, {Kewley}, {Sutherland}, \&
  {Nicholls}}]{Dopita+2016}
{Dopita}, M.~A., {Kewley}, L.~J., {Sutherland}, R.~S., \& {Nicholls}, D.~C.
  2016, \apss, 361, 61, \dodoi{10.1007/s10509-016-2657-8}

\bibitem[{{Dopita} {et~al.}(2013){Dopita}, {Sutherland}, {Nicholls}, {Kewley},
  \& {Vogt}}]{Dopita+2013}
{Dopita}, M.~A., {Sutherland}, R.~S., {Nicholls}, D.~C., {Kewley}, L.~J., \&
  {Vogt}, F. P.~A. 2013, \apjs, 208, 10, \dodoi{10.1088/0067-0049/208/1/10}

\bibitem[{{Eldridge} \& {Stanway}(2016)}]{Eldridge+2016}
{Eldridge}, J.~J., \& {Stanway}, E.~R. 2016, \mnras, 462, 3302,
  \dodoi{10.1093/mnras/stw1772}

\bibitem[{{Eldridge} {et~al.}(2017){Eldridge}, {Stanway}, {Xiao}, {McClelland},
  {Taylor}, {Ng}, {Greis}, \& {Bray}}]{Eldridge+2017}
{Eldridge}, J.~J., {Stanway}, E.~R., {Xiao}, L., {et~al.} 2017, \pasa, 34,
  e058, \dodoi{10.1017/pasa.2017.51}

\bibitem[{{Ellison} {et~al.}(2008){Ellison}, {Patton}, {Simard}, \&
  {McConnachie}}]{Ellison+2008}
{Ellison}, S.~L., {Patton}, D.~R., {Simard}, L., \& {McConnachie}, A.~W. 2008,
  \apjl, 672, L107, \dodoi{10.1086/527296}

\bibitem[{{Erb} {et~al.}(2010){Erb}, {Pettini}, {Shapley}, {Steidel}, {Law}, \&
  {Reddy}}]{Erb+2010}
{Erb}, D.~K., {Pettini}, M., {Shapley}, A.~E., {et~al.} 2010, \apj, 719, 1168,
  \dodoi{10.1088/0004-637X/719/2/1168}

\bibitem[{{Erb} {et~al.}(2006{\natexlab{a}}){Erb}, {Shapley}, {Pettini},
  {Steidel}, {Reddy}, \& {Adelberger}}]{Erb+2006}
{Erb}, D.~K., {Shapley}, A.~E., {Pettini}, M., {et~al.} 2006{\natexlab{a}},
  \apj, 644, 813, \dodoi{10.1086/503623}

\bibitem[{{Erb} {et~al.}(2018){Erb}, {Steidel}, \& {Chen}}]{Erb+2018}
{Erb}, D.~K., {Steidel}, C.~C., \& {Chen}, Y. 2018, \apjl, 862, L10,
  \dodoi{10.3847/2041-8213/aacff6}

\bibitem[{{Erb} {et~al.}(2006{\natexlab{b}}){Erb}, {Steidel}, {Shapley},
  {Pettini}, {Reddy}, \& {Adelberger}}]{Erb+2006a}
{Erb}, D.~K., {Steidel}, C.~C., {Shapley}, A.~E., {et~al.} 2006{\natexlab{b}},
  \apj, 647, 128, \dodoi{10.1086/505341}

\bibitem[{{Esteban} {et~al.}(2009){Esteban}, {Bresolin}, {Peimbert},
  {Garc{\'\i}a-Rojas}, {Peimbert}, \& {Mesa-Delgado}}]{Esteban+2009}
{Esteban}, C., {Bresolin}, F., {Peimbert}, M., {et~al.} 2009, \apj, 700, 654,
  \dodoi{10.1088/0004-637X/700/1/654}

\bibitem[{{Esteban} {et~al.}(2014){Esteban}, {Garc{\'\i}a-Rojas}, {Carigi},
  {Peimbert}, {Bresolin}, {L{\'o}pez-S{\'a}nchez}, \&
  {Mesa-Delgado}}]{Esteban+2014}
{Esteban}, C., {Garc{\'\i}a-Rojas}, J., {Carigi}, L., {et~al.} 2014, \mnras,
  443, 624, \dodoi{10.1093/mnras/stu1177}

\bibitem[{{Esteban} {et~al.}(2004){Esteban}, {Peimbert}, {Garc{\'\i}a-Rojas},
  {Ruiz}, {Peimbert}, \& {Rodr{\'\i}guez}}]{Esteban+2004}
{Esteban}, C., {Peimbert}, M., {Garc{\'\i}a-Rojas}, J., {et~al.} 2004, \mnras,
  355, 229, \dodoi{10.1111/j.1365-2966.2004.08313.x}

\bibitem[{{Feltre} {et~al.}(2016){Feltre}, {Charlot}, \&
  {Gutkin}}]{Feltre+2016}
{Feltre}, A., {Charlot}, S., \& {Gutkin}, J. 2016, \mnras, 456, 3354,
  \dodoi{10.1093/mnras/stv2794}

\bibitem[{{Ferrara} \& {Ricotti}(2006)}]{Ferrara_Ricotti+2006}
{Ferrara}, A., \& {Ricotti}, M. 2006, \mnras, 373, 571,
  \dodoi{10.1111/j.1365-2966.2006.10978.x}

\bibitem[{{Garc{\'\i}a-Rojas} \& {Esteban}(2007)}]{GarciaRojas+2007}
{Garc{\'\i}a-Rojas}, J., \& {Esteban}, C. 2007, \apj, 670, 457,
  \dodoi{10.1086/521871}

\bibitem[{{Garc{\'\i}a-Rojas} {et~al.}(2005){Garc{\'\i}a-Rojas}, {Esteban},
  {Peimbert}, {Peimbert}, {Rodr{\'\i}guez}, \& {Ruiz}}]{GarciaRojas+2005}
{Garc{\'\i}a-Rojas}, J., {Esteban}, C., {Peimbert}, A., {et~al.} 2005, \mnras,
  362, 301, \dodoi{10.1111/j.1365-2966.2005.09302.x}

\bibitem[{{Garc{\'\i}a-Rojas} {et~al.}(2004){Garc{\'\i}a-Rojas}, {Esteban},
  {Peimbert}, {Rodr{\'\i}guez}, {Ruiz}, \& {Peimbert}}]{GarciaRojas+2004}
{Garc{\'\i}a-Rojas}, J., {Esteban}, C., {Peimbert}, M., {et~al.} 2004, \apjs,
  153, 501, \dodoi{10.1086/421909}

\bibitem[{{Garnett} {et~al.}(1999){Garnett}, {Shields}, {Peimbert},
  {Torres-Peimbert}, {Skillman}, {Dufour}, {Terlevich}, \&
  {Terlevich}}]{Garnett+1999}
{Garnett}, D.~R., {Shields}, G.~A., {Peimbert}, M., {et~al.} 1999, \apj, 513,
  168, \dodoi{10.1086/306860}

\bibitem[{{Gburek} {et~al.}(2019){Gburek}, {Siana}, {Alavi}, {Emami},
  {Richard}, {Freeman}, {Stark}, {Snapp-Kolas}, \& {Lucero}}]{Gburek+2019}
{Gburek}, T., {Siana}, B., {Alavi}, A., {et~al.} 2019, \apj, 887, 168,
  \dodoi{10.3847/1538-4357/ab5713}

\bibitem[{{Genzel} {et~al.}(2015){Genzel}, {Tacconi}, {Lutz}, {Saintonge},
  {Berta}, {Magnelli}, {Combes}, {Garc{\'\i}a-Burillo}, {Neri}, {Bolatto},
  {Contini}, {Lilly}, {Boissier}, {Boone}, {Bouch{\'e}}, {Bournaud}, {Burkert},
  {Carollo}, {Colina}, {Cooper}, {Cox}, {Feruglio}, {F{\"o}rster Schreiber},
  {Freundlich}, {Gracia-Carpio}, {Juneau}, {Kovac}, {Lippa}, {Naab}, {Salome},
  {Renzini}, {Sternberg}, {Walter}, {Weiner}, {Weiss}, \&
  {Wuyts}}]{Genzel+2015}
{Genzel}, R., {Tacconi}, L.~J., {Lutz}, D., {et~al.} 2015, \apj, 800, 20,
  \dodoi{10.1088/0004-637X/800/1/20}

\bibitem[{{Gonz{\'a}lez Delgado} {et~al.}(2014){Gonz{\'a}lez Delgado}, {Cid
  Fernandes}, {Garc{\'\i}a-Benito}, {P{\'e}rez}, {de Amorim},
  {Cortijo-Ferrero}, {Lacerda}, {L{\'o}pez Fern{\'a}ndez}, {S{\'a}nchez}, {Vale
  Asari}, {Alves}, {Bland-Hawthorn}, {Galbany}, {Gallazzi}, {Husemann},
  {Bekeraite}, {Jungwiert}, {L{\'o}pez-S{\'a}nchez}, {de Lorenzo-C{\'a}ceres},
  {Marino}, {Mast}, {Moll{\'a}}, {del Olmo}, {S{\'a}nchez-Bl{\'a}zquez}, {van
  de Ven}, {V{\'\i}lchez}, {Walcher}, {Wisotzki}, {Ziegler}, \& {CALIFA
  Collaboration}}]{GonzalezDelgad0+2014}
{Gonz{\'a}lez Delgado}, R.~M., {Cid Fernandes}, R., {Garc{\'\i}a-Benito}, R.,
  {et~al.} 2014, \apjl, 791, L16, \dodoi{10.1088/2041-8205/791/1/L16}

\bibitem[{{Gordon} {et~al.}(2003){Gordon}, {Clayton}, {Misselt}, {Landolt}, \&
  {Wolff}}]{Gordon+2003}
{Gordon}, K.~D., {Clayton}, G.~C., {Misselt}, K.~A., {Landolt}, A.~U., \&
  {Wolff}, M.~J. 2003, \apj, 594, 279, \dodoi{10.1086/376774}

\bibitem[{{Green} {et~al.}(2018){Green}, {Schlafly}, {Finkbeiner}, {Rix},
  {Martin}, {Burgett}, {Draper}, {Flewelling}, {Hodapp}, {Kaiser}, {Kudritzki},
  {Magnier}, {Metcalfe}, {Tonry}, {Wainscoat}, \& {Waters}}]{Green+2018}
{Green}, G.~M., {Schlafly}, E.~F., {Finkbeiner}, D., {et~al.} 2018, \mnras,
  478, 651, \dodoi{10.1093/mnras/sty1008}

\bibitem[{{Gribel} {et~al.}(2017){Gribel}, {Miranda}, \& {Williams
  Vilas-Boas}}]{Gribel+2017}
{Gribel}, C., {Miranda}, O.~D., \& {Williams Vilas-Boas}, J. 2017, \apj, 849,
  108, \dodoi{10.3847/1538-4357/aa921a}

\bibitem[{{Groves} {et~al.}(2006){Groves}, {Heckman}, \&
  {Kauffmann}}]{Groves+2006}
{Groves}, B.~A., {Heckman}, T.~M., \& {Kauffmann}, G. 2006, \mnras, 371, 1559,
  \dodoi{10.1111/j.1365-2966.2006.10812.x}

\bibitem[{{Guo} {et~al.}(2016){Guo}, {Koo}, {Lu}, {Forbes}, {Rafelski},
  {Trump}, {Amor{\'\i}n}, {Barro}, {Dav{\'e}}, {Faber}, {Hathi}, {Yesuf},
  {Cooper}, {Dekel}, {Guhathakurta}, {Kirby}, {Koekemoer},
  {P{\'e}rez-Gonz{\'a}lez}, {Lin}, {Newman}, {Primack}, {Rosario}, {Willmer},
  \& {Yan}}]{Guo+2016}
{Guo}, Y., {Koo}, D.~C., {Lu}, Y., {et~al.} 2016, \apj, 822, 103,
  \dodoi{10.3847/0004-637X/822/2/103}

\bibitem[{{Heintz} {et~al.}(2023){Heintz}, {Gim{\'e}nez-Arteaga}, {Fujimoto},
  {Brammer}, {Espada}, {Gillman}, {Gonz{\'a}lez-L{\'o}pez}, {Greve},
  {Harikane}, {Hatsukade}, {Knudsen}, {Koekemoer}, {Kohno}, {Kokorev}, {Lee},
  {Magdis}, {Nelson}, {Rizzo}, {Sanders}, {Schaerer}, {Shapley}, {Strait},
  {Toft}, {Valentino}, {van der Wel}, {Vijayan}, {Watson}, {Bauer},
  {Christiansen}, \& {Wilson}}]{Heintz+2023}
{Heintz}, K.~E., {Gim{\'e}nez-Arteaga}, C., {Fujimoto}, S., {et~al.} 2023,
  \apjl, 944, L30, \dodoi{10.3847/2041-8213/acb2cf}

\bibitem[{{Henry} {et~al.}(2021){Henry}, {Rafelski}, {Sunnquist}, {Pirzkal},
  {Pacifici}, {Atek}, {Bagley}, {Baronchelli}, {Barro}, {Bunker}, {Colbert},
  {Dai}, {Elmegreen}, {Elmegreen}, {Finkelstein}, {Kocevski}, {Koekemoer},
  {Malkan}, {Martin}, {Mehta}, {Pahl}, {Papovich}, {Rutkowski}, {S{\'a}nchez
  Almeida}, {Scarlata}, {Snyder}, \& {Teplitz}}]{Henry+2021}
{Henry}, A., {Rafelski}, M., {Sunnquist}, B., {et~al.} 2021, \apj, 919, 143,
  \dodoi{10.3847/1538-4357/ac1105}

\bibitem[{{Henry} {et~al.}(2000{\natexlab{a}}){Henry}, {Edmunds}, \&
  {K{\"o}ppen}}]{Henry+2000b}
{Henry}, R.~B.~C., {Edmunds}, M.~G., \& {K{\"o}ppen}, J. 2000{\natexlab{a}},
  \apj, 541, 660, \dodoi{10.1086/309471}

\bibitem[{{Henry} {et~al.}(2000{\natexlab{b}}){Henry}, {Kwitter}, \&
  {Bates}}]{Henry+2000a}
{Henry}, R.~B.~C., {Kwitter}, K.~B., \& {Bates}, J.~A. 2000{\natexlab{b}},
  \apj, 531, 928, \dodoi{10.1086/308509}

\bibitem[{{Hernandez} {et~al.}(2020){Hernandez}, {Aloisi}, {James}, {Ferland},
  {Fox}, {Tosi}, \& {Tumlinson}}]{Hernandez+2020}
{Hernandez}, S., {Aloisi}, A., {James}, B.~L., {et~al.} 2020, \apj, 892, 19,
  \dodoi{10.3847/1538-4357/ab77c6}

\bibitem[{{Hirschmann} {et~al.}(2017){Hirschmann}, {Charlot}, {Feltre}, {Naab},
  {Choi}, {Ostriker}, \& {Somerville}}]{Hirschmann+2017}
{Hirschmann}, M., {Charlot}, S., {Feltre}, A., {et~al.} 2017, \mnras, 472,
  2468, \dodoi{10.1093/mnras/stx2180}

\bibitem[{{Hunt} {et~al.}(2016){Hunt}, {Dayal}, {Magrini}, \&
  {Ferrara}}]{Hunt+2016}
{Hunt}, L., {Dayal}, P., {Magrini}, L., \& {Ferrara}, A. 2016, \mnras, 463,
  2002, \dodoi{10.1093/mnras/stw1993}

\bibitem[{{Iani} {et~al.}(2023){Iani}, {Zanella}, {Vernet}, {Richard},
  {Gronke}, {Arrigoni-Battaia}, {Bolamperti}, {Caputi}, {Humphrey},
  {Rodighiero}, {Rinaldi}, \& {Vanzella}}]{Iani+2023}
{Iani}, E., {Zanella}, A., {Vernet}, J., {et~al.} 2023, \mnras, 518, 5018,
  \dodoi{10.1093/mnras/stac3198}

\bibitem[{{Isobe} {et~al.}(2023){Isobe}, {Ouchi}, {Tominaga}, {Watanabe},
  {Nakajima}, {Umeda}, {Yajima}, {Harikane}, {Fukushima}, {Xu}, {Ono}, \&
  {Zhang}}]{Isobe+2023}
{Isobe}, Y., {Ouchi}, M., {Tominaga}, N., {et~al.} 2023, \apj, 959, 100,
  \dodoi{10.3847/1538-4357/ad09be}

\bibitem[{{James} {et~al.}(2018){James}, {Auger}, {Pettini}, {Stark},
  {Belokurov}, \& {Carniani}}]{James+2018}
{James}, B.~L., {Auger}, M., {Pettini}, M., {et~al.} 2018, \mnras, 476, 1726,
  \dodoi{10.1093/mnras/sty315}

\bibitem[{{James} {et~al.}(2014){James}, {Pettini}, {Christensen}, {Auger},
  {Becker}, {King}, {Quider}, {Shapley}, \& {Steidel}}]{James+2014}
{James}, B.~L., {Pettini}, M., {Christensen}, L., {et~al.} 2014, \mnras, 440,
  1794, \dodoi{10.1093/mnras/stu287}

\bibitem[{{Jaskot} \& {Ravindranath}(2016)}]{Jaskot_Ravindranath2016}
{Jaskot}, A.~E., \& {Ravindranath}, S. 2016, \apj, 833, 136,
  \dodoi{10.3847/1538-4357/833/2/136}

\bibitem[{{Jiang} {et~al.}(2019){Jiang}, {Malhotra}, {Rhoads}, \&
  {Yang}}]{Jiang+2019}
{Jiang}, T., {Malhotra}, S., {Rhoads}, J.~E., \& {Yang}, H. 2019, \apj, 872,
  145, \dodoi{10.3847/1538-4357/aaee8a}

\bibitem[{{Johnson} {et~al.}(2021){Johnson}, {Leja}, {Conroy}, \&
  {Speagle}}]{Johnson+2021}
{Johnson}, B.~D., {Leja}, J., {Conroy}, C., \& {Speagle}, J.~S. 2021, \apjs,
  254, 22, \dodoi{10.3847/1538-4365/abef67}

\bibitem[{{Jones} {et~al.}(2023){Jones}, {Sanders}, {Chen}, {Wang},
  {Morishita}, {Roberts-Borsani}, {Treu}, {Dressler}, {Merlin}, {Paris},
  {Santini}, {Bergamini}, {Huntzinger}, {Nanayakkara}, {Boyett}, {Bradac},
  {Brammer}, {Calabro}, {Glazebrook}, {Grasha}, {Mascia}, {Pentericci},
  {Trenti}, \& {Vulcani}}]{Jones+2023}
{Jones}, T., {Sanders}, R., {Chen}, Y., {et~al.} 2023, arXiv e-prints,
  arXiv:2301.07126, \dodoi{10.48550/arXiv.2301.07126}

\bibitem[{{Kashino} {et~al.}(2022){Kashino}, {Lilly}, {Renzini}, {Daddi},
  {Zamorani}, {Silverman}, {Ilbert}, {Peng}, {Mainieri}, {Bardelli}, {Zucca},
  {Kartaltepe}, \& {Sanders}}]{Kashino+2022}
{Kashino}, D., {Lilly}, S.~J., {Renzini}, A., {et~al.} 2022, \apj, 925, 82,
  \dodoi{10.3847/1538-4357/ac399e}

\bibitem[{{Kehrig} {et~al.}(2018){Kehrig}, {V{\'\i}lchez}, {Guerrero},
  {Iglesias-P{\'a}ramo}, {Hunt}, {Duarte-Puertas}, \&
  {Ramos-Larios}}]{Kehrig+2018}
{Kehrig}, C., {V{\'\i}lchez}, J.~M., {Guerrero}, M.~A., {et~al.} 2018, \mnras,
  480, 1081, \dodoi{10.1093/mnras/sty1920}

\bibitem[{{Kelson}(2003)}]{Kelson2003}
{Kelson}, D.~D. 2003, \pasp, 115, 688, \dodoi{10.1086/375502}

\bibitem[{{Kewley} \& {Ellison}(2008)}]{kewley_ellison2008}
{Kewley}, L.~J., \& {Ellison}, S.~L. 2008, \apj, 681, 1183,
  \dodoi{10.1086/587500}

\bibitem[{{Khusanova} {et~al.}(2020){Khusanova}, {Le F{\`e}vre}, {Cassata},
  {Cucciati}, {Lemaux}, {Tasca}, {Thomas}, {Garilli}, {Le Brun}, {Maccagni},
  {Pentericci}, {Zamorani}, {Amor{\'\i}n}, {Bardelli}, {Castellano},
  {Cassar{\`a}}, {Cimatti}, {Giavalisco}, {Hathi}, {Ilbert}, {Koekemoer},
  {Marchi}, {Pforr}, {Ribeiro}, {Schaerer}, {Tresse}, {Vergani}, \&
  {Zucca}}]{Khusanova+2020}
{Khusanova}, Y., {Le F{\`e}vre}, O., {Cassata}, P., {et~al.} 2020, \aap, 634,
  A97, \dodoi{10.1051/0004-6361/201935400}

\bibitem[{{Kirby} {et~al.}(2013){Kirby}, {Cohen}, {Guhathakurta}, {Cheng},
  {Bullock}, \& {Gallazzi}}]{Kirby+2013}
{Kirby}, E.~N., {Cohen}, J.~G., {Guhathakurta}, P., {et~al.} 2013, \apj, 779,
  102, \dodoi{10.1088/0004-637X/779/2/102}

\bibitem[{{Kobulnicky} \& {Kewley}(2004)}]{Kobulnicky_Kewley2004}
{Kobulnicky}, H.~A., \& {Kewley}, L.~J. 2004, \apj, 617, 240,
  \dodoi{10.1086/425299}

\bibitem[{{Kudritzki} {et~al.}(2015){Kudritzki}, {Ho}, {Schruba}, {Burkert},
  {Zahid}, {Bresolin}, \& {Dima}}]{Kudritzki+2015}
{Kudritzki}, R.-P., {Ho}, I.~T., {Schruba}, A., {et~al.} 2015, \mnras, 450,
  342, \dodoi{10.1093/mnras/stv522}

\bibitem[{{Kudritzki} {et~al.}(2014){Kudritzki}, {Urbaneja}, {Bresolin},
  {Hosek}, \& {Przybilla}}]{Kudritzki+2014}
{Kudritzki}, R.-P., {Urbaneja}, M.~A., {Bresolin}, F., {Hosek}, Matthew~W., J.,
  \& {Przybilla}, N. 2014, \apj, 788, 56, \dodoi{10.1088/0004-637X/788/1/56}

\bibitem[{{Langeroodi} {et~al.}(2022){Langeroodi}, {Hjorth}, {Chen}, {Kelly},
  {Williams}, {Lin}, {Scarlata}, {Zitrin}, {Broadhurst}, {Diego}, {Huang},
  {Filippenko}, {Foley}, {Jha}, {Koekemoer}, {Oguri}, {Perez-Fournon},
  {Pierel}, {Poidevin}, \& {Strolger}}]{Langeroodi+2022}
{Langeroodi}, D., {Hjorth}, J., {Chen}, W., {et~al.} 2022, arXiv e-prints,
  arXiv:2212.02491, \dodoi{10.48550/arXiv.2212.02491}

\bibitem[{{Langeroodi} {et~al.}(2023){Langeroodi}, {Hjorth}, {Chen}, {Kelly},
  {Williams}, {Lin}, {Scarlata}, {Zitrin}, {Broadhurst}, {Diego}, {Huang},
  {Filippenko}, {Foley}, {Jha}, {Koekemoer}, {Oguri}, {Perez-Fournon},
  {Pierel}, {Poidevin}, \& {Strolger}}]{Langeroodi+2023c}
---. 2023, \apj, 957, 39, \dodoi{10.3847/1538-4357/acdbc1}

\bibitem[{{Lara-L{\'o}pez} {et~al.}(2010){Lara-L{\'o}pez}, {Cepa},
  {Bongiovanni}, {P{\'e}rez Garc{\'\i}a}, {Ederoclite}, {Casta{\~n}eda},
  {Fern{\'a}ndez Lorenzo}, {Povi{\'c}}, \&
  {S{\'a}nchez-Portal}}]{LaraLopez+2010}
{Lara-L{\'o}pez}, M.~A., {Cepa}, J., {Bongiovanni}, A., {et~al.} 2010, \aap,
  521, L53, \dodoi{10.1051/0004-6361/201014803}

\bibitem[{{Laseter} {et~al.}(2023){Laseter}, {Maseda}, {Curti}, {Maiolino},
  {D'Eugenio}, {Cameron}, {Looser}, {Arribas}, {Baker}, {Bhatawdekar},
  {Boyett}, {Bunker}, {Carniani}, {Charlot}, {Chevallard}, {Curtis-lake},
  {Egami}, {Eisenstein}, {Hainline}, {Hausen}, {Ji}, {Kumari}, {Perna},
  {Rawle}, {Rix}, {Robertson}, {Rodr{\'\i}guez Del Pino}, {Sandles}, {Scholtz},
  {Smit}, {Tacchella}, {{\"U}bler}, {Williams}, {Willott}, \&
  {Witstok}}]{Laseter+2023}
{Laseter}, I.~H., {Maseda}, M.~V., {Curti}, M., {et~al.} 2023, arXiv e-prints,
  arXiv:2306.03120, \dodoi{10.48550/arXiv.2306.03120}

\bibitem[{{Leclercq} {et~al.}(2017){Leclercq}, {Bacon}, {Wisotzki}, {Mitchell},
  {Garel}, {Verhamme}, {Blaizot}, {Hashimoto}, {Herenz}, {Conseil},
  {Cantalupo}, {Inami}, {Contini}, {Richard}, {Maseda}, {Schaye}, {Marino},
  {Akhlaghi}, {Brinchmann}, \& {Carollo}}]{Leclercq+2017}
{Leclercq}, F., {Bacon}, R., {Wisotzki}, L., {et~al.} 2017, \aap, 608, A8,
  \dodoi{10.1051/0004-6361/201731480}

\bibitem[{{Leclercq} {et~al.}(2020){Leclercq}, {Bacon}, {Verhamme}, {Garel},
  {Blaizot}, {Brinchmann}, {Cantalupo}, {Claeyssens}, {Conseil}, {Contini},
  {Hashimoto}, {Herenz}, {Kusakabe}, {Marino}, {Maseda}, {Matthee}, {Mitchell},
  {Pezzulli}, {Richard}, {Schmidt}, \& {Wisotzki}}]{Leclercq+2020}
{Leclercq}, F., {Bacon}, R., {Verhamme}, A., {et~al.} 2020, \aap, 635, A82,
  \dodoi{10.1051/0004-6361/201937339}

\bibitem[{{Lecroq} {et~al.}(2024){Lecroq}, {Charlot}, {Bressan}, {Bruzual},
  {Costa}, {Iorio}, {Spera}, {Mapelli}, {Chen}, {Chevallard}, \&
  {Dall'Amico}}]{Lecroq+2024}
{Lecroq}, M., {Charlot}, S., {Bressan}, A., {et~al.} 2024, \mnras, 527, 9480,
  \dodoi{10.1093/mnras/stad3838}

\bibitem[{{Lee} {et~al.}(2006){Lee}, {Skillman}, {Cannon}, {Jackson}, {Gehrz},
  {Polomski}, \& {Woodward}}]{Lee+2006}
{Lee}, H., {Skillman}, E.~D., {Cannon}, J.~M., {et~al.} 2006, \apj, 647, 970,
  \dodoi{10.1086/505573}

\bibitem[{{Lehnert} {et~al.}(2015){Lehnert}, {van Driel}, {Le Tiran}, {Di
  Matteo}, \& {Haywood}}]{Lehnert+2015}
{Lehnert}, M.~D., {van Driel}, W., {Le Tiran}, L., {Di Matteo}, P., \&
  {Haywood}, M. 2015, \aap, 577, A112, \dodoi{10.1051/0004-6361/201322630}

\bibitem[{{Leitherer} {et~al.}(2011){Leitherer}, {Tremonti}, {Heckman}, \&
  {Calzetti}}]{Leitherer+2011}
{Leitherer}, C., {Tremonti}, C.~A., {Heckman}, T.~M., \& {Calzetti}, D. 2011,
  \aj, 141, 37, \dodoi{10.1088/0004-6256/141/2/37}

\bibitem[{{Leja} {et~al.}(2019){Leja}, {Carnall}, {Johnson}, {Conroy}, \&
  {Speagle}}]{Leja+2019}
{Leja}, J., {Carnall}, A.~C., {Johnson}, B.~D., {Conroy}, C., \& {Speagle},
  J.~S. 2019, \apj, 876, 3, \dodoi{10.3847/1538-4357/ab133c}

\bibitem[{{Lian} {et~al.}(2017){Lian}, {Thomas}, {Goddard}, \&
  {Marston}}]{Lian+2017}
{Lian}, J., {Thomas}, D., {Goddard}, D., \& {Marston}, C. 2017, in Galaxy
  Evolution Across Time, 36, \dodoi{10.5281/zenodo.807550}

\bibitem[{{Llerena} {et~al.}(2022){Llerena}, {Amor{\'\i}n}, {Cullen},
  {Pentericci}, {Calabr{\`o}}, {McLure}, {Carnall}, {P{\'e}rez-Montero},
  {Marchi}, {Bongiorno}, {Castellano}, {Fontana}, {McLeod}, {Talia}, {Hathi},
  {Hibon}, {Mannucci}, {Saxena}, {Schaerer}, \& {Zamorani}}]{Llerena+2022}
{Llerena}, M., {Amor{\'\i}n}, R., {Cullen}, F., {et~al.} 2022, \aap, 659, A16,
  \dodoi{10.1051/0004-6361/202141651}

\bibitem[{{L{\'o}pez-S{\'a}nchez} {et~al.}(2007){L{\'o}pez-S{\'a}nchez},
  {Esteban}, {Garc{\'\i}a-Rojas}, {Peimbert}, \&
  {Rodr{\'\i}guez}}]{LopezSanchez+2007}
{L{\'o}pez-S{\'a}nchez}, {\'A}.~R., {Esteban}, C., {Garc{\'\i}a-Rojas}, J.,
  {Peimbert}, M., \& {Rodr{\'\i}guez}, M. 2007, \apj, 656, 168,
  \dodoi{10.1086/510112}

\bibitem[{{Luridiana} {et~al.}(2015){Luridiana}, {Morisset}, \&
  {Shaw}}]{Luridiana+2015}
{Luridiana}, V., {Morisset}, C., \& {Shaw}, R.~A. 2015, \aap, 573, A42,
  \dodoi{10.1051/0004-6361/201323152}

\bibitem[{{Madau} \& {Dickinson}(2014)}]{Madau_Dickinson2014}
{Madau}, P., \& {Dickinson}, M. 2014, \araa, 52, 415,
  \dodoi{10.1146/annurev-astro-081811-125615}

\bibitem[{{Mainali} {et~al.}(2020){Mainali}, {Stark}, {Tang}, {Chevallard},
  {Charlot}, {Sharon}, {Coe}, {Salmon}, {Bradley}, {Johnson}, {Frye}, {Avila},
  {Ogaz}, {Zitrin}, {Brada{\v{c}}}, {Lemaux}, {Mahler}, {Paterno-Mahler},
  {Strait}, \& {Andrade-Santos}}]{Mainali+2020}
{Mainali}, R., {Stark}, D.~P., {Tang}, M., {et~al.} 2020, \mnras, 494, 719,
  \dodoi{10.1093/mnras/staa751}

\bibitem[{{Maiolino} \& {Mannucci}(2019)}]{MaiolinoMannucci2019}
{Maiolino}, R., \& {Mannucci}, F. 2019, \aapr, 27, 3,
  \dodoi{10.1007/s00159-018-0112-2}

\bibitem[{{Maiolino} {et~al.}(2008){Maiolino}, {Nagao}, {Grazian}, {Cocchia},
  {Marconi}, {Mannucci}, {Cimatti}, {Pipino}, {Ballero}, {Calura}, {Chiappini},
  {Fontana}, {Granato}, {Matteucci}, {Pastorini}, {Pentericci}, {Risaliti},
  {Salvati}, \& {Silva}}]{Maiolino+2008}
{Maiolino}, R., {Nagao}, T., {Grazian}, A., {et~al.} 2008, \aap, 488, 463,
  \dodoi{10.1051/0004-6361:200809678}

\bibitem[{{Mannucci} {et~al.}(2010){Mannucci}, {Cresci}, {Maiolino}, {Marconi},
  \& {Gnerucci}}]{Mannucci+2010}
{Mannucci}, F., {Cresci}, G., {Maiolino}, R., {Marconi}, A., \& {Gnerucci}, A.
  2010, \mnras, 408, 2115, \dodoi{10.1111/j.1365-2966.2010.17291.x}

\bibitem[{{Mannucci} {et~al.}(2009){Mannucci}, {Cresci}, {Maiolino}, {Marconi},
  {Pastorini}, {Pozzetti}, {Gnerucci}, {Risaliti}, {Schneider}, {Lehnert}, \&
  {Salvati}}]{Mannucci+2009}
{Mannucci}, F., {Cresci}, G., {Maiolino}, R., {et~al.} 2009, \mnras, 398, 1915,
  \dodoi{10.1111/j.1365-2966.2009.15185.x}

\bibitem[{{Marino} {et~al.}(2013){Marino}, {Rosales-Ortega}, {S{\'a}nchez},
  {Gil de Paz}, {V{\'\i}lchez}, {Miralles-Caballero}, {Kehrig},
  {P{\'e}rez-Montero}, {Stanishev}, {Iglesias-P{\'a}ramo}, {D{\'\i}az},
  {Castillo-Morales}, {Kennicutt}, {L{\'o}pez-S{\'a}nchez}, {Galbany},
  {Garc{\'\i}a-Benito}, {Mast}, {Mendez-Abreu}, {Monreal-Ibero}, {Husemann},
  {Walcher}, {Garc{\'\i}a-Lorenzo}, {Masegosa}, {Del Olmo Orozco},
  {Mour{\~a}o}, {Ziegler}, {Moll{\'a}}, {Papaderos},
  {S{\'a}nchez-Bl{\'a}zquez}, {Gonz{\'a}lez Delgado}, {Falc{\'o}n-Barroso},
  {Roth}, {van de Ven}, \& {CALIFA Team}}]{Marino+2013}
{Marino}, R.~A., {Rosales-Ortega}, F.~F., {S{\'a}nchez}, S.~F., {et~al.} 2013,
  \aap, 559, A114, \dodoi{10.1051/0004-6361/201321956}

\bibitem[{{Marques-Chaves} {et~al.}(2020){Marques-Chaves}, {P{\'e}rez-Fournon},
  {Shu}, {Colina}, {Bolton}, {{\'A}lvarez-M{\'a}rquez}, {Brownstein},
  {Cornachione}, {Geier}, {Jim{\'e}nez-{\'A}ngel}, {Kojima}, {Mao},
  {Montero-Dorta}, {Oguri}, {Ouchi}, {Poidevin}, {Shirley}, \&
  {Zheng}}]{MarquesChaves+2020}
{Marques-Chaves}, R., {P{\'e}rez-Fournon}, I., {Shu}, Y., {et~al.} 2020,
  \mnras, 492, 1257, \dodoi{10.1093/mnras/stz3500}

\bibitem[{{Marques-Chaves} {et~al.}(2024){Marques-Chaves}, {Schaerer},
  {Kuruvanthodi}, {Korber}, {Prantzos}, {Charbonnel}, {Weibel}, {Izotov},
  {Messa}, {Brammer}, {Dessauges-Zavadsky}, \& {Oesch}}]{MarquesChaves2024}
{Marques-Chaves}, R., {Schaerer}, D., {Kuruvanthodi}, A., {et~al.} 2024, \aap,
  681, A30, \dodoi{10.1051/0004-6361/202347411}

\bibitem[{{Martin}(2005)}]{Martin+2005}
{Martin}, C.~L. 2005, \apj, 621, 227, \dodoi{10.1086/427277}

\bibitem[{{Matthee} {et~al.}(2021){Matthee}, {Sobral}, {Hayes}, {Pezzulli},
  {Gronke}, {Schaerer}, {Naidu}, {R{\"o}ttgering}, {Calhau}, {Paulino-Afonso},
  {Santos}, \& {Amor{\'\i}n}}]{Matthee+2021}
{Matthee}, J., {Sobral}, D., {Hayes}, M., {et~al.} 2021, \mnras, 505, 1382,
  \dodoi{10.1093/mnras/stab1304}

\bibitem[{{Mattsson}(2010)}]{Mattsson2010}
{Mattsson}, L. 2010, \aap, 515, A68, \dodoi{10.1051/0004-6361/200913315}

\bibitem[{{Micheva} {et~al.}(2020){Micheva}, {{\"O}stlin}, {Melinder}, {Hayes},
  {Oey}, {Inoue}, {Iwata}, {Adamo}, {Wisotzki}, \& {Nakajima}}]{Micheva+2020}
{Micheva}, G., {{\"O}stlin}, G., {Melinder}, J., {et~al.} 2020, \apj, 903, 123,
  \dodoi{10.3847/1538-4357/abbdff}

\bibitem[{{Mingozzi} {et~al.}(2022){Mingozzi}, {James}, {Arellano-C{\'o}rdova},
  {Berg}, {Senchyna}, {Chisholm}, {Brinchmann}, {Aloisi}, {Amor{\'\i}n},
  {Charlot}, {Feltre}, {Hayes}, {Heckman}, {Henry}, {Hernandez}, {Kumari},
  {Leitherer}, {Llerena}, {Martin}, {Nanayakkara}, {Ravindranath}, {Skillman},
  {Sugahara}, {Wofford}, \& {Xu}}]{Mingozzi+2022}
{Mingozzi}, M., {James}, B.~L., {Arellano-C{\'o}rdova}, K.~Z., {et~al.} 2022,
  \apj, 939, 110, \dodoi{10.3847/1538-4357/ac952c}

\bibitem[{{Moustakas} {et~al.}(2010){Moustakas}, {Kennicutt}, {Tremonti},
  {Dale}, {Smith}, \& {Calzetti}}]{Moustakas+2010}
{Moustakas}, J., {Kennicutt}, Robert~C., J., {Tremonti}, C.~A., {et~al.} 2010,
  \apjs, 190, 233, \dodoi{10.1088/0067-0049/190/2/233}

\bibitem[{{Murray} {et~al.}(2011){Murray}, {M{\'e}nard}, \&
  {Thompson}}]{Murray+2011}
{Murray}, N., {M{\'e}nard}, B., \& {Thompson}, T.~A. 2011, \apj, 735, 66,
  \dodoi{10.1088/0004-637X/735/1/66}

\bibitem[{{Nagao} {et~al.}(2006){Nagao}, {Maiolino}, \& {Marconi}}]{Nagao+2006}
{Nagao}, T., {Maiolino}, R., \& {Marconi}, A. 2006, \aap, 459, 85,
  \dodoi{10.1051/0004-6361:20065216}

\bibitem[{{Nakajima} {et~al.}(2023){Nakajima}, {Ouchi}, {Isobe}, {Harikane},
  {Zhang}, {Ono}, {Umeda}, \& {Oguri}}]{Nakajima+2023}
{Nakajima}, K., {Ouchi}, M., {Isobe}, Y., {et~al.} 2023, arXiv e-prints,
  arXiv:2301.12825, \dodoi{10.48550/arXiv.2301.12825}

\bibitem[{{Nakajima} {et~al.}(2018){Nakajima}, {Schaerer}, {Le F{\`e}vre},
  {Amor{\'\i}n}, {Talia}, {Lemaux}, {Tasca}, {Vanzella}, {Zamorani},
  {Bardelli}, {Grazian}, {Guaita}, {Hathi}, {Pentericci}, \&
  {Zucca}}]{Nakajima+2018}
{Nakajima}, K., {Schaerer}, D., {Le F{\`e}vre}, O., {et~al.} 2018, \aap, 612,
  A94, \dodoi{10.1051/0004-6361/201731935}

\bibitem[{{Nanayakkara} {et~al.}(2019){Nanayakkara}, {Brinchmann}, {Boogaard},
  {Bouwens}, {Cantalupo}, {Feltre}, {Kollatschny}, {Marino}, {Maseda},
  {Matthee}, {Paalvast}, {Richard}, \& {Verhamme}}]{Nanayakkara+2019}
{Nanayakkara}, T., {Brinchmann}, J., {Boogaard}, L., {et~al.} 2019, \aap, 624,
  A89, \dodoi{10.1051/0004-6361/201834565}

\bibitem[{{Nicholls} {et~al.}(2017){Nicholls}, {Sutherland}, {Dopita},
  {Kewley}, \& {Groves}}]{Nicholls+2017}
{Nicholls}, D.~C., {Sutherland}, R.~S., {Dopita}, M.~A., {Kewley}, L.~J., \&
  {Groves}, B.~A. 2017, \mnras, 466, 4403, \dodoi{10.1093/mnras/stw3235}

\bibitem[{{Nomoto} {et~al.}(2013){Nomoto}, {Kobayashi}, \&
  {Tominaga}}]{Nomoto+2013}
{Nomoto}, K., {Kobayashi}, C., \& {Tominaga}, N. 2013, \araa, 51, 457,
  \dodoi{10.1146/annurev-astro-082812-140956}

\bibitem[{{Onodera} {et~al.}(2016){Onodera}, {Carollo}, {Lilly}, {Renzini},
  {Arimoto}, {Capak}, {Daddi}, {Scoville}, {Tacchella}, {Tatehora}, \&
  {Zamorani}}]{Onodera+2016}
{Onodera}, M., {Carollo}, C.~M., {Lilly}, S., {et~al.} 2016, \apj, 822, 42,
  \dodoi{10.3847/0004-637X/822/1/42}

\bibitem[{{Pe{\~n}a-Guerrero} {et~al.}(2017){Pe{\~n}a-Guerrero}, {Leitherer},
  {de Mink}, {Wofford}, \& {Kewley}}]{PenaGuerrero+2017}
{Pe{\~n}a-Guerrero}, M.~A., {Leitherer}, C., {de Mink}, S., {Wofford}, A., \&
  {Kewley}, L. 2017, \apj, 847, 107, \dodoi{10.3847/1538-4357/aa88bf}

\bibitem[{{Peimbert} {et~al.}(2005){Peimbert}, {Peimbert}, \&
  {Ruiz}}]{Peimbert+2005}
{Peimbert}, A., {Peimbert}, M., \& {Ruiz}, M.~T. 2005, \apj, 634, 1056,
  \dodoi{10.1086/444557}

\bibitem[{{Pettini} \& {Pagel}(2004)}]{Pettini_Pagel2004}
{Pettini}, M., \& {Pagel}, B. E.~J. 2004, \mnras, 348, L59,
  \dodoi{10.1111/j.1365-2966.2004.07591.x}

\bibitem[{{Pettini} {et~al.}(2002){Pettini}, {Rix}, {Steidel}, {Adelberger},
  {Hunt}, \& {Shapley}}]{Pettini+2002}
{Pettini}, M., {Rix}, S.~A., {Steidel}, C.~C., {et~al.} 2002, \apj, 569, 742,
  \dodoi{10.1086/339355}

\bibitem[{{Pettini} {et~al.}(2001){Pettini}, {Shapley}, {Steidel}, {Cuby},
  {Dickinson}, {Moorwood}, {Adelberger}, \& {Giavalisco}}]{Pettini+2001}
{Pettini}, M., {Shapley}, A.~E., {Steidel}, C.~C., {et~al.} 2001, \apj, 554,
  981, \dodoi{10.1086/321403}

\bibitem[{{Pilyugin} \& {Grebel}(2016)}]{Pilyugin_Grebel2016}
{Pilyugin}, L.~S., \& {Grebel}, E.~K. 2016, \mnras, 457, 3678,
  \dodoi{10.1093/mnras/stw238}

\bibitem[{{Planck Collaboration} {et~al.}(2020){Planck Collaboration},
  {Aghanim}, {Akrami}, {Ashdown}, {Aumont}, {Baccigalupi}, {Ballardini},
  {Banday}, {Barreiro}, {Bartolo}, {Basak}, {Battye}, {Benabed}, {Bernard},
  {Bersanelli}, {Bielewicz}, {Bock}, {Bond}, {Borrill}, {Bouchet}, {Boulanger},
  {Bucher}, {Burigana}, {Butler}, {Calabrese}, {Cardoso}, {Carron},
  {Challinor}, {Chiang}, {Chluba}, {Colombo}, {Combet}, {Contreras}, {Crill},
  {Cuttaia}, {de Bernardis}, {de Zotti}, {Delabrouille}, {Delouis}, {Di
  Valentino}, {Diego}, {Dor{\'e}}, {Douspis}, {Ducout}, {Dupac}, {Dusini},
  {Efstathiou}, {Elsner}, {En{\ss}lin}, {Eriksen}, {Fantaye}, {Farhang},
  {Fergusson}, {Fernandez-Cobos}, {Finelli}, {Forastieri}, {Frailis},
  {Fraisse}, {Franceschi}, {Frolov}, {Galeotta}, {Galli}, {Ganga},
  {G{\'e}nova-Santos}, {Gerbino}, {Ghosh}, {Gonz{\'a}lez-Nuevo}, {G{\'o}rski},
  {Gratton}, {Gruppuso}, {Gudmundsson}, {Hamann}, {Handley}, {Hansen},
  {Herranz}, {Hildebrandt}, {Hivon}, {Huang}, {Jaffe}, {Jones}, {Karakci},
  {Keih{\"a}nen}, {Keskitalo}, {Kiiveri}, {Kim}, {Kisner}, {Knox},
  {Krachmalnicoff}, {Kunz}, {Kurki-Suonio}, {Lagache}, {Lamarre}, {Lasenby},
  {Lattanzi}, {Lawrence}, {Le Jeune}, {Lemos}, {Lesgourgues}, {Levrier},
  {Lewis}, {Liguori}, {Lilje}, {Lilley}, {Lindholm}, {L{\'o}pez-Caniego},
  {Lubin}, {Ma}, {Mac{\'\i}as-P{\'e}rez}, {Maggio}, {Maino}, {Mandolesi},
  {Mangilli}, {Marcos-Caballero}, {Maris}, {Martin}, {Martinelli},
  {Mart{\'\i}nez-Gonz{\'a}lez}, {Matarrese}, {Mauri}, {McEwen}, {Meinhold},
  {Melchiorri}, {Mennella}, {Migliaccio}, {Millea}, {Mitra},
  {Miville-Desch{\^e}nes}, {Molinari}, {Montier}, {Morgante}, {Moss}, {Natoli},
  {N{\o}rgaard-Nielsen}, {Pagano}, {Paoletti}, {Partridge}, {Patanchon},
  {Peiris}, {Perrotta}, {Pettorino}, {Piacentini}, {Polastri}, {Polenta},
  {Puget}, {Rachen}, {Reinecke}, {Remazeilles}, {Renzi}, {Rocha}, {Rosset},
  {Roudier}, {Rubi{\~n}o-Mart{\'\i}n}, {Ruiz-Granados}, {Salvati}, {Sandri},
  {Savelainen}, {Scott}, {Shellard}, {Sirignano}, {Sirri}, {Spencer},
  {Sunyaev}, {Suur-Uski}, {Tauber}, {Tavagnacco}, {Tenti}, {Toffolatti},
  {Tomasi}, {Trombetti}, {Valenziano}, {Valiviita}, {Van Tent}, {Vibert},
  {Vielva}, {Villa}, {Vittorio}, {Wandelt}, {Wehus}, {White}, {White},
  {Zacchei}, \& {Zonca}}]{Planck+2020}
{Planck Collaboration}, {Aghanim}, N., {Akrami}, Y., {et~al.} 2020, \aap, 641,
  A6, \dodoi{10.1051/0004-6361/201833910}

\bibitem[{{Quider} {et~al.}(2010){Quider}, {Shapley}, {Pettini}, {Steidel}, \&
  {Stark}}]{Quider+2010}
{Quider}, A.~M., {Shapley}, A.~E., {Pettini}, M., {Steidel}, C.~C., \& {Stark},
  D.~P. 2010, \mnras, 402, 1467, \dodoi{10.1111/j.1365-2966.2009.16005.x}

\bibitem[{{Ravindranath} {et~al.}(2020){Ravindranath}, {Monroe}, {Jaskot},
  {Ferguson}, \& {Tumlinson}}]{Ravindranath+2020}
{Ravindranath}, S., {Monroe}, T., {Jaskot}, A., {Ferguson}, H.~C., \&
  {Tumlinson}, J. 2020, \apj, 896, 170, \dodoi{10.3847/1538-4357/ab91a5}

\bibitem[{{Reddy} {et~al.}(2008){Reddy}, {Steidel}, {Pettini}, {Adelberger},
  {Shapley}, {Erb}, \& {Dickinson}}]{Reddy+2008}
{Reddy}, N.~A., {Steidel}, C.~C., {Pettini}, M., {et~al.} 2008, \apjs, 175, 48,
  \dodoi{10.1086/521105}

\bibitem[{{Rigby} {et~al.}(2015){Rigby}, {Bayliss}, {Gladders}, {Sharon},
  {Wuyts}, {Dahle}, {Johnson}, \& {Pe{\~n}a-Guerrero}}]{Rigby+2015}
{Rigby}, J.~R., {Bayliss}, M.~B., {Gladders}, M.~D., {et~al.} 2015, \apjl, 814,
  L6, \dodoi{10.1088/2041-8205/814/1/L6}

\bibitem[{{Rogers} {et~al.}(2023){Rogers}, {Scarlata}, {Skillman}, {Eggen},
  {Jaskot}, {Mehta}, \& {Cannon}}]{Rogers+2023}
{Rogers}, N. S.~J., {Scarlata}, C.~M., {Skillman}, E.~D., {et~al.} 2023, \apj,
  955, 112, \dodoi{10.3847/1538-4357/acf294}

\bibitem[{{Romano}(2022)}]{Romano2022}
{Romano}, D. 2022, \aapr, 30, 7, \dodoi{10.1007/s00159-022-00144-z}

\bibitem[{{Salpeter}(1955)}]{Salpeter+1955}
{Salpeter}, E.~E. 1955, \apj, 121, 161, \dodoi{10.1086/145971}

\bibitem[{{Sancisi} {et~al.}(2008){Sancisi}, {Fraternali}, {Oosterloo}, \& {van
  der Hulst}}]{Sancisi+2008}
{Sancisi}, R., {Fraternali}, F., {Oosterloo}, T., \& {van der Hulst}, T. 2008,
  \aapr, 15, 189, \dodoi{10.1007/s00159-008-0010-0}

\bibitem[{{Sanders} {et~al.}(2023){Sanders}, {Shapley}, {Topping}, {Reddy}, \&
  {Brammer}}]{Sanders+2023c}
{Sanders}, R.~L., {Shapley}, A.~E., {Topping}, M.~W., {Reddy}, N.~A., \&
  {Brammer}, G.~B. 2023, arXiv e-prints, arXiv:2303.08149,
  \dodoi{10.48550/arXiv.2303.08149}

\bibitem[{{Sanders} {et~al.}(2024){Sanders}, {Shapley}, {Topping}, {Reddy}, \&
  {Brammer}}]{Sanders+2024}
---. 2024, \apj, 962, 24, \dodoi{10.3847/1538-4357/ad15fc}

\bibitem[{{Sanders} {et~al.}(2015){Sanders}, {Shapley}, {Kriek}, {Reddy},
  {Freeman}, {Coil}, {Siana}, {Mobasher}, {Shivaei}, {Price}, \& {de
  Groot}}]{Sanders+2015}
{Sanders}, R.~L., {Shapley}, A.~E., {Kriek}, M., {et~al.} 2015, \apj, 799, 138,
  \dodoi{10.1088/0004-637X/799/2/138}

\bibitem[{{Sanders} {et~al.}(2018){Sanders}, {Shapley}, {Kriek}, {Freeman},
  {Reddy}, {Siana}, {Coil}, {Mobasher}, {Dav{\'e}}, {Shivaei}, {Azadi},
  {Price}, {Leung}, {Fetherolf}, {de Groot}, {Zick}, {Fornasini}, \&
  {Barro}}]{Sanders+2018}
---. 2018, \apj, 858, 99, \dodoi{10.3847/1538-4357/aabcbd}

\bibitem[{{Sanders} {et~al.}(2020){Sanders}, {Shapley}, {Reddy}, {Kriek},
  {Siana}, {Coil}, {Mobasher}, {Shivaei}, {Freeman}, {Azadi}, {Price}, {Leung},
  {Fetherolf}, {de Groot}, {Zick}, {Fornasini}, \& {Barro}}]{Sanders+2020a}
{Sanders}, R.~L., {Shapley}, A.~E., {Reddy}, N.~A., {et~al.} 2020, \mnras, 491,
  1427, \dodoi{10.1093/mnras/stz3032}

\bibitem[{{Sanders} {et~al.}(2021){Sanders}, {Shapley}, {Jones}, {Reddy},
  {Kriek}, {Siana}, {Coil}, {Mobasher}, {Shivaei}, {Dav{\'e}}, {Azadi},
  {Price}, {Leung}, {Freeman}, {Fetherolf}, {de Groot}, {Zick}, \&
  {Barro}}]{Sanders+2021}
{Sanders}, R.~L., {Shapley}, A.~E., {Jones}, T., {et~al.} 2021, \apj, 914, 19,
  \dodoi{10.3847/1538-4357/abf4c1}

\bibitem[{{Savage} \& {Sembach}(1991)}]{SavageSembach1991}
{Savage}, B.~D., \& {Sembach}, K.~R. 1991, \apj, 379, 245,
  \dodoi{10.1086/170498}

\bibitem[{{Saxena} {et~al.}(2020){Saxena}, {Pentericci}, {Mirabelli},
  {Schaerer}, {Schneider}, {Cullen}, {Amorin}, {Bolzonella}, {Bongiorno},
  {Carnall}, {Castellano}, {Cucciati}, {Fontana}, {Fynbo}, {Garilli},
  {Gargiulo}, {Guaita}, {Hathi}, {Hutchison}, {Koekemoer}, {Marchi}, {McLeod},
  {McLure}, {Papovich}, {Pozzetti}, {Talia}, \& {Zamorani}}]{Saxena+2020}
{Saxena}, A., {Pentericci}, L., {Mirabelli}, M., {et~al.} 2020, \aap, 636, A47,
  \dodoi{10.1051/0004-6361/201937170}

\bibitem[{{Schlegel} {et~al.}(1998){Schlegel}, {Finkbeiner}, \&
  {Davis}}]{Schlegel+1998}
{Schlegel}, D.~J., {Finkbeiner}, D.~P., \& {Davis}, M. 1998, \apj, 500, 525,
  \dodoi{10.1086/305772}

\bibitem[{{Senchyna} {et~al.}(2021){Senchyna}, {Stark}, {Charlot},
  {Chevallard}, {Bruzual}, \& {Vidal-Garc{\'\i}a}}]{Senchyna+2021}
{Senchyna}, P., {Stark}, D.~P., {Charlot}, S., {et~al.} 2021, \mnras, 503,
  6112, \dodoi{10.1093/mnras/stab884}

\bibitem[{{Senchyna} {et~al.}(2019){Senchyna}, {Stark}, {Chevallard},
  {Charlot}, {Jones}, \& {Vidal-Garc{\'\i}a}}]{Senchyna+2019}
{Senchyna}, P., {Stark}, D.~P., {Chevallard}, J., {et~al.} 2019, \mnras, 488,
  3492, \dodoi{10.1093/mnras/stz1907}

\bibitem[{{Senchyna} {et~al.}(2017){Senchyna}, {Stark}, {Vidal-Garc{\'\i}a},
  {Chevallard}, {Charlot}, {Mainali}, {Jones}, {Wofford}, {Feltre}, \&
  {Gutkin}}]{Senchyna+2017}
{Senchyna}, P., {Stark}, D.~P., {Vidal-Garc{\'\i}a}, A., {et~al.} 2017, \mnras,
  472, 2608, \dodoi{10.1093/mnras/stx2059}

\bibitem[{{Shapley} {et~al.}(2005){Shapley}, {Steidel}, {Erb}, {Reddy},
  {Adelberger}, {Pettini}, {Barmby}, \& {Huang}}]{Shapley+2005}
{Shapley}, A.~E., {Steidel}, C.~C., {Erb}, D.~K., {et~al.} 2005, \apj, 626,
  698, \dodoi{10.1086/429990}

\bibitem[{{Shapley} {et~al.}(2003){Shapley}, {Steidel}, {Pettini}, \&
  {Adelberger}}]{Shapley+2003}
{Shapley}, A.~E., {Steidel}, C.~C., {Pettini}, M., \& {Adelberger}, K.~L. 2003,
  \apj, 588, 65, \dodoi{10.1086/373922}

\bibitem[{{Shapley} {et~al.}(2015){Shapley}, {Reddy}, {Kriek}, {Freeman},
  {Sanders}, {Siana}, {Coil}, {Mobasher}, {Shivaei}, {Price}, \& {de
  Groot}}]{Shapley+2015}
{Shapley}, A.~E., {Reddy}, N.~A., {Kriek}, M., {et~al.} 2015, \apj, 801, 88,
  \dodoi{10.1088/0004-637X/801/2/88}

\bibitem[{{Sharma} \& {Nath}(2012)}]{Sharma_Nath2012}
{Sharma}, M., \& {Nath}, B.~B. 2012, \apj, 750, 55,
  \dodoi{10.1088/0004-637X/750/1/55}

\bibitem[{{Speagle}(2020)}]{Speagle2020}
{Speagle}, J.~S. 2020, \mnras, 493, 3132, \dodoi{10.1093/mnras/staa278}

\bibitem[{{Stark} {et~al.}(2014){Stark}, {Richard}, {Siana}, {Charlot},
  {Freeman}, {Gutkin}, {Wofford}, {Robertson}, {Amanullah}, {Watson}, \&
  {Milvang-Jensen}}]{Stark+2014}
{Stark}, D.~P., {Richard}, J., {Siana}, B., {et~al.} 2014, \mnras, 445, 3200,
  \dodoi{10.1093/mnras/stu1618}

\bibitem[{{Stark} {et~al.}(2017){Stark}, {Ellis}, {Charlot}, {Chevallard},
  {Tang}, {Belli}, {Zitrin}, {Mainali}, {Gutkin}, {Vidal-Garc{\'\i}a},
  {Bouwens}, \& {Oesch}}]{Stark+2017}
{Stark}, D.~P., {Ellis}, R.~S., {Charlot}, S., {et~al.} 2017, \mnras, 464, 469,
  \dodoi{10.1093/mnras/stw2233}

\bibitem[{{Stasi{\'n}ska}(1982)}]{Stasinska1982}
{Stasi{\'n}ska}, G. 1982, \aaps, 48, 299

\bibitem[{{Steidel} {et~al.}(1999){Steidel}, {Adelberger}, {Giavalisco},
  {Dickinson}, \& {Pettini}}]{Steidel+1999}
{Steidel}, C.~C., {Adelberger}, K.~L., {Giavalisco}, M., {Dickinson}, M., \&
  {Pettini}, M. 1999, \apj, 519, 1, \dodoi{10.1086/307363}

\bibitem[{{Steidel} {et~al.}(2011){Steidel}, {Bogosavljevi{\'c}}, {Shapley},
  {Kollmeier}, {Reddy}, {Erb}, \& {Pettini}}]{Steidel+2011}
{Steidel}, C.~C., {Bogosavljevi{\'c}}, M., {Shapley}, A.~E., {et~al.} 2011,
  \apj, 736, 160, \dodoi{10.1088/0004-637X/736/2/160}

\bibitem[{{Steidel} {et~al.}(2010){Steidel}, {Erb}, {Shapley}, {Pettini},
  {Reddy}, {Bogosavljevi{\'c}}, {Rudie}, \& {Rakic}}]{Steidel+2010}
{Steidel}, C.~C., {Erb}, D.~K., {Shapley}, A.~E., {et~al.} 2010, \apj, 717,
  289, \dodoi{10.1088/0004-637X/717/1/289}

\bibitem[{{Steidel} {et~al.}(1996){Steidel}, {Giavalisco}, {Pettini},
  {Dickinson}, \& {Adelberger}}]{Steidel+1996}
{Steidel}, C.~C., {Giavalisco}, M., {Pettini}, M., {Dickinson}, M., \&
  {Adelberger}, K.~L. 1996, \apjl, 462, L17, \dodoi{10.1086/310029}

\bibitem[{{Steidel} {et~al.}(2016){Steidel}, {Strom}, {Pettini}, {Rudie},
  {Reddy}, \& {Trainor}}]{Steidel+2016}
{Steidel}, C.~C., {Strom}, A.~L., {Pettini}, M., {et~al.} 2016, \apj, 826, 159,
  \dodoi{10.3847/0004-637X/826/2/159}

\bibitem[{{Steidel} {et~al.}(2014){Steidel}, {Rudie}, {Strom}, {Pettini},
  {Reddy}, {Shapley}, {Trainor}, {Erb}, {Turner}, {Konidaris}, {Kulas}, {Mace},
  {Matthews}, \& {McLean}}]{Steidel+2014}
{Steidel}, C.~C., {Rudie}, G.~C., {Strom}, A.~L., {et~al.} 2014, \apj, 795,
  165, \dodoi{10.1088/0004-637X/795/2/165}

\bibitem[{{Stiavelli} {et~al.}(2023){Stiavelli}, {Morishita}, {Chiaberge},
  {Grillo}, {Leethochawalit}, {Rosati}, {Schuldt}, {Trenti}, \&
  {Treu}}]{Stiavelli+2023}
{Stiavelli}, M., {Morishita}, T., {Chiaberge}, M., {et~al.} 2023, \apjl, 957,
  L18, \dodoi{10.3847/2041-8213/ad0159}

\bibitem[{{Sugahara} {et~al.}(2017){Sugahara}, {Ouchi}, {Lin}, {Martin}, {Ono},
  {Harikane}, {Shibuya}, \& {Yan}}]{Sugahara+2017}
{Sugahara}, Y., {Ouchi}, M., {Lin}, L., {et~al.} 2017, \apj, 850, 51,
  \dodoi{10.3847/1538-4357/aa956d}

\bibitem[{{Suzuki} {et~al.}(2017){Suzuki}, {Kodama}, {Onodera}, {Shimakawa},
  {Hayashi}, {Tadaki}, {Koyama}, {Tanaka}, {Sobral}, {Smail}, {Best},
  {Khostovan}, {Minowa}, \& {Yamamoto}}]{Suzuki+2017}
{Suzuki}, T.~L., {Kodama}, T., {Onodera}, M., {et~al.} 2017, \apj, 849, 39,
  \dodoi{10.3847/1538-4357/aa8df3}

\bibitem[{{Tacconi} {et~al.}(2018){Tacconi}, {Genzel}, {Saintonge}, {Combes},
  {Garc{\'\i}a-Burillo}, {Neri}, {Bolatto}, {Contini}, {F{\"o}rster Schreiber},
  {Lilly}, {Lutz}, {Wuyts}, {Accurso}, {Boissier}, {Boone}, {Bouch{\'e}},
  {Bournaud}, {Burkert}, {Carollo}, {Cooper}, {Cox}, {Feruglio}, {Freundlich},
  {Herrera-Camus}, {Juneau}, {Lippa}, {Naab}, {Renzini}, {Salome}, {Sternberg},
  {Tadaki}, {{\"U}bler}, {Walter}, {Weiner}, \& {Weiss}}]{Tacconi+2018}
{Tacconi}, L.~J., {Genzel}, R., {Saintonge}, A., {et~al.} 2018, \apj, 853, 179,
  \dodoi{10.3847/1538-4357/aaa4b4}

\bibitem[{{Theios} {et~al.}(2019){Theios}, {Steidel}, {Strom}, {Rudie},
  {Trainor}, \& {Reddy}}]{Theios+2019}
{Theios}, R.~L., {Steidel}, C.~C., {Strom}, A.~L., {et~al.} 2019, \apj, 871,
  128, \dodoi{10.3847/1538-4357/aaf386}

\bibitem[{{Tinsley}(1980)}]{Tinsley1980}
{Tinsley}, B.~M. 1980, \fcp, 5, 287, \dodoi{10.48550/arXiv.2203.02041}

\bibitem[{{Topping} {et~al.}(2020{\natexlab{a}}){Topping}, {Shapley}, {Reddy},
  {Sanders}, {Coil}, {Kriek}, {Mobasher}, \& {Siana}}]{Topping+2020a}
{Topping}, M.~W., {Shapley}, A.~E., {Reddy}, N.~A., {et~al.}
  2020{\natexlab{a}}, \mnras, 495, 4430, \dodoi{10.1093/mnras/staa1410}

\bibitem[{{Topping} {et~al.}(2020{\natexlab{b}}){Topping}, {Shapley}, {Reddy},
  {Sanders}, {Coil}, {Kriek}, {Mobasher}, \& {Siana}}]{Topping+2020b}
---. 2020{\natexlab{b}}, \mnras, 499, 1652, \dodoi{10.1093/mnras/staa2941}

\bibitem[{{Topping} {et~al.}(2021){Topping}, {Shapley}, {Sanders}, {Kriek},
  {Reddy}, {Coil}, {Mobasher}, {Siana}, {Freeman}, {Shivaei}, {Azadi}, {Price},
  {Leung}, {Fetherolf}, {de Groot}, {Zick}, {Fornasini}, {Barro}, \&
  {Runco}}]{Topping+2021}
{Topping}, M.~W., {Shapley}, A.~E., {Sanders}, R.~L., {et~al.} 2021, \mnras,
  506, 1237, \dodoi{10.1093/mnras/stab1793}

\bibitem[{{Toribio San Cipriano} {et~al.}(2017){Toribio San Cipriano},
  {Dom{\'\i}nguez-Guzm{\'a}n}, {Esteban}, {Garc{\'\i}a-Rojas}, {Mesa-Delgado},
  {Bresolin}, {Rodr{\'\i}guez}, \&
  {Sim{\'o}n-D{\'\i}az}}]{ToribioSanCipriano+2017}
{Toribio San Cipriano}, L., {Dom{\'\i}nguez-Guzm{\'a}n}, G., {Esteban}, C.,
  {et~al.} 2017, \mnras, 467, 3759, \dodoi{10.1093/mnras/stx328}

\bibitem[{{Toribio San Cipriano} {et~al.}(2016){Toribio San Cipriano},
  {Garc{\'\i}a-Rojas}, {Esteban}, {Bresolin}, \&
  {Peimbert}}]{ToribioSanCipriano+2016}
{Toribio San Cipriano}, L., {Garc{\'\i}a-Rojas}, J., {Esteban}, C., {Bresolin},
  F., \& {Peimbert}, M. 2016, \mnras, 458, 1866, \dodoi{10.1093/mnras/stw397}

\bibitem[{{Torrey} {et~al.}(2018){Torrey}, {Vogelsberger}, {Hernquist},
  {McKinnon}, {Marinacci}, {Simcoe}, {Springel}, {Pillepich}, {Naiman},
  {Pakmor}, {Weinberger}, {Nelson}, \& {Genel}}]{Torrey+2018}
{Torrey}, P., {Vogelsberger}, M., {Hernquist}, L., {et~al.} 2018, \mnras, 477,
  L16, \dodoi{10.1093/mnrasl/sly031}

\bibitem[{{Trainor} {et~al.}(2015){Trainor}, {Steidel}, {Strom}, \&
  {Rudie}}]{Trainor+2015}
{Trainor}, R.~F., {Steidel}, C.~C., {Strom}, A.~L., \& {Rudie}, G.~C. 2015,
  \apj, 809, 89, \dodoi{10.1088/0004-637X/809/1/89}

\bibitem[{{Tremonti} {et~al.}(2004){Tremonti}, {Heckman}, {Kauffmann},
  {Brinchmann}, {Charlot}, {White}, {Seibert}, {Peng}, {Schlegel}, {Uomoto},
  {Fukugita}, \& {Brinkmann}}]{Tremonti+2004}
{Tremonti}, C.~A., {Heckman}, T.~M., {Kauffmann}, G., {et~al.} 2004, \apj, 613,
  898, \dodoi{10.1086/423264}

\bibitem[{{Troncoso} {et~al.}(2014){Troncoso}, {Maiolino}, {Sommariva},
  {Cresci}, {Mannucci}, {Marconi}, {Meneghetti}, {Grazian}, {Cimatti},
  {Fontana}, {Nagao}, \& {Pentericci}}]{Troncoso+2014}
{Troncoso}, P., {Maiolino}, R., {Sommariva}, V., {et~al.} 2014, \aap, 563, A58,
  \dodoi{10.1051/0004-6361/201322099}

\bibitem[{{Tsamis} {et~al.}(2003){Tsamis}, {Barlow}, {Liu}, {Danziger}, \&
  {Storey}}]{Tsamis+2003}
{Tsamis}, Y.~G., {Barlow}, M.~J., {Liu}, X.~W., {Danziger}, I.~J., \& {Storey},
  P.~J. 2003, \mnras, 338, 687, \dodoi{10.1046/j.1365-8711.2003.06081.x}

\bibitem[{{Vegetti} \& {Koopmans}(2009)}]{Vegetti+2009}
{Vegetti}, S., \& {Koopmans}, L.~V.~E. 2009, \mnras, 392, 945,
  \dodoi{10.1111/j.1365-2966.2008.14005.x}

\bibitem[{{Weiner} {et~al.}(2009){Weiner}, {Coil}, {Prochaska}, {Newman},
  {Cooper}, {Bundy}, {Conselice}, {Dutton}, {Faber}, {Koo}, {Lotz}, {Rieke}, \&
  {Rubin}}]{Weiner+2009}
{Weiner}, B.~J., {Coil}, A.~L., {Prochaska}, J.~X., {et~al.} 2009, \apj, 692,
  187, \dodoi{10.1088/0004-637X/692/1/187}

\bibitem[{{Weldon} {et~al.}(2022){Weldon}, {Reddy}, {Topping}, {Shapley},
  {Sanders}, {Du}, {Price}, {Kriek}, {Coil}, {Siana}, {Mobasher}, {Fetherolf},
  {Shivaei}, \& {Rezaee}}]{Weldon+2022}
{Weldon}, A., {Reddy}, N.~A., {Topping}, M.~W., {et~al.} 2022, \mnras, 515,
  841, \dodoi{10.1093/mnras/stac1822}

\bibitem[{{Xu} {et~al.}(2022){Xu}, {Heckman}, {Henry}, {Berg}, {Chisholm},
  {James}, {Martin}, {Stark}, {Aloisi}, {Amor{\'\i}n}, {Arellano-C{\'o}rdova},
  {Bordoloi}, {Charlot}, {Chen}, {Hayes}, {Mingozzi}, {Sugahara}, {Kewley},
  {Ouchi}, {Scarlata}, \& {Steidel}}]{Xu+2022}
{Xu}, X., {Heckman}, T., {Henry}, A., {et~al.} 2022, \apj, 933, 222,
  \dodoi{10.3847/1538-4357/ac6d56}

\bibitem[{{Yates} {et~al.}(2012){Yates}, {Kauffmann}, \& {Guo}}]{Yates+2012}
{Yates}, R.~M., {Kauffmann}, G., \& {Guo}, Q. 2012, \mnras, 422, 215,
  \dodoi{10.1111/j.1365-2966.2012.20595.x}

\bibitem[{{York} {et~al.}(2000){York}, {Adelman}, {Anderson}, {Anderson},
  {Annis}, {Bahcall}, {Bakken}, {Barkhouser}, {Bastian}, {Berman}, {Boroski},
  {Bracker}, {Briegel}, {Briggs}, {Brinkmann}, {Brunner}, {Burles}, {Carey},
  {Carr}, {Castander}, {Chen}, {Colestock}, {Connolly}, {Crocker}, {Csabai},
  {Czarapata}, {Davis}, {Doi}, {Dombeck}, {Eisenstein}, {Ellman}, {Elms},
  {Evans}, {Fan}, {Federwitz}, {Fiscelli}, {Friedman}, {Frieman}, {Fukugita},
  {Gillespie}, {Gunn}, {Gurbani}, {de Haas}, {Haldeman}, {Harris}, {Hayes},
  {Heckman}, {Hennessy}, {Hindsley}, {Holm}, {Holmgren}, {Huang}, {Hull},
  {Husby}, {Ichikawa}, {Ichikawa}, {Ivezi{\'c}}, {Kent}, {Kim}, {Kinney},
  {Klaene}, {Kleinman}, {Kleinman}, {Knapp}, {Korienek}, {Kron}, {Kunszt},
  {Lamb}, {Lee}, {Leger}, {Limmongkol}, {Lindenmeyer}, {Long}, {Loomis},
  {Loveday}, {Lucinio}, {Lupton}, {MacKinnon}, {Mannery}, {Mantsch}, {Margon},
  {McGehee}, {McKay}, {Meiksin}, {Merelli}, {Monet}, {Munn}, {Narayanan},
  {Nash}, {Neilsen}, {Neswold}, {Newberg}, {Nichol}, {Nicinski}, {Nonino},
  {Okada}, {Okamura}, {Ostriker}, {Owen}, {Pauls}, {Peoples}, {Peterson},
  {Petravick}, {Pier}, {Pope}, {Pordes}, {Prosapio}, {Rechenmacher}, {Quinn},
  {Richards}, {Richmond}, {Rivetta}, {Rockosi}, {Ruthmansdorfer}, {Sandford},
  {Schlegel}, {Schneider}, {Sekiguchi}, {Sergey}, {Shimasaku}, {Siegmund},
  {Smee}, {Smith}, {Snedden}, {Stone}, {Stoughton}, {Strauss}, {Stubbs},
  {SubbaRao}, {Szalay}, {Szapudi}, {Szokoly}, {Thakar}, {Tremonti}, {Tucker},
  {Uomoto}, {Vanden Berk}, {Vogeley}, {Waddell}, {Wang}, {Watanabe},
  {Weinberg}, {Yanny}, {Yasuda}, \& {SDSS Collaboration}}]{York+2000}
{York}, D.~G., {Adelman}, J., {Anderson}, John~E., J., {et~al.} 2000, \aj, 120,
  1579, \dodoi{10.1086/301513}

\bibitem[{{Yuan} \& {Kewley}(2009)}]{Yuan_Kewley2009}
{Yuan}, T.~T., \& {Kewley}, L.~J. 2009, \apjl, 699, L161,
  \dodoi{10.1088/0004-637X/699/2/L161}

\bibitem[{{Zahid} {et~al.}(2014){Zahid}, {Dima}, {Kudritzki}, {Kewley},
  {Geller}, {Hwang}, {Silverman}, \& {Kashino}}]{Zahid+2014}
{Zahid}, H.~J., {Dima}, G.~I., {Kudritzki}, R.-P., {et~al.} 2014, \apj, 791,
  130, \dodoi{10.1088/0004-637X/791/2/130}

\bibitem[{{Zahid} {et~al.}(2011){Zahid}, {Kewley}, \& {Bresolin}}]{Zahid+2011}
{Zahid}, H.~J., {Kewley}, L.~J., \& {Bresolin}, F. 2011, \apj, 730, 137,
  \dodoi{10.1088/0004-637X/730/2/137}

\bibitem[{{Zahid} {et~al.}(2017){Zahid}, {Kudritzki}, {Conroy}, {Andrews}, \&
  {Ho}}]{Zahid+2017}
{Zahid}, H.~J., {Kudritzki}, R.-P., {Conroy}, C., {Andrews}, B., \& {Ho}, I.~T.
  2017, \apj, 847, 18, \dodoi{10.3847/1538-4357/aa88ae}

\end{thebibliography}

\end{document}